\documentclass[a4paper,12pt]{article}
\usepackage{amssymb}

\usepackage[dvips]{graphicx}
\usepackage[usenames]{color}
\usepackage{amsmath}
\usepackage{fullpage}
\usepackage{boxedminipage}
\usepackage{listings}
\usepackage{minitoc}
\usepackage{wrapfig}
\usepackage{latexsym}

\makeatletter \@addtoreset{equation}{section} \makeatother

\input epsf

\begin{document}

\renewcommand{\[}{\begin{equation}} \renewcommand{\]}{\end{equation}} %
\renewcommand{\>}{\rangle}

\begin{titlepage}

    \thispagestyle{empty}
    \begin{flushright}
        \hfill{CERN-PH-TH/2008-002}\\\hfill{UCLA/08/TEP/11}
    \end{flushright}

    \vspace{5pt}
    \begin{center}
        { \Huge{\textbf{Erice Lectures\\\vspace{10pt}on Black Holes and Attractors}}}\vspace{25pt}
        \vspace{55pt}

         {\large{\bf Sergio Ferrara$^{\diamondsuit\clubsuit}$, Kuniko Hayakawa$^{\heartsuit\clubsuit}$ and\ Alessio Marrani$^{\heartsuit\clubsuit}$}}

        \vspace{15pt}

        {$\diamondsuit$ \it Physics Department,Theory Unit, CERN, \\
        CH 1211, Geneva 23, Switzerland\\
        \texttt{sergio.ferrara@cern.ch}}

        \vspace{10pt}

        {$\heartsuit$ \it Museo Storico della Fisica e\\
        Centro Studi e Ricerche ``Enrico Fermi"\\
        Via Panisperna 89A, 00184 Roma, Italy}

        \vspace{10pt}

        {$\clubsuit$ \it INFN - Laboratori Nazionali di Frascati, \\
        Via Enrico Fermi 40,00044 Frascati, Italy\\
        \texttt{hayakawa, marrani@lnf.infn.it}}

 \vspace{30pt}

        \noindent \textit{Contribution to the Proceedings of the International School of Subnuclear
        Physics,\\45th Course: Search for the ``Totally Unexpected" in the
        LHC era,\\Erice, Italy, 29 August -- 7 September 2007, Directors: G. 't Hooft -- A. Zichichi}
\end{center}

\vspace{30pt}

\begin{abstract}
These lectures give an elementary introduction to the subject of
four dimensional black holes (BHs) in supergravity and the Attractor
Mechanism in the extremal case. Some thermodynamical properties are
discussed and some relevant formul{\ae} for the critical points of
the BH effective potential are given. The case of
Maxwell-Einstein-axion-dilaton (super)gravity is discussed in
detail.

Analogies among BH entropy and multipartite entanglement of qubits
in quantum information theory, as well moduli spaces of extremal BH
attractors, are also discussed.
\end{abstract}

\end{titlepage}
\newpage\tableofcontents
\newpage

\section{Introduction}

The aim of the present lecture notes is to give an elementary introduction
to some aspects of black hole (BH) physics \cite{witt}--\nocite
{moore,duff1,hawking1,penrose}\cite{gibbons1} in $d=4$ space-time
dimensions, as well as an overview of some recent developments on the
so-called \textit{Attractor Mechanism} \cite{ferrara1}--\nocite{ferrara2}
\cite{strominger2} for extremal BHs, as they appear in $d=4$ supergravity
with a number $\mathcal{N}$ of local supersymmetry \cite{Sen-old1}--\nocite
{GIJT,Sen-old2,K1,TT,G,GJMT,Ebra1,K2,Ira1,Tom,
BFM,AoB-book,FKlast,Ebra2,bellucci1,rotating-attr,K3,Misra1,Lust2,Morales,
BFMY,Astefa,CdWMa,
DFT07-1,BFM-SIGRAV06,Cer-Dal,ADFT-2,Saraikin-Vafa-1,Ferrara-Marrani-1, TT2,
ADOT-1,ferrara4,CCDOP,Misra2,Astefanesei,Anber,Myung1,Ceresole,BMOS-1,Hotta,
Gao,
PASCOS07,Sen-review,Belhaj1,AFMT1,Gaiotto1,BFMS1,GLS1,ANYY1,bellucci2,Cai-Pang,Vaula, Li,BFMY2,Saidi2,Saidi3,Saidi4}
\cite{Valencia-RTN-07} (for further developments, see also \textit{e.g.}
\cite{OSV}--\nocite{OVV,ANV}\cite{GSV}).

Supergravity may be regarded as the low-energy limit (in a small curvature
expansion) of some candidates for a quantum theory of gravity such as
\textit{superstring theory} \cite{maldacena}--\nocite{schwarz1,schwarz2}\cite
{gasperini} or \textit{M-theory} \cite{witten,schwarz3}. In situations where
higher curvature effects may be neglected, the gravity part of the theory
reduces to the Einstein-Hilbert action coupled to a certain number of matter
fields, whose specific content depends on the particular low-energy theory.
Typically, these fields are (\textit{moduli}) scalars, spin $1/2$ fermions,
spin $1$ gauge fields (Abelian in our examples) and spin $3/2$ fermions, the
gravitinos. The latter ones, $\mathcal{N}$ in numbers, are the gauge fields
of local supersymmetry.

In this situation, asymptotically flat charged BH solutions, within a static
and spherically symmetric \textit{Ansatz}, can be regarded as a
generalization of the famous Schwarzschild BH. However, the presence of
additional quantum numbers (such as charges and scalar hair) make their
properties change drastically, and new phenomena appear. A novel important
feature of electrically (and/or magnetically) charged BHs \cite{nordstrom}
as well as rotating ones \cite{smarr} is a somewhat unconventional
thermodynamical property named \textit{extremality} (\textit{i.e.} zero
temperature, as we will see below) \cite{gibbons1,bellucci1,kallosh1}.
Extremal BHs are possibly stable gravitational objects with finite entropy
but vanishing temperature, in which case the contribution to the
gravitational energy entirely comes from the electromagnetic (charges) and
rotational \cite{rotating-attr} (angular momentum/spin) attributes. \textit{%
Extremality} also means that the inner (Cauchy) and outer (event) horizons
do coincide, thus implying vanishing surface gravity (see Sects. \ref{Sect2}
and \ref{Sect4}).

The extremal situation entails a particular relation between the entropy,
charges and spin, since in this case the gravitational mass is not an
independent quantity. Four dimensional stationary and spherically symmetric
BHs in an environment of scalar fields (typically described by a non-linear
sigma model), have scalar hair (\textit{scalar charges}), corresponding to
the values of the scalars at (asymptotically flat) spatial infinity. These
values may continuously vary, being an arbitrary point in the moduli space
of the theory or, in a more geometrical language, a point in the target
manifold of the scalar non-linear Lagrangian \cite{ferrara1,gibbons3}.
Nevertheless, the BH entropy, as given by the Bekenstein-Hawking
entropy-area formula \cite{hawking2}, is also in this case independent on
the scalar charges (\textit{``no scalar hair''}) and it only depends on the
asymptotic (generally \textit{dyonic}) BH charges (see Sect. \ref{Sect3}).

This apparent puzzle can be resolved thanks to the so-called \textit{%
Attractor Mechanism }(see Sect. \ref{Sect2}), a fascinating phenomenon that
combines extremal BHs, dynamical systems, algebraic geometry and number
theory \cite{moore}. It was firstly discovered in the context of
supergravity; in a few words, in constructing extremal dyonic BHs of $%
\mathcal{N}=2$, $d=4$ supergravity coupled to vector and hypermultiplets
(with no $d=4$ scalar potential ), two phenomena occur: the hyperscalars can
take arbitrary constant values, while the radial evolution of the vector
multiplets' scalars is described by a dynamical system \cite
{ferrara2,strominger2}. Under some mild assumptions, the scalar trajectory
flows to a ``fixed point'', located at the BH event horizon, in the target
(moduli) space. The ``fixed point'' (\textit{i.e.} a point of vanishing
\textit{phase velocity}) represents the system in equilibrium, and it is the
analogue of an \textit{attractor} in the dynamical flow of dissipative
systems. In approaching such an \textit{attractor}, the orbits lose
practically all memory of initial conditions (\textit{i.e.} of the scalar
hair), even though the dynamics is fully deterministic. The scalars at the
BH horizon turn out to depend only on the dyonic (asymptotic) charges.

For $\frac{1}{2}-$BPS (\textit{i.e.} supersymmetric), $\mathcal{N}=2$
attractors all the scalars are fixed \cite{ferrara3}, and one deals with the
\textit{attractor varieties}, which have their own interest in algebraic
geometry and number theory \cite{moore}. For the so-called ``large'' BHs,
within the Einstein approximation, the entropy can be shown to be
proportional to a suitably defined \textit{BH effective potential }(positive
definite function in the moduli space) computed (for fixed BH electric and
magnetic charges) at its critical point(s), reached at the horizon. All
extremal static, spherically symmetric and asymptotically flat BHs in $d=4$
have a Bertotti-Robinson \cite{bertotti} $AdS_{2}\times S^{2}$ near-horizon
geometry, with vanishing scalar curvature and conformally flat; in
particular, the radius of $AdS_{2}$ coincides with the radius of $S^{2}$,
and it is proportional to the (square root of the) BH entropy (in turn
proportional, through the Bekenstein-Hawking formula \cite{hawking2}, to the
area of the event horizon). Non-BPS (\textit{i.e.} non-supersymmetric) (see
\textit{e.g.} \cite{bellucci1,TT2,ferrara4,Gaiotto1,GLS1,ANYY1}) extremal
BHs exist as well, and they also exhibit an attractor behavior. However, in
this case not all vector multiplets' scalars are stabilized in terms of BH
conserved charges at the event horizon, but rather some of them remain, at
least at the classical level, ``flat'' directions, as is the case for
hyperscalars \cite{ferrara4}. In spite of this, the entropy of non-BPS BHs
enjoys the same property of their supersymmetric counterparts, namely it
only depends on the dyonic BH charges \cite{TT2,ferrara4}.

When the scalar manifold is an homogeneous symmetric space \cite
{bellucci1,FG} (as it is always the case for all $\mathcal{N}>2$, $d=4$
extended supergravities), the theory of extremal BH attractors has a
beautiful connection to group theory and differential geometry. In this
framework, the BPS or non-BPS nature of BH attractors can be related to the
theory of orbits of the dyonic (asymptotic) charge vector \cite
{FG,FG2,Ferrara-Maldacena}: different orbits correspond to different
supersymmetry-preserving features of the fixed points of the scalar
dynamics. All ``non-flat'' directions are \textit{attractive}, which means
that the Hessian matrix the BH effective potential is semi-positive definite
\cite{bellucci2,ADFT}.

All issues mentioned above will be reviewed in Sects. \ref{Sect5}-\ref{Sect8}%
. Thence, in Sect. \ref{Sect9} we report on some stunning relations recently
found between pure states of multipartite entanglement of qubits in quantum
information theory and extremal BH in superstring theory \cite{duff1}. A
particularly striking example of a tripartite entanglement of seven qubits,
related to the (particular non-compact, real form of the) exceptional Lie
group $E_{7(7)}$, the octonions and the Fano plane is considered in detail.
The final Sect. \ref{Sect10} deals with some recent results on the
classifications of extremal BH attractors in $\mathcal{N}>2$-extended, $d=4$
supergravity, in which case ``flat'' directions of the effective BH
potential in the target moduli-space appear for both $\frac{1}{\mathcal{N}}$%
-BPS and non-BPS attractors. \setcounter{equation}0

\section{\label{Sect2}Attractor Mechanism and the Bekenstein-Hawking
Entropy-Area formula}

Extremal BHs with electric and magnetic charges and scalar hair are
solitonic objects whose scalar degrees of freedom describe trajectories (in
the radial evolution parameter) with \textit{fixed points} (within the
corresponding \textit{basin of attraction}) \cite
{ferrara1,ferrara2,strominger2}:

\begin{equation}
\left\{
\begin{array}{l}
\lim_{r\rightarrow r_{H}^{+}}\phi ^{a}(r)=\phi _{H}^{a}(p,q)\equiv \phi
_{fix}^{a}; \\
\\
\lim_{r\rightarrow r_{H}^{+}}\frac{d\phi ^{a}(r)}{dr}=0.
\end{array}
\right.
\end{equation}
The orbits lose all memory of the initial conditions (\textit{i.e.} the
asymptotic values $\phi _{\infty }^{a}\equiv \lim_{r\rightarrow \infty }\phi
^{a}(r)$), and the fixed (attractor) point $\phi _{H}^{a}(p,q)$ only depends
on the BH charges $p$, $q$.

\begin{figure}[h!]
\caption{\textbf{Realization of the Attractor Mechanism in the $\frac{1}{2}$%
-BPS\ dilatonic BH} \protect\cite{ferrara2,kallosh1}. Independently on the
set of asymptotical ($r\rightarrow \infty $) scalar configurations, the
near-horizon evolution of the dilatonic function $e^{-2\protect\phi }$
converges towards a fixed \textit{attractor} value, which is purely
dependent on the (ratio of the) quantized conserved charges of the BH.
\protect\cite{ferrara2}}
\begin{center}
\includegraphics[width=0.48\textwidth,height=0.3\textheight]{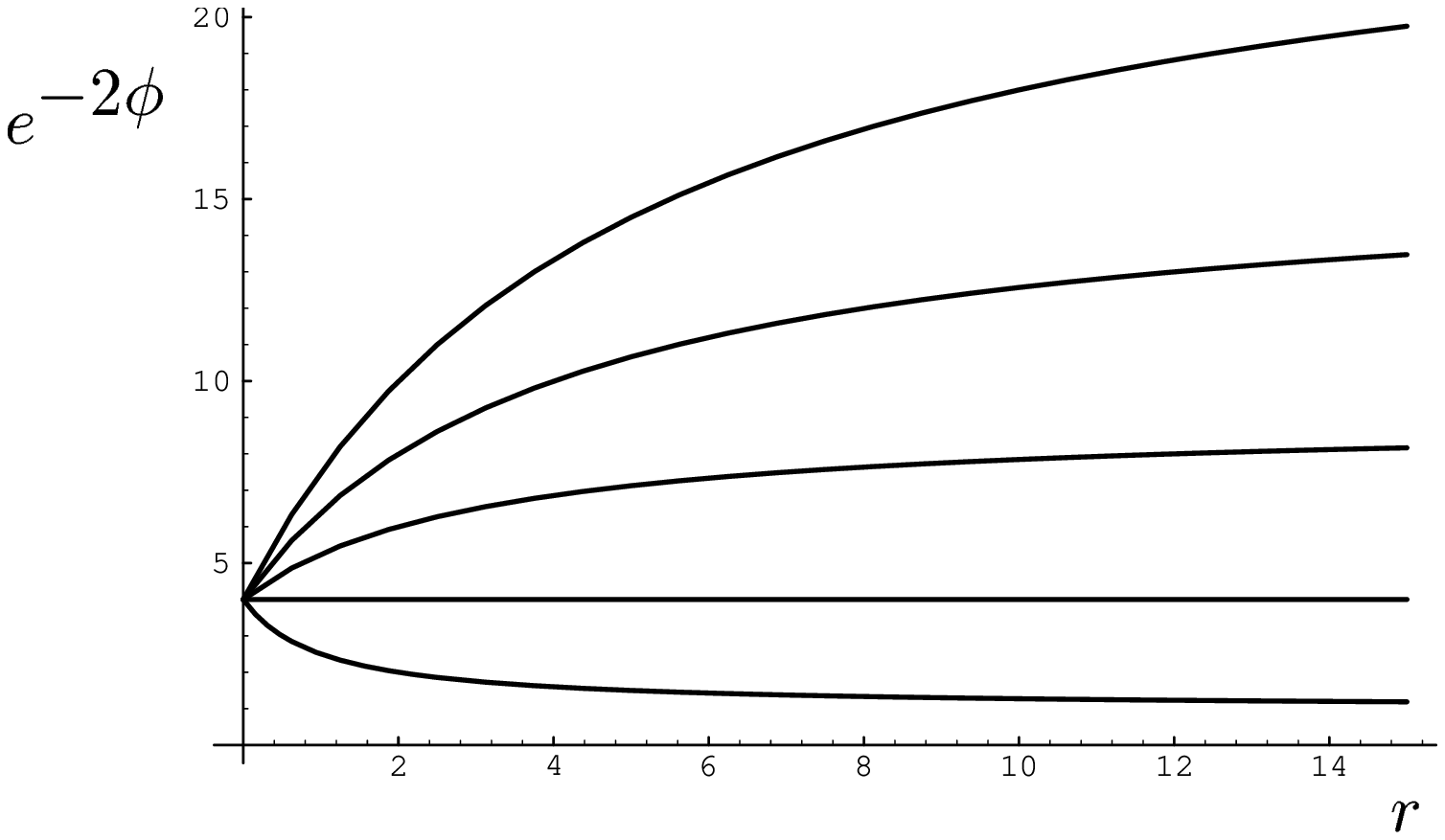}
\end{center}
\par
\end{figure}

\begin{figure}[h!]
\begin{center}
\includegraphics[width=0.48\textwidth,height=0.3\textheight]{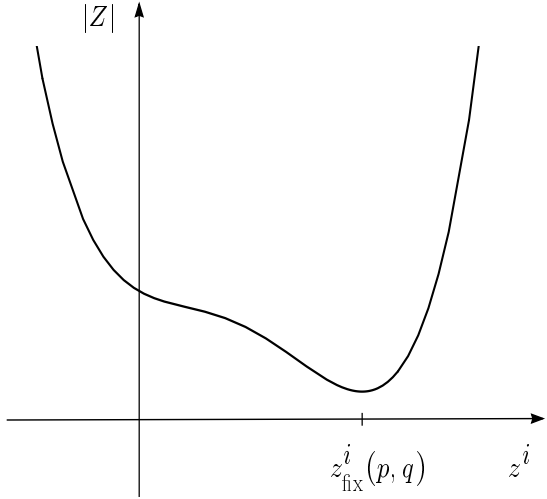}
\end{center}
\par
\caption{\textbf{Minimization of the absolute value of the ``central
charge'' function $Z$ in $\mathcal{M}$.} In the picture $z_{fix}^{i}\left(
p,q\right) $ stands ~for ~the ~\textit{attractor}, purely charge-dependent
value of the scalars at the event horizon of the considered $\frac{1}{2}$%
-BPS\ extremal BH. The Attractor Mechanism fixes the extrema of the central
charge to correspond to the discrete \textit{fixed} points of the
corresponding \textit{attractor variety} \protect\cite{moore}. Of course,
the dependence of the central charge on scalars is shown for a given
supporting BH charge configuration. \protect\cite{ferrara2}}
\end{figure}

The trajectories are solutions of the field equations derived from a
one-dimensional effective Lagrangian of almost-geodesic form, whose
potential (computed at spatial infinity) is related to the
Arnowitt-Deser-Misner (ADM) gravitational mass \cite{arnowitt,ferrara3} of
the BH (see Eq. (\ref{Sun-1}) below). The entropy, as given by the
Bekenstein-Hawking entropy-area formula \cite{hawking2}

\begin{equation}
S_{BH}=\frac{A_{H}}{4}=\pi V_{BH}(\phi _{H}(p,q),p,q),  \label{sunday0}
\end{equation}
is the value of the \textit{effective BH potential} $V_{BH}$ \cite{ferrara3}
at the fixed attractor point $\phi _{H}(p,q)$ at the BH event horizon. Note
that $S_{BH}$ is independent on $\phi _{\infty }^{a}$ (\textit{initial data}
of the radial scalar dynamics), which are continuous parameters. This is
reasonable in view of the microscopic interpretation of $S_{BH}$ as derived
from a microstate counting \cite{strominger1} (see also \cite
{maldacena,Sen-review}):
\begin{equation}
S_{BH}\sim \ln N_{ms},
\end{equation}
$N_{ms}$ being the number of microscopic states realizing the considered
macroscopic BH configuration. In general, \textit{within the Ansatz of
spherical symmetry}, the Bekenstein-Hawking entropy-area formula reads

\begin{equation}
S_{BH}=\frac{A_{H}}{4}\equiv \pi R_{+}^{2},  \label{sunday1}
\end{equation}
where the \textit{effective radius} $R_{+}$ was introduced.

Notice that, in presence of scalars and/or for angular momentum $J\neq 0$
(which is constant, in the stationary rotating regime), $A_{H}$ is an
\textit{effective} quantity, \textit{i.e.} (within the spherical symmetry
\textit{Ansatz}) it is not given by $4\pi r_{+}^{2}$ (where $r_{+}$ is the
radius of the - \textit{outer} - \textit{event} horizon), but rather by $\pi
R_{+}^{2}$. For instance, in the static ($J=0$) \textit{%
Maxwell-Einstein-(axion-)dilaton} BH (\cite{K3}, treated in Sects. \ref
{Sect6} and \ref{Sect7}) it holds (both in \textit{non-extremal} and \textit{%
extremal} cases; see below, as well as Eq. (72) of the first of Refs. \cite
{K3}) that
\begin{equation}
R_{+}^{2}=\left. R^{2}\right| _{r=\Sigma }^{r=r_{+}}=r_{+}^{2}-\Sigma
^{2}\leqslant r_{+}^{2},  \label{Lun-1}
\end{equation}
where $\Sigma $ denotes the \textit{dilaton scalar charge}, and the \textit{%
(squared) effective radial coordinate} \cite{K3}
\begin{equation}
R^{2}\equiv r^{2}-\Sigma ^{2}\leqslant r^{2}  \label{CERN-sunny-1}
\end{equation}
was introduced (see also Eqs. (\ref{CERN-solo-1}) and (\ref{CERN-solo-2})
further below).

Eq. (\ref{Lun-1}) can be generalized to the \textit{multi-dilaton system
(i.e. }to $\mathcal{N}=2$, $d=4$ ungauged supergravity \textit{minimally
coupled} to a number $n_{V}$ of Abelian vector multiplets \cite{Luciani}),
where (for asymptotically flat, spherically symmetric, static BHs) the
\textit{squared} \textit{effective radius} reads
\begin{equation}
R_{+}^{2}=\left. R^{2}\right| _{r=\sqrt{\frac{1}{2}G_{ab}\left( \phi
_{\infty }\right) \Sigma ^{a}\Sigma ^{b}}}^{r=r_{+}}=r_{+}^{2}-\frac{1}{2}%
G_{ab}\left( \phi _{\infty }\right) \Sigma ^{a}\Sigma ^{b}\leqslant
r_{+}^{2},  \label{sunday-Night-2}
\end{equation}
where $G_{ab}$ and $\Sigma ^{a}$ respectively are the (non-singular) metric
of the scalar manifold and the so-called \textit{scalar charges} (see
definition (\ref{monday5}) below); in this case the \textit{(squared)}
\textit{effective radial coordinate }$R$ can be defined as
\begin{equation}
R^{2}\equiv r^{2}-\frac{1}{2}G_{ab}\left( \phi _{\infty }\right) \Sigma
^{a}\Sigma ^{b}\leqslant r^{2},  \label{CERN-sunny-2}
\end{equation}
which is nothing but the \textit{many-moduli generalization} of Eq. (\ref
{CERN-sunny-1}) (see Eq. (\ref{CERN-solo-3}) below). In the \textit{extremal}
case ($c=0$; see Eq. (\ref{corr1}) below) Eq. (\ref{sunday-Night-2}) yields
the $U$-duality invariant $\mathcal{I}_{2}$, \textit{quadratic} in the
electric and magnetic BH charges, and \textit{independent} on the scalar
fields (see Sect. \ref{Sect7}). For examples showing how $J\neq 0$ affects
the expression of the \textit{effective radius} in the \textit{extremal}
case, see Sect. \ref{Sect8} and Refs. therein).

Furthermore, the temperature of the BH is given by the following formula
\cite{hawking2}:
\begin{equation}
T_{BH}=\frac{\kappa }{{2}\pi },  \label{sunday2}
\end{equation}
where
\begin{equation}
\kappa \equiv -\frac{1}{2}\left[ \nabla {_{\mu }}\xi {_{\nu }}\nabla {^{\mu }%
}\xi {^{\nu }}\right] _{r=R_{+}}^{1/2}=\frac{{r_{0}}}{R_{+}^{2}}%
,~~~~r_{0}\equiv {\frac{1}{2}(}r{_{+}-}r{_{-}),\label{sunday2-bis}}
\end{equation}
is the so-called \textit{surface gravity}, and $r_{-}$, $\nabla {_{\mu }}$
and $\xi {_{\nu }}$ respectively denote the radius of the inner (Cauchy) BH
horizon, the Christoffel covariant derivatives and a Killing vector, which
for any static metric simply reads (see the first of \cite{K3}, and Refs.
therein)

\begin{equation}
\xi {_{\nu }}=\delta _{\nu \,t}g_{tt}(x),
\end{equation}
$x$ standing for the spatial coordinates. Eqs. (\ref{sunday1}) and (\ref
{sunday2}) yield\footnote{%
Eq. (\ref{corr1}) fixes a typo in \cite{ferrara3} and in the whole treatment
of \cite{AoB-book} and \cite{ADFT}, where $c^{2}\equiv 2S_{BH}T_{BH}$.}
\begin{equation}
r_{0}=2S_{BH}T_{BH}\equiv c.  \label{corr1}
\end{equation}
$c$ is the so-called \textit{extremality parameter}. It vanishes for \textit{%
extremal} BHs, which indeed have $r_{+}=r_{-}\equiv r_{H}$ (the Cauchy and
event BH horizons do coincide). The \textit{extremal} limit ($c=0$) of Eqs. (%
\ref{sunday1}), (\ref{Lun-1}) and (\ref{sunday-Night-2}) respectively reads
\begin{eqnarray}
S_{BH} &=&\frac{A_{H}}{4}\equiv \pi R_{H}^{2};  \label{CERN-sunny-3} \\
&&  \notag \\
R_{H}^{2} &=&\left. R^{2}\right| _{r=\Sigma }^{r=r_{H}}=r_{H}^{2}-\Sigma
^{2}\leqslant r_{H}^{2};  \label{CERN-sunny-4} \\
&&  \notag \\
R_{H}^{2} &=&\left. R^{2}\right| _{r=\sqrt{\frac{1}{2}G_{ab}\left( \phi
_{\infty }\right) \Sigma ^{a}\Sigma ^{b}}}^{r=r_{H}}=r_{H}^{2}-\frac{1}{2}%
G_{ab}\left( \phi _{\infty }\right) \Sigma ^{a}\Sigma ^{b}\leqslant
r_{H}^{2},
\end{eqnarray}
where the notation $R_{+,c=0}^{2}\equiv R_{H}^{2}$ has been introduced, and
will be used throughout the following treatment.

In the case in which $S_{BH}\neq 0$ (which, through Eqs. (\ref{sunday0}) and
(\ref{sunday1}), expresses the non-vanishing of $A_{H}$, and thus the
absence of a \textit{naked singularity}), it necessarily implies $T_{BH}=0$.
Thus, \textit{large} (\textit{i.e.} with non-vanishing $A_{H}$) extremal BHs
necessarily have $T_{BH}=0$ and they are thermodynamically \textit{stable}.
\setcounter{equation}0

\section{\label{Sect3}Black Holes and their Horizon Geometry}

Extremal BHs can be described as supersymmetric solitons \cite{stelle,duff2}
which interpolate between maximally symmetric geometries at spatial infinity
and at the BH event horizon \cite{gib-town}. Concerning the near-horizon
geometry of asymptotically flat, extremal $p$-(black) branes in $d$
space-time dimensions, it holds that \cite{gib-town}
\begin{equation}
\lim_{r\rightarrow r_{H}^{+}}ds^{2}=AdS_{p+2}\times S^{d-p-2}=\frac{SO(p+1,2)%
}{SO(p+1,1)}\times \frac{SO(d-p-1)}{SO(d-p-2)}\equiv ds_{H}^{2}.
\end{equation}
The related isometries at the various relevant regimes are as follows:

\begin{itemize}
\item  horizon isometry ($r\rightarrow r_{H}^{+}$): $SO(p+1,2)\times
SO(d-p-1)$;

\item  generic isometry ($r_{H}<r<\infty $): $ISO(p,1)\times SO(d-p-1)$;

\item  asymptotic isometry ($r\rightarrow \infty $) : $ISO(d-1,1)$,

where $ISO(n,1)$ denotes the inhomogeneous Lorentz group (\textit{i.e.} the
Poincar\'{e} group) in $n+1$ dimensions.

Some relevant examples are

\item  $p=3$, $d=10$: $ds_{H}^{2}$ $=$ $AdS_{5}$ $\times $$S^{5}$
(Maldacena's $AdS/CFT$ background), with isometry
\begin{equation}
SO(4,2)\times SO(6)\sim SU(2,2)\times SU(4)\sim Spin\left( 4,2\right) \times
Spin\left( 6\right) ;
\end{equation}

\item  $p=0$, $d=4$: $ds_{H}^{2}$ $=$ $AdS_{2}$ $\times $$S^{2}$
(Bertotti-Robinson (BR) BH metric), with isometry
\begin{equation}
SO(1,2)\times SO(3)\sim SU(1,1)\times SU(2)\sim Spin\left( 1,2\right) \times
Spin\left( 3\right) ;
\end{equation}

\item  $p=0,1$, $d=5$: $ds_{H,p=0}^{2}$ $=$ $AdS_{2}$ $\times $$S^{3}$
(Tangherlini BH metric), $ds_{H,p=1}^{2}$ $=$ $AdS_{3}$ $\times $$S^{2}$ (%
\textit{black string}), with isometry
\begin{equation}
\begin{array}{l}
p=0:SO(1,2)\times SO(4)\sim SU(1,1)\times SU(2)\times SU(2)\sim Spin\left(
1,2\right) \times Spin\left( 4\right) , \\
\\
p=1:SO(2,2)\times SO(3)\sim \left( SU(1,1)\right) ^{2}\times SU(2)\sim
Spin\left( 2,2\right) \times Spin\left( 3\right) ;
\end{array}
\end{equation}

\item  $p=1$, $d=6$: $ds_{H}^{2}$ $=$ $AdS_{3}$ $\times $$S^{3}$ (\textit{%
self-dual dyonic black string}), with isometry
\begin{equation}
SO(2,2)\times SO(4)\sim \left( SU(1,1)\right) ^{2}\times SU(2)\times
SU\left( 2\right) \sim Spin\left( 2,2\right) \times Spin\left( 4\right) .
\end{equation}

Notice that, introducing rotational degrees of freedom (in the stationary
regime) affects the near-horizon geometry. For example $p=0$, $d=4$ with a
constant angular momentum $J\neq 0$ yields an horizon geometry $AdS_{2}$ $%
\times $$S^{1}$, with isometry $SO(1,2)\times SO(2)$.
\end{itemize}

Considering regular space-times with \textit{physically reasonable} matter,
\textit{i.e.} with matter whose stress-energy tensor satisfies the so-called
\textit{dominant energy condition} \cite{hawking3}
\begin{equation}
T_{\mu \nu }^{matter}U^{\mu }V^{\nu }\geqslant 0
\end{equation}
for any pair of not space-like vector $U^{\mu }$ and $V^{\nu }$, the
long-standing conjecture that they must have positive ADM or Bondi \cite
{arnowitt,deser} mass $M_{ADM}\equiv M$ has been proven in the 80's by using
spinor techniques suggested by supergravity (see e.g. \cite
{hawking3,deser,nester}, as well as \cite{gibbons2} and Refs. therein):
\begin{equation}
M\geqslant 0\newline
,  \label{sunday!1}
\end{equation}
\newline
holding for asymptotically flat space-times, and with $M_{ADM}=0$ \textit{%
only in the case of global flatness}.

The bound (\ref{sunday!1}) implies that the Schwarzschild (Schw) BH,
described by the metric\footnote{%
We use \textit{Planck units} in which the Newton gravitational constant $G$,
the speed of light in vacuum $c$, the Boltzmann constant $K_{B}$ and the
Planck constant $\hbar $ are all put equal to $1$. Moreover, $M$ is shortcut
for $M_{ADM}$, and $d\Omega {^{2}=}d\theta {^{2}+}sin{^{2}}\theta d\phi {^{2}%
}$ throughout.}

\begin{equation}
ds_{Schw}^{2}=-\,\left( {1-\frac{{2}M}{r}}\right) dt^{2}+\left( {1-\frac{{2}M%
}{r}}\right) ^{-1}dr^{2}+r^{2}d\Omega ^{2}  \label{Schw1}
\end{equation}
has no \textit{naked singularity}, \textit{i.e.} the singularity at $r=0$ is
\textit{covered} by the event horizon at $r_{H}=2M$. Indeed, for
Schwarzschild BH $g_{tt}$ $=0$ at $r=r_{H}$ and $g_{tt}<0$ inside the
horizon ($0<r<r_{H}$). This means that the light-cone is inward and a light
signal, emitted inside the horizon can never reach an outside observer.
Thus, no communication is possible between the region $r<r_{H}$ and the
region $r>r_{H}$, so that the physical singularity at $r=0$ is \textit{%
covered} by the BH event horizon. It is here worth observing that this
phenomenon in General Relativity is a consequence of the necessity of
extending the Newtonian potential to a relativistic theory of gravity. The
BH interpretation is thus a necessary outcome of such an extension.

By defining the Nester antisymmetric tensor (two-form) \cite{nester}
\begin{equation}
E^{\alpha \sigma }\equiv 2\left( \overline{\nabla {_{\beta }}\epsilon }%
\Gamma {^{\sigma \alpha \beta }}\epsilon {-}\overline{\epsilon }\Gamma {%
^{\sigma \alpha \beta }}\nabla {_{\beta }\epsilon }\right) ,
\label{sunday!3}
\end{equation}
where $\Gamma {^{\sigma \alpha \beta }=g}^{\beta \zeta }{g}^{\alpha \xi
}\Gamma _{\zeta \xi }^{~~\sigma }$ is the Christoffel connection, and
following \cite{nester,hawking3,deser} and \cite{gibbons2}, the application
of the Gauss law yields (see \textit{e.g.} \cite{gibbons3}):

\begin{equation}
M=P_{\Lambda }u^{\Lambda }=-\frac{1}{2}\int_{S=\partial \Sigma }E{^{\sigma
\alpha }}dS{_{\sigma \alpha }}=\int_{\Sigma }{\nabla _{\alpha }}E{^{\alpha
\sigma }}d\Sigma {_{\sigma }}\Rightarrow \;M\geqslant 0.  \label{sunday!2}
\end{equation}
\newline
As shown in \cite{gibbons2} the generalization of the bound (\ref{sunday!2})
to include BHs with electric and magnetic charges $p$ and $q$ is nothing but
the so-called \textit{BPS}(-like) \textit{bound}

\begin{equation}
M\geqslant \left( q{^{2}+}p{^{2}}\right) ^{1/2}\newline
,  \label{sunday!1-gen}
\end{equation}
holding under the covariant generalization of the \textit{dominant energy
condition} \cite{gibbons2}

\begin{equation}
T_{\mu \nu }^{matter}U^{\mu }V^{\nu }\geqslant \left[ {\left( J{_{\alpha
}^{e}}V{^{\alpha }}\right) ^{2}+\left( J{_{\alpha }^{m}}V{^{\alpha }}\right)
^{2}}\right] ^{1/2}  \label{sunday-eve1}
\end{equation}
where $J{_{\alpha }^{e}}$ and $J{_{\alpha }^{m}}$ respectively are the
electric and magnetic current vectors. The physical meaning of condition (%
\ref{sunday-eve1}) is the requirement that the local charge density does not
exceed the local matter density. The bound (\ref{sunday!1-gen}),
generalization of the bound (\ref{sunday!1}), is the so-called
Bogomol'ny-Prasad-Sommerfeld (BPS) bound, and it is saturated iff a Dirac
spinor field $\epsilon $ exists satisfying the covariant constancy condition
\cite{gibbons2}
\begin{equation}
{\hat{\nabla}_{\mu }\epsilon \equiv \nabla _{\mu }\epsilon -}\frac{1}{4}%
F_{\alpha \beta }\gamma ^{\alpha }\gamma ^{\beta }\gamma _{\mu }\epsilon {=0,%
\label{as1}}
\end{equation}
thus making the supercovariant version of the Nester tensor \cite{gibbons2}
\begin{equation}
\widehat{E}^{\alpha \sigma }\equiv 2\left( \overline{{\hat{\nabla}_{\beta }}%
\epsilon }\Gamma {^{\sigma \alpha \beta }}\epsilon {-}\overline{\epsilon }%
\Gamma {_{\beta }^{\sigma \alpha \beta }\hat{\nabla}\epsilon }\right)
\end{equation}
vanish. Notice that ${\hat{\nabla}_{\mu }\epsilon }$ is the gravitino
supersymmetry variation; thus, the vanishing condition (\ref{as1}) implies
the corresponding geometric background to preserve some supersymmetry.

Let us now move to consider the Reissner-Nordstr\"{o}m (RN) electrically
charged\footnote{%
The generalization to include a magnetic charge $p$ can be straightforwardly
implemented by performing the shift $q^{2}\rightarrow q^{2}+p^{2}$.} BH,
described by the metric

\begin{equation}
ds_{RN}^{2}=-\;\left( {1-\frac{{2}M}{r}+\frac{q{^{2}}}{r{{}^{2}}}}\right)
dt^{2}+\;\left( {1-\frac{{2}M}{r}+\frac{q{^{2}}}{r{{}^{2}}}}\right)
^{-1}dr^{2}+r^{2}d\Omega ^{2}.  \label{RN-RN-1}
\end{equation}
Such a metric exhibits two horizons, with radii
\begin{equation}
r_{\pm }=M\pm \sqrt{M^{2}-q^{2}}=M\pm r_{0}.  \label{sunday-Night-1}
\end{equation}
In the \textit{neutral limit} $q=0$ the RN metric (\ref{RN-RN-1}) reduces to
the Schwarzchild one, given by Eq. (\ref{Schw1}). On the other hand, in the
same limit Eq. (\ref{sunday-Night-1}) yields $r_{-}=0$ (\textit{physical
singularity}) and $r_{+}=2M$, which is usually named \textit{Schwarzchild BH
event horizon}, and noted, with a slight abuse of notation, as
\begin{equation}
r_{+,Schw}=r_{H}=R_{H}.  \label{CERN-mart-1}
\end{equation}

The well definiteness of both the event horizon ($r_{+}$) and Cauchy horizon
($r_{-}$) radii requires the condition $M^{2}\geqslant q^{2}$, which is
nothing but the \textit{BPS bound} (\ref{sunday!1-gen}) for $p=0$, and it is
the implementation of the \textit{Cosmic Censorship Principle}.

In the extremal case \cite{gibbons1}
\begin{equation}
c=0\Leftrightarrow r_{+}=r_{-}\equiv r_{H}=M=R_{H}\Leftrightarrow M{=\left|
q\right| }.  \label{sunday-eve2}
\end{equation}
\newline
Thus, for RN BHs the \textit{extremality is a necessary and sufficient
condition for the saturation of the BPS bound (\ref{sunday!1-gen})}. In
other words, for RN BHs \textit{extremality}$\Leftrightarrow $\textit{%
supersymmetry}\footnote{%
Recently, the name \textit{``BPS''} has been associated to BHs admitting a
first-order (thus \textit{BPS-like}) formulation. In such a generalized
sense, also extremal BHs not preserving any supersymmetry or certain
non-extremal BHs (\textit{e.g.} the ones with frozen scalars, or with no
scalars at all, such as RN ones), can be called \textit{''BPS''} (\cite
{Miller}; see also \cite{Cer-Dal,ADOT-1}\textbf{).}}. However, as it will be
clear from the treatment below, this is no more the case in presence of
scalars. It is here also worth pointing out that, as yielded by the \textit{%
neutral limit} $q=0$ of Eqs. (\ref{sunday-eve2}) and (\ref{sunday-Night-1}),
the \textit{extremality parameter} for Scharzchild BH is nothing but its
mass $M$:
\begin{equation}
c_{Schw}=M,  \label{CERN-mart-2}
\end{equation}
and thus \textit{extremal} Schwarzchild BHs are necessarily \textit{``small''%
}, \textit{i.e.} with vanishing Bekenstein-Hawking entropy.

Anyway, (\textit{at least}) for static, spherically symmetric,
asymptotically flat, dyonic, \textit{extremal} BHs (\textit{i.e.} for the
class we consider in the present lectures, excluding the treatment of Sect.
\ref{Sect8}) the BH event horizon radius $r_{H}$ and the ADM mass $M_{\left(
ADM\right) }$ do coincide:
\begin{equation}
r_{H}=M.  \label{LNF-1}
\end{equation}

When Eqs. (\ref{sunday-eve2}) and (\ref{LNF-1}) hold, by defining
\begin{equation}
\rho \equiv r-M=r-r_{H}  \label{Thu-3}
\end{equation}
one obtains
\begin{equation}
1-\frac{M}{r}=\left( {1+\frac{M}{\rho }}\right) ^{-1},
\end{equation}
and thus the extremal RN metric can be rewritten as follows:
\begin{equation}
ds_{RN,c=0}^{2}=-\,\left( {1+\frac{M}{\rho }}\right) ^{-2}dt^{2}+\left( {1+%
\frac{M}{\rho }}\right) ^{2}\left( {d\rho ^{2}+\rho ^{2}d\Omega ^{2}}\right)
.=-\,e^{2U}dt^{2}+e^{-2U}d\vec{x}^{2},  \label{extr-RN-1}
\end{equation}
In such a coordinate system, the relevant regimes of the metric reads

\begin{itemize}
\item  near-horizon: $r\rightarrow r_{H}^{+}$ ${\Leftrightarrow }{\rho
\rightarrow 0}^{+}$;

\item  physical singularity: ${r\rightarrow 0}^{+}{\Leftrightarrow \rho
\rightarrow -M}^{+}$;

\item  Asymptotic regime: ${r\rightarrow \infty \Leftrightarrow \rho
\rightarrow \infty }$.

Moreover, $ds_{RN,c=0}^{2}$ acquires the general static Papapetrou-Majumdar
\cite{papapetrou} form
\begin{equation}
ds^{2}=-\,e^{2U}dt^{2}+e^{-2U}d\vec{x}^{2},  \label{papa1}
\end{equation}
\newline
with $U=U\left( \overrightarrow{x}\right) $ satisfying the $3$-d. D'Alembert
equation
\begin{equation}
\Delta e^{-U\left( \overrightarrow{x}\right) }=0.  \label{papa2}
\end{equation}
The general class of (static) Papapetrou, Majumdar metrics is known to admit
a covariantly constant spinor \cite{tod}. Thus, extremal RN BH metric (\ref
{extr-RN-1}), for which
\begin{equation}
U\left( \overrightarrow{x}\right) =U\left( r\right) =-\ln \left( {1+\frac{M}{%
\rho }}\right) =\ln \left( {1-\frac{M}{r}}\right) =\ln \left( {1-\frac{%
\left| q\right| }{r}}\right) ,  \label{sunday-finn1}
\end{equation}
is a maximally supersymmetric background of $\mathcal{N}=2$, $d=4$
supergravity (see \textit{e.g.} \cite{DFF}), preserving $4$ out of the $8$
supersymmetries pertaining to the asymptotic $d=4$ Minkowski background.
Thus, the extremal RN BH is a $\frac{1}{2}$-BPS solution of $\mathcal{N}=2$,
$d=4$ supergravity.
\end{itemize}

As for the Schwarzschild case, also for the RN BH the physical singularity
at $r=0$ is covered by an event horizon. Let us consider the radial geodesic
dynamics of a point-like massless probe falling into the RN BH; in the
reference frame of a distant observer, such a massless probe will travel
from a radius $r_{0}$ to a radius $r$ (both bigger than $r_{+}$) in a time
given by the following formula \cite{gibbons1}:
\begin{equation}
\Delta t(r)=\int_{r_{0}}^{r}\frac{dt}{dr}dr=\int_{r_{0}}^{r}\sqrt{\frac{%
g_{rr}}{g_{tt}}}dr\rightarrow \infty \text{ for }r\rightarrow r_{+}\text{.}
\end{equation}

In order to determine the near-horizon geometry of an extremal RN BH, let us
define a new radial coordinate as $\tau =-\frac{1}{\rho }=\frac{1}{r_{H}-r}$%
. Thus, the relevant regimes of the metric now reads

\begin{itemize}
\item  near-horizon: $r\rightarrow r_{H}^{+}$ ${\Leftrightarrow \tau
\rightarrow -\infty }$;

\item  physical singularity: ${r\rightarrow 0}^{+}{\Leftrightarrow \tau
\rightarrow }\frac{{1}}{M}^{+}$;

\item  asymptotic regime: ${r\rightarrow \infty \Leftrightarrow \tau
\rightarrow 0}^{-}$.

In the near-horizon limit $\rho \rightarrow 0^{+}$, and thus $1+\frac{M}{%
\rho }\sim \frac{M}{\rho }$; consequently, from Eq. (\ref{extr-RN-1}) one
obtains
\begin{equation}
\lim_{\rho \rightarrow 0^{+}}ds_{RN,c=0}^{2}=-\frac{1}{M^{2}\tau ^{2}}dt^{2}+%
\frac{M^{2}}{\tau ^{2}}\left( {d\tau ^{2}+\tau ^{2}d\Omega ^{2}}\right) .
\label{sunday-fin1}
\end{equation}
\end{itemize}

By further rescaling $\tau \rightarrow \frac{\tau }{M^{2}}$, one finally
gets
\begin{equation}
\lim_{\rho \rightarrow 0^{+}}ds_{RN,c=0}^{2}=\frac{M^{2}}{\tau ^{2}}\left(
-dt^{2}+{d\tau ^{2}+\tau ^{2}d\Omega ^{2}}\right) ,  \label{BR-BR-1}
\end{equation}
which is nothing but the ${AdS_{2}\times S_{2}}$ BR metric \cite{bertotti},
both flat and conformally flat. Such a result is consistent with the fact
that an extremal RN BH is nothing but an extremal $0$-brane in $d=4$.

It is also worth observing that, by introducing the physical distance
coordinate $\omega \equiv \ln \rho $, the metric (\ref{sunday-fin1}) can be
rewritten as
\begin{equation}
\lim_{{\omega \rightarrow -\infty }}ds_{RN,c=0}^{2}=-\frac{1}{{M^{2}}}%
e^{2\omega }dt^{2}+M^{2}d\omega ^{2}+M^{2}d\Omega ^{2}.
\end{equation}
Thus, the physical distance from any finite $\omega _{0}$ to the BH event
horizon is \textit{infinite}, because

\begin{equation}
\rho \rightarrow 0^{+}\Leftrightarrow {\omega \rightarrow -\infty .}
\end{equation}
As we will point out further below, this crucially discriminates the
extremal case from the non-extremal one.

The extremal RN single-center metric function (\ref{sunday-finn1}) can be
generalized to the multi-center case as follows:
\begin{equation}
e^{-U(r)}=1+\sum\limits_{s=1}^{n}{\frac{{M}_{s}}{{\left| \overrightarrow{x}{-%
}\overrightarrow{x}{_{s}}\right| }},}
\end{equation}
and the corresponding solution (\ref{papa1}) can be interpreted as many
non-rotating RN extremal BHs, with centers (\textit{i.e.} physical
singularities) at $\overrightarrow{x}{_{s}}$ and masses ${M}_{s}{%
=(q_{s}^{2}+p_{s}^{2})^{1/2}}$, saturating the BPS bound (\ref{sunday!1-gen}%
). The saturation of the BPS bound allows for the gravitational attraction
to equal the electro-magnetic repulsion, determining a static neutral
equilibrium. Such a phenomenon is sometimes called \textit{antigravity}. In
particular, the additive nature of the solution is related to the BPS nature
of such a force-free environments of (single-center) BHs.

As pointed out above, for RN BHs (in which no scalars are present)
extremality is equivalent to the saturation of the BPS bound, which also
implies that a Killing spinor exists \cite{tod}. As mentioned above, the
near-horizon BR geometry (\ref{BR-BR-1}) is conformally flat, satisfying
\cite{gibbons1}
\begin{equation}
\begin{array}{l}
\mathcal{R}=0~~\text{\textit{(vanishing scalar curvature)}}; \\
\\
C{_{\mu \nu \rho \sigma }=0~~}\text{\textit{(vanishing Weyl tensor)}}; \\
\\
D{_{\mu }}F{_{\rho \sigma }=0}\newline
~~\text{\textit{(covariantly constant Abelian vector field strength)}},
\end{array}
\end{equation}
\newline
implying that a \textit{doubling} of supersymmetry occurs at the event
horizon of the extremal RN BH (see \textit{e.g.} Kallosh's paper in \cite
{gibbons1}, and \cite{K3}). Notice that the \textit{doubling} also occurs at
${\rho \rightarrow \infty }$: indeed, within the asymptotic flatness \textit{%
Ansatz} the asymptotic spatial limit yields the $\mathcal{N}=2$, $d=4$
super-Poincar\'{e} superalgebra, pertaining to the $d=4$ Minkowski space.

Thus, it can be stated that the \textit{extremal} RN BH is nothing but an
\textit{extremal }$0$\textit{-black brane solitonic solution} of $\mathcal{N}%
=2$, $d=4$ (ungauged) supergravity, interpolating between two \textit{%
maximally supersymmetric} space-time geometries \cite{gib-town}:
\begin{equation}
c=0:\left\{
\begin{array}{l}
\text{\textit{at event horizon} (}r\rightarrow r_{H}^{+}\text{): \textit{%
algebra} }\mathfrak{psu}(1,1\left| 2\right) \text{\textit{, BR }}%
AdS_{2}\times S^{2}\text{~\textit{metric}}; \\
\\
\text{\textit{at spatial infinity }(}r\rightarrow \infty \text{): \textit{%
algebra} }\mathcal{N}=2,d=4~\text{\textit{super-Poincar\'{e}, Minkowski
metric}}.
\end{array}
\right.  \label{Thu-19}
\end{equation}
Beside $\mathcal{N}=2,d=4$ super-Poincar\'{e} algebra, $\mathfrak{psu}%
(1,1\left| 2\right) $ is a maximal $\mathcal{N}=2$, $d=4$ superalgebra,
which does not contain Poincar\'{e} nor other semisimple Lie algebra, but
direct sum of simple Lie algebra as maximal bosonic subalgebra. Indeed, in
this case the maximal bosonic subalgebra is $\mathfrak{so}(1,2)\oplus %
\mathfrak{so}(3)$ (with related maximal spin bosonic subalgebra $%
\mathfrak{su}(1,1)\oplus \mathfrak{su}(2)$), matching the corresponding
bosonic isometry group of the BR metric (see above). \newline
\setcounter{equation}0

\section{\label{Sect4}Thermodynamical Properties of Black Holes}

BHs have an entropy $S_{BH}$ and a temperature $T_{BH}$; in absence of other
attributes, they obey the first law of thermodynamics
\begin{equation}
{dM=T}_{BH}{dS}_{BH}{+...}\newline
,
\end{equation}
where $M$ is the gravitational (ADM) mass. By recalling Eqs. (\ref{sunday0}%
), (\ref{sunday1}), (\ref{sunday2}) and (\ref{sunday2-bis}), the temperature
and entropy are related to geometric quantities of the BH background, namely
to the surface gravity $\kappa $ and the to the horizon area $A_{H}$ \cite
{gibbons1}.

Thus, for a Schwarzschild BH, having $r_{H}=R_{H}=2M$ (see Eqs. (\ref
{CERN-mart-1}) and (\ref{CERN-mart-2})), it holds that ($r_{0}=\frac{r_{H}}{2%
}=M$) \cite{smarr}

\begin{equation}
\begin{array}{l}
dM=\frac{1}{{8\pi }}\kappa dA_{H}\Longrightarrow M=\frac{1}{{4\pi }}\kappa
A_{H}; \\
\\
A_{H}=4\pi r_{H}^{2}=16\pi M^{2}\,; \\
\\
S_{BH}=4\pi M^{2}\,; \\
\\
T_{BH}=\frac{1}{{8\pi M}}.
\end{array}
\quad
\end{equation}

Next, one might also consider (stationary) BHs with electric charge $q$ (RN,
mentioned above), and with (constant) angular momentum $J$, namely the
Kerr-Newman BHs, whose \textit{neutral limit} ($q=0$) is the Kerr BH (see
also the treatment of Sect. \ref{Sect8}; for recent advances, see \textit{%
e.g.} \cite{BH-Thermo,BH-Thermo2} and Refs. therein).

For such cases, the \textit{first law of BH thermodynamics} gets modified as
\cite{smarr}\cite{ruppeiner}

\begin{equation}
dM=T_{BH}dS_{BH}+\Omega dJ+\psi dq,  \label{sunday-night1}
\end{equation}
where $\Omega $ is the (constant, within the assumed stationary \textit{%
Ansatz}) angular velocity and $\psi $ is the electric potential (at the BH
event horizon) for the considered charged (possibly rotating) BH. The
generalization to include a magnetic charge $p$ is straightforward:
\begin{equation}
dM=T_{BH}dS_{BH}+\Omega dJ+\psi dq+\chi dp,  \label{sunday-night1-bis!}
\end{equation}
where $\chi $ is the magnetic potential (at the BH event horizon).

At a finite level, $M$, $S_{BH}$, $q$, $p$ and $J$ are related, in the
(semi)classical Einstein approximation, by the\textit{\ Smarr-integrated}
form of the (generalized) first law of the thermodynamics given by Eq. (\ref
{sunday-night1-bis!}),\textit{\ i.e.} by \cite{smarr}

\begin{equation}
M=2T_{BH}S_{BH}+2\Omega J+\psi q+\chi p,  \label{tuesday-after-3}
\end{equation}
which, by recalling the definition (\ref{corr1}), can be rewritten as
\begin{equation}
M=c+2\Omega J+\psi q+\chi p.  \label{tuesday-after-3-bis}
\end{equation}

As a further generalization\footnote{%
Interesting considerations can be found in \cite{Astefa}.}, one can couple a
set of (real) scalars $\phi ^{a}$ ($a=1,....,n$), parameterizing a scalar
manifold $\mathcal{M}_{\phi }$ with non-singular metric $G_{ab}$. The
scalars are associated to the so-called \textit{scalar charges} $\Sigma ^{a}$%
, defined through an expansion at $r\rightarrow \infty $ as \cite{gibbons3}
\begin{equation}
\lim_{r\rightarrow \infty }\phi ^{a}(r)\equiv \phi _{\infty }^{a}+\frac{1}{r}%
\Sigma ^{a}+\mathcal{O}\left( \frac{1}{r^{2}}\right) ,~\Sigma ^{a}=\Sigma
^{a}\left( p^{\Lambda },q_{\Lambda },A_{H},J,\phi _{\infty }^{b}\right) .
\label{monday5}
\end{equation}
Thus, $\Sigma ^{a}$ can be equivalently defined as follows \cite{ferrara3}
(see also \cite{AoB-book}):
\begin{equation}
\Sigma ^{a}\left( p^{\Lambda },q_{\Lambda },A_{H},J,\phi _{\infty
}^{b}\right) \equiv \lim_{\tau \rightarrow 0^{-}}\frac{d\phi ^{a}(\tau )}{%
d\tau }.  \label{tuesday-fin-4}
\end{equation}
The BH charges are defined as the fluxes at spatial infinity of the Abelian
field strengths:
\begin{equation}
p^{\Lambda }\equiv \frac{1}{4\pi }\int_{S_{\infty }^{2}}\mathcal{F}^{\Lambda
},~~~q_{\Lambda }\equiv \frac{1}{4\pi }\int_{S_{\infty }^{2}}\mathcal{G}%
_{\Lambda },~~~\Lambda =1,2,...,n+1,  \label{Gamma-tilde}
\end{equation}
where $\mathcal{F}^{\Lambda }=dA^{\Lambda }$ and $\mathcal{G}_{\Lambda }$ is
the ``dual'' field-strength two-form \cite{4}.

Thus, Eq. (\ref{sunday-night1}) becomes \cite{gibbons3}
\begin{eqnarray}
dM &=&T_{BH}dS_{BH}+\Omega dJ+\psi ^{\Lambda }dq_{\Lambda }+\chi _{\Lambda
}dp^{\Lambda }+\left. \frac{\partial M}{\partial \phi ^{a}}\right|
_{S_{BH},J,q_{\Lambda },p^{\Lambda }}d\phi ^{a}=  \notag \\
&&  \notag \\
&=&T_{BH}dS_{BH}+\Omega dJ+\psi ^{\Lambda }dq_{\Lambda }+\chi _{\Lambda
}dp^{\Lambda }-G_{ab}(\phi _{\infty })\Sigma ^{b}d\phi ^{a}.
\label{tuesday-after-2}
\end{eqnarray}
It is worth remarking that, despite the presence of the extra term $%
-G_{ab}(\phi _{\infty })\Sigma ^{b}d\phi ^{a}$ in the generalized first law
of BH thermodynamics (\ref{tuesday-after-2}), the integrated version,
\textit{i.e.} the corresponding Smarr formula, remains (see P.
Breitenlohner, Maison, and Gibbons' paper in \cite{gibbons1}; see also \cite
{gibbons3})
\begin{equation}
M=2T_{BH}S_{BH}+2\Omega J+\psi ^{\Lambda }q_{\Lambda }+\chi _{\Lambda
}p^{\Lambda }=c+2\Omega J+\psi ^{\Lambda }q_{\Lambda }+\chi _{\Lambda
}p^{\Lambda }.  \label{tuesday-after-4}
\end{equation}

The thermodynamical quantities $S_{BH}$ (or, equivalently by the
Bekenstein-Hawking relation (\ref{sunday0}) or (\ref{sunday1}), $A_{H}$), $J$%
, ${q_{\Lambda }}$, ${p^{\Lambda }}$, $\phi {_{\infty }^{a}}$ are
coordinates on the state space (in the - semi - classical limit of large BH
charges)

\begin{equation}
\mathbb{R}_{0}^{+}\times \mathbb{R}^{+}\times \mathbb{R}^{2n+2}\times
\mathcal{M}_{\phi _{\infty }},
\end{equation}
where the first factor $\mathbb{R}_{0}^{+}$ disregards the so-called \textit{%
small} BHs, which in the classical (Einstein) approximation represent
\textit{naked singularities} in the space-time, and for which the Attractor
Mechanism does not hold. Furthermore, $\mathcal{M}_{\phi _{\infty }}$ is the
asymptotic ($r\rightarrow \infty $) scalar manifold.

It is worth remarking that, due to the key result \cite{gibbons3}
\begin{equation}
\left. \frac{\partial M}{\partial \phi ^{a}}\right| _{S_{BH},J,q_{\Lambda
},p^{\Lambda }}=-G_{ab}(\phi _{\infty })\Sigma ^{b}\left(
S_{BH,}J,p^{\Lambda },q_{\Lambda },\phi _{\infty }^{b}\right) ,
\end{equation}
for non-singular metric of the scalar manifold, the scalar charges $\Sigma
^{a}$ vanish iff $\phi _{\infty }$, and hence the vacuum state, is chosen to
extremize the ADM mass $M$ at fixed BH entropy $S_{BH}=A_{H}/4$, angular
momentum $J$, and conserved electric and magnetic charges $q_{\Lambda }$ and
$p^{\Lambda }$.

Let us now consider the static case ($J=0$), described by the bosonic
Lagrangian density \cite{ferrara3}

\begin{equation}
-{}\frac{\mathcal{L}}{\sqrt{-g}}=\frac{1}{2}\mathcal{R}-\frac{1}{2}%
G_{ab}\partial _{\mu }\phi ^{a}\partial _{\nu }\phi ^{b}g^{\mu \nu }-\left(
\text{Im}\mathcal{N}{}_{\Lambda \Sigma }\right) F^{\Lambda }F^{\Sigma
}-\left( \text{Re}\mathcal{N}_{\Lambda \Sigma }\right) F^{\Lambda }\tilde{F}%
^{\Sigma }.  \label{tuesday-fin-7}
\end{equation}

Under the convexity condition\footnote{$\nabla _{a}$ is the Christoffel
covariant derivative, which acquires also a K\"{a}hler connection component
when the scalars parameterize a (special) K\"{a}hler manifold, such as in $%
\mathcal{N}=2$, $d=4$ supergravity coupled to Abelian vector supermultiplets
(see \textit{e.g.} \cite{4} and Refs. therein).}
\begin{equation}
\nabla _{a}\partial _{b}V_{BH}\geqslant 0,  \label{tuesday-after-5}
\end{equation}
requiring the (semi-)positive definiteness of the covariant Hessian of the
so-called \textit{BH effective potential} \cite{ferrara3}

\begin{equation}
\begin{array}{l}
V_{BH}\left( \phi ,p,q\right) \equiv -\frac{1}{2}\Gamma ^{T}M\Gamma ; \\
\\
\Gamma ^{T}\equiv (p^{\Lambda },q_{\Lambda }); \\
\\
M\equiv \left(
\begin{array}{ccc}
\text{Im}\mathcal{N}+\text{Re}\mathcal{N}\left( \text{Im}\mathcal{N}\right)
^{-1}\text{Re}\mathcal{N} &  & -\text{Re}\mathcal{N}\left( \text{Im}\mathcal{%
N}\right) ^{-1} \\
&  &  \\
-\left( \text{Im}\mathcal{N}\right) ^{-1}\text{Re}\mathcal{N} &  & \left(
\text{Im}\mathcal{N}\right) ^{-1}
\end{array}
\right) ,
\end{array}
\end{equation}
in \cite{gibbons3} it was proved that
\begin{equation}
\Sigma ^{a}=0\Leftrightarrow \left. \frac{\partial M}{\partial \phi ^{a}}%
\right| _{S_{BH},J,q_{\Lambda },p^{\Lambda }}=0\Leftrightarrow \left\{
\begin{array}{l}
\frac{d\phi ^{a}\left( \tau \right) }{d\tau }=0~\forall \tau \in \mathbb{R}%
^{-}\Leftrightarrow \phi ^{a}\neq \phi ^{a}\left( \tau \right) ; \\
\\
\phi ^{a}=\phi _{\infty }^{a}=\phi _{H}(p,q):\frac{\partial V_{BH}\left(
\phi ,p,q\right) }{\partial \phi ^{a}}=0.
\end{array}
\right.  \label{Wed-9}
\end{equation}
In words, the scalar charges vanish iff the extremal BH is actually
double-extremal,\textit{\ i.e.} if the scalars are \textit{constant} (%
\textit{independent} on $\tau $); moreover, their constant values are not
arbitrary, but they are chosen to extremize $V_{BH}\left( \phi ,p,q\right) $
(at fixed, supporting BH charge configuration). Thus (at fixed, supporting
BH charge configuration $p^{\Lambda },q_{\Lambda }$, corresponding to a
fixed value of the BH entropy $S_{BH}\left( p,q\right) $), under the
condition (\ref{tuesday-after-5}) it holds that
\begin{equation}
\Sigma ^{a}=0\Leftrightarrow \phi _{\infty }^{a}=\phi _{H}(p,q),~\forall
a=1,...,n,~\forall \tau \in \mathbb{R}^{-},
\end{equation}
with $\phi _{H}(p,q)$ fulfilling the \textit{criticality conditions (}$%
\partial _{a}\equiv \frac{\partial }{\partial \phi ^{a}}$\textit{)}
\begin{equation}
\left. \partial _{a}V_{BH}\left( \phi ,p,q\right) \right| _{\phi =\phi
_{H}\left( p,q\right) }=0.
\end{equation}
When evaluated at the critical point of $V_{BH}$ given by purely
charge-dependent scalar configuration $\phi _{H}(p,q)$, condition (\ref
{tuesday-after-5}) is nothing but the requirement that the $\phi _{H}(p,q)$
actually corresponds to a \textit{stable} (up to some \textit{massless
Hessian modes}) critical point of $V_{BH}$, and thus that it gives raise to
an \textit{attractor} in a strict sense.

\subsection*{First Order Formalism}

However, one can go a step further beyond the proof of \cite{gibbons3}
(which however, beside implying some loss of generality, holds also in the
non extremal case), and, in the extremal case, relate directly the scalar
charges to the critical points of a suitably generalized real, positive
superpotential $\mathcal{W}$ (\cite{ADOT-1}; see below, Eqs. (\ref{tuesday8}%
) and (\ref{tuesday8-bis})).

Indeed, in \cite{Cer-Dal,ADOT-1} a first order formalism for static,
spherically symmetric and asymptotically flat extremal BHs was introduced,
based on the \textit{generalized (fake) superpotential} $\mathcal{W}$. For $%
c=0$ the second order ordinary differential equation satisfied by the
scalars $\phi ^{a}$, \textit{i.e.} \cite{ferrara3}
\begin{equation}
\frac{\nabla ^{2}\phi ^{a}\left( \tau \right) }{\nabla \tau ^{2}}=\frac{%
d^{2}\phi ^{a}\left( \tau \right) }{d\tau ^{2}}+\Gamma _{bc}^{~~a}\left(
\phi \left( \tau \right) \right) \frac{d\phi ^{b}\left( \tau \right) }{d\tau
}\frac{d\phi ^{c}\left( \tau \right) }{d\tau }=G^{ab}\left( \phi \left( \tau
\right) \right) \frac{\partial V_{BH}\left( \phi ,p,q\right) }{\partial \phi
^{b}}e^{2U\left( \tau \right) }
\end{equation}
have been shown to reduce to the following first order one\footnote{%
Some subtleties (however immaterial for $c=0$) are given in footnote 3 of
\cite{ADOT-1}.} \cite{ADOT-1}:
\begin{equation}
\frac{d\phi ^{a}\left( \tau \right) }{d\tau }=2e^{U\left( \tau
\right)
}G^{ab}\left( \phi \left( \tau \right) \right) \frac{\partial \mathcal{W}%
\left( \phi ,p,q\right) }{\partial \phi ^{b}}.  \label{tuesday-fin-1}
\end{equation}
The relation between $V_{BH}$ and $\mathcal{W}$ is given by \cite
{Cer-Dal,ADOT-1}
\begin{equation}
V_{BH}=\mathcal{W}^{2}+2G^{ab}\left( \partial _{a}\mathcal{W}\right)
\partial _{b}\mathcal{W},  \label{tuesday-fin-2}
\end{equation}
yielding that
\begin{equation}
\partial _{a}V_{BH}=2\left[ \delta _{a}^{b}\mathcal{W}+2G^{bc}\left( \nabla
_{a}\partial _{c}\mathcal{W}\right) \right] \partial _{b}\mathcal{W}=2\left[
\delta _{a}^{b}\mathcal{W}+2G^{bc}\left( \partial _{a}\partial _{c}\mathcal{W%
}-\Gamma _{ac}^{~~f}\partial _{f}\mathcal{W}\right) \right] \partial _{b}%
\mathcal{W}.  \label{tuesday-fin-3}
\end{equation}
Eq. (\ref{tuesday-fin-1}) is \textit{BPS-like}; for a given $\mathcal{W}$,
it relates the evolution of the scalar fields to the partial derivative of $%
\mathcal{W}$ itself with respect to the scalar fields. For a fixed,
supporting BH charge configuration, the extrema of $\mathcal{W}$ in $%
\mathcal{M}_{\phi }$ are fixed points in the radial evolution of $\phi ^{a}$%
. Furthermore, Eq. (\ref{tuesday-fin-3}) implies that for a fixed supporting
BH charge configuration, the extrema of $\mathcal{W}$ in $\mathcal{M}_{\phi
} $ are also extrema of $V_{BH}$ in $\mathcal{M}_{\phi }$.

Now, by recalling the definition (\ref{tuesday-fin-4}) and performing the
asymptotic limit of the first order differential Eq. (\ref{tuesday-fin-1}),
one obtains the general expression of the \textit{contravariant} scalar
charges in the extremal environment under consideration\footnote{%
We assume
\begin{equation*}
\left( \partial _{a}\mathcal{W}\left( \phi ,p,q\right) \right) _{\infty
}\equiv \lim_{\tau \rightarrow 0^{-}}\frac{\partial \mathcal{W}\left( \phi
\left( \tau \right) ,p,q\right) }{\partial \phi ^{a}\left( \tau \right) }=%
\frac{\partial \mathcal{W}\left( \phi _{\infty },p,q\right) }{\partial \phi
_{\infty }^{a}}.
\end{equation*}
}:
\begin{equation}
\Sigma ^{a}\left( \phi _{\infty },p,q\right) =2\lim_{\tau
\rightarrow 0^{-}}e^{U\left( \tau \right) }G^{ab}\left( \phi \left(
\tau \right) \right) \partial _{b}\mathcal{W}\left( \phi \left( \tau
\right) ,p,q\right) =2G^{ab}\left( \phi _{\infty }\right)
\frac{\partial \mathcal{W}\left( \phi _{\infty },p,q\right)
}{\partial \phi _{\infty }^{b}},  \label{tuesday-fin-6}
\end{equation}
where minimal regularity conditions in order to split the asymptotical limit
have been understood, and in the last step the asymptotic boundary condition
for $U\left( \tau \right) $ have been used. Similarly, \textit{covariant}
scalar charges read
\begin{equation}
\Sigma _{a}\left( \phi _{\infty },p,q\right) =2\lim_{\tau
\rightarrow 0^{-}}e^{U\left( \tau \right) }\partial
_{a}\mathcal{W}\left( \phi \left( \tau \right) ,p,q\right)
=2\frac{\partial \mathcal{W}\left( \phi _{\infty },p,q\right)
}{\partial \phi _{\infty }^{a}},  \label{tuesday-fin-6-bis}
\end{equation}
Eqs. (\ref{tuesday-fin-6}) and (\ref{tuesday-fin-6-bis}), holding true for
extremal, static, spherically symmetric and asymptotically flat BHs (without
any assumption on the Hessian matrix of $V_{BH}$), relates the (vanishing of
the) scalar charges $\Sigma ^{a}$ to (the critical points of) the
generalized superpotential $\mathcal{W}$.

Now, since the explicit forms of $\mathcal{W}$ and of its critical points in
$\mathcal{M}_{\phi _{\left( \infty \right) }}$ have been explicitly
determined for some $\mathcal{N}>2$-extended, $d=4$ supergravities \cite
{ADOT-1} and for some examples of $\mathcal{N}=2$, $d=4$ supergravity%
\footnote{%
Namely, the $\mathcal{N}=2$, $d=4$ models treated so far are:
\par
1) the so-called $t^{3}$ model, in the \textit{electric} BH charge
configuration $\left( p^{0},p^{1}=0,q_{0}=0,q_{1}\right) $ \cite{Cer-Dal};
\par
2) the so-called $stu$ model, in the \textit{electric} BH charge
configuration $\left( p^{0},p^{i}=0,q_{0}=0,q_{i}\right) $ ($i=1,2,3$) \cite
{Cer-Dal};
\par
3) the models based on the sequence $\frac{SU\left( 1,n_{V}\right) }{U\left(
1\right) \times SU\left( n_{V}\right) }$ of irreducible homogeneous
symmetric special K\"{a}hler manifolds with quadratic prepotential \cite
{ADOT-1}\textbf{\ }(for $n_{V}=1$, see also \cite{Cer-Dal});
\par
4) the model related to the degree three Jordan algebra on the quaternions $%
J_{3}^{\mathbb{H}}$ (dual to the $\mathcal{N}=6$, $d=4$ case) \cite{ADOT-1}.
\par
{}}, the explicit forms of $\Sigma ^{a}$ and of the geometrical \textit{loci}
of their zeros in $\mathcal{M}_{\phi _{\left( \infty \right) }}$ (for a
given, supporting BH charge configuration) for all such cases are currently
known. It is worth noticing that $\mathcal{W}$ varies depending on the class
of extremal BH attractors under consideration (and it is not unique inside
the same class, too), thus the expression of $\Sigma ^{a}$ will be dependent
on the considered class of attractor flows.

It is then clear from the reasoning above that the scalar charges $\Sigma
^{a}$ can vanish also in \textit{non-double-extremal} BHs (\textit{i.e.} in
BHs with non-trivial scalar dynamics), when a fine-tuning of the
asymptotical values $\phi _{\infty \text{ }}^{a}$of the scalars is performed
such that $\phi _{\infty \text{ }}^{a}=\phi _{H}^{a}\left( p,q\right) $,
with $\phi _{H}^{a}\left( p,q\right) $ satisfying the \textit{criticality
conditions} of $\mathcal{W}$ (and thus, through Eq. (\ref{tuesday-fin-3}),
of $V_{BH}$) in $\mathcal{M}_{\phi _{\left( \infty \right) }}$ (for a given,
supporting BH charge configuration). In other words, in order to make the $%
\Sigma ^{a}$'s vanish, the \textit{initial data} of the radial scalar
dynamics must be fine-tuned to coincide with the near-horizon, \textit{%
attracted} scalar configurations $\phi _{H}^{a}\left( p,q\right) $ (at least
within the same \textit{basin of attraction}).\bigskip\ \textit{\ }

As anticipated above, an interesting feature of such an environment with
scalars and/or electric and magnetic charges concerns extremality: $M$ and $%
S_{BH}$ can be finite and non-vanishing, with $T_{BH}=0$. In other words,
extremal \textit{large} BHs with non-vanishing ADM mass and entropy can
exist!

The simplest example is provided by the RN BH treated above, in which $n=0$.
Let us now focus on its thermodynamical properties; by recalling Eqs. (\ref
{sunday1}), (\ref{sunday2}) and (\ref{sunday2-bis}), and putting $\Sigma =0$%
, one obtains that
\begin{equation}
T_{BH}=\frac{\kappa }{{2\pi }}=\frac{1}{{2\pi }}\frac{{r_{0}}}{{R_{+}^{2}}}=%
\frac{c}{{2S}_{BH}},
\end{equation}
where (recalling Eqs. (\ref{corr1}), (\ref{sunday1}) and (\ref
{sunday-Night-1}) and perform the shift $q^{2}\rightarrow q^{2}+p^{2}$)
\begin{equation}
\begin{array}{c}
c\equiv r_{0}=\sqrt{M^{2}-q^{2}-p^{2}}; \\
\\
S_{BH}=\pi (2M^{2}-q^{2}-p^{2}+2M\sqrt{M^{2}-p^{2}-q^{2}}); \\
\Updownarrow \\
r_{+}^{2}=R_{+}^{2}=2M^{2}-q^{2}-p^{2}+2M\sqrt{M^{2}-p^{2}-q^{2}}.
\end{array}
\label{sunday-Night-3}
\end{equation}
Thus, as stated above, for RN BHs the \textit{extremality} (\textit{i.e.} $%
c=0$) is a \textit{necessary} and \textit{sufficient} condition for the
\textit{saturation} of the BPS bound (\ref{sunday!1-gen}), and thus for the (%
\textit{maximally}, in the present case) \textit{supersymmetric nature} of
the considered background. Indeed, Eqs. (\ref{sunday-Night-3}) yield
\begin{equation}
c=0\Leftrightarrow R_{H}=r_{H}=M=\left( q^{2}+p^{2}\right)
^{1/2}\Longrightarrow \left. S_{BH}\right| _{c=0}=\pi (q^{2}+p^{2}).
\label{sunday-Night-4}
\end{equation}
Consequently, \textit{large }extremal RN BHs are ($\frac{1}{2}$-)BPS, and
they have $T_{BH}=0$. As given by Eq. (\ref{sunday-Night-4}), the \textit{%
small BH limit} is reached in the degenerate limit $p\rightarrow 0$ and $%
q\rightarrow 0$.

By further introducing (constant) angular momentum $J$, one may consider the
stationary, rotating uncharged (Kerr) or charged (Kerr-Newman) BHs. At a
finite level, $M$, $S_{BH}$, $q$, $p$ and $J$ are related, in the
(semi)classical Einstein approximation, by the\textit{\ Smarr-integrated}
form of the (generalized) first law of the thermodynamics given by Eq. (\ref
{sunday-night1-bis!}),\textit{\ i.e.} by Eqs. (\ref{tuesday-after-3}) and (%
\ref{tuesday-after-3-bis}). Eqs. (\ref{sunday-Night-1}) and (\ref{sunday1})
are generalized as
\begin{equation}
\begin{array}{l}
r_{\pm }=M\pm \sqrt{M^{2}-q^{2}-p^{2}-\frac{J^{2}}{M^{2}}}; \\
\\
R_{+}^{2}=r_{+}^{2}+\frac{J^{2}}{M^{2}}\geqslant r_{+}^{2}; \\
\\
S_{BH}=\frac{A_{H}}{4}=\pi R_{+}^{2}=\pi \left( r_{+}^{2}+\frac{J^{2}}{M^{2}}%
\right) ,
\end{array}
\label{monday-after-1}
\end{equation}

and thus
\begin{gather}
S_{BH}=\pi \left[ 2M^{2}-q^{2}-p^{2}+2M\left( M^{2}-q^{2}-p^{2}-\frac{J^{2}}{%
M^{2}}\right) ^{1/2}\right] ;  \notag \\
\Updownarrow  \notag \\
R_{+}^{2}=2M^{2}-q^{2}-p^{2}+2M\left( M^{2}-q^{2}-p^{2}-\frac{J^{2}}{M^{2}}%
\right) ^{1/2},  \label{monday-after-2}
\end{gather}
whose inversion reads
\begin{eqnarray}
M^{2} &=&\frac{\pi J^{2}}{S_{BH}}+\frac{\pi \left( q^{2}+p^{2}\right) ^{2}}{%
4S_{BH}}+\frac{S_{BH}}{4\pi }+\frac{\left( q^{2}+p^{2}\right) }{2}=  \notag
\\
&=&\left( \frac{J}{R_{+}}\right) ^{2}+\left( \frac{q^{2}+p^{2}}{2R_{+}}%
\right) ^{2}+\left( \frac{R_{+}}{2}\right) ^{2}+\frac{\left(
q^{2}+p^{2}\right) }{2}.
\end{eqnarray}

Notice that for (constant, in the considered stationary rotating regime)
angular momentum $J\neq 0$, $A_{H}$ is an \textit{effective} quantity,
\textit{i.e.} (within the spherical symmetry \textit{Ansatz}) it is not
given by $\pi r_{+}^{2}$, but rather by $\pi R_{+}^{2}$. In particular, the
second of Eqs. (\ref{monday-after-1}) and Eq. (\ref{monday-after-2}) yields
that $R_{+}\geqslant r_{+}$. Thus, the (stationary) rotating regime actually
\textit{increases} the BH radius relevant for the computation of the entropy
through the Bekenstein-Hawking (semi)classical formula (\ref{sunday1}).

Other useful relations among the various geometric and thermodynamical
quantities read \cite{smarr}
\begin{eqnarray}
T_{BH} &=&\frac{1}{{32M}}\left[ \frac{1}{\pi }-\frac{{4\pi J^{2}}}{{%
S_{BH}^{2}}}-\frac{{\pi }\left( q^{2}+p^{2}\right) ^{2}}{{S_{BH}^{2}}}\right]
=\frac{1}{{32\pi M}}\left[ 1-\left( \frac{2{J}}{R_{+}^{2}}\right)
^{2}-\left( \frac{q^{2}+p^{2}}{R_{+}^{2}}\right) ^{2}\right] ;  \notag \\
&& \\
\Omega &=&\frac{{\pi J}}{{MS}_{BH}}=\frac{{J}}{{MR}_{+}^{2}}; \\
&&  \notag \\
\psi &=&\frac{1}{2M}\left( q+\frac{\pi q^{3}}{S_{BH}}\right) =\frac{q}{2M}%
\left[ 1+\left( \frac{q}{R_{+}}\right) ^{2}\right] .
\end{eqnarray}

The expression of the \textit{extremality parameter} given by the first of
Eqs. (\ref{sunday-Night-3}) is suitably generalized as
\begin{equation}
c\equiv r_{0}=\sqrt{M^{2}-q^{2}-p^{2}-\frac{J^{2}}{M^{2}}},
\label{sunday-Night-fin-1}
\end{equation}
and thus extremality implies (see \cite{rotating-attr} and Refs. therein)
\begin{gather}
c=0\Leftrightarrow \left\{
\begin{array}{l}
M^{2}=r_{H}^{2}=\frac{\left( q^{2}+p^{2}\right) }{2}+\sqrt{\frac{\left(
q^{2}+p^{2}\right) ^{2}}{4}+J^{2}}; \\
\\
J=M\sqrt{M^{2}-q^{2}-p^{2}};
\end{array}
\right.  \notag \\
\Downarrow  \notag \\
S_{BH,c=0}=\pi \left( r_{H}^{2}+\frac{J^{2}}{M^{2}}\right) =\pi \sqrt{\left(
q^{2}+p^{2}\right) ^{2}+4J^{2}},
\end{gather}
and thus the \textit{(squared) effective radius} for extremal Kerr-Newman BH
reads
\begin{equation}
R_{+,c=0}^{2}\equiv R_{H}^{2}=r_{H}^{2}+\frac{J^{2}}{M^{2}}=\sqrt{\left(
q^{2}+p^{2}\right) ^{2}+4J^{2}}.
\end{equation}
\setcounter{equation}0

\section{\label{Sect5}Geodesic Action with a Constraint\newline
and Critical Points of the Black Hole Effective Potential}

Let us now reconsider the system described by the (bosonic) Lagrangian
density (\ref{tuesday-fin-7}). In such a framework, the expression (\ref
{sunday-Night-fin-1}) of the extremality parameter gets modified as follows
(see below) \cite{ferrara3}:

\begin{equation}
M^{2}+\frac{1}{2}G_{ab}\left( \phi _{\infty }\right) \Sigma ^{a}\Sigma
^{b}-V_{BH}(\phi _{\infty }^{b},p^{\Lambda },q_{\Lambda })=c^{2},
\label{monday4}
\end{equation}
which in the extremal case ($c=0$) becomes
\begin{equation}
M^{2}=V_{BH}(\phi _{\infty }^{b},p^{\Lambda },q_{\Lambda })-\frac{1}{2}%
G_{ab}\left( \phi _{\infty }\right) \Sigma ^{a}\Sigma ^{b}.
\label{monday4-bis}
\end{equation}
Let us also recall that $c$ is defined by (\ref{corr1}) to be $%
c=2S_{BH}T_{BH}$. Thus, in presence of scalars it follows that in general it
depends on the asymptotical values $\phi _{\infty }^{a}$ of the scalars
(through the dependence of both $S_{BH}$ and $T_{BH}$ on $\phi _{\infty
}^{a} $'s; consistently with the holding of Eq. (\ref{monday4}), we consider
the case $J=0$):
\begin{equation}
c=2S_{BH}\left( \phi _{\infty },p,q\right) T_{BH}\left( \phi _{\infty
},p,q\right) ,
\end{equation}
and Eq. (\ref{monday4}) can be rewritten as follows:
\begin{equation}
M^{2}+\frac{1}{2}G_{ab}\left( \phi _{\infty }\right) \Sigma ^{a}\Sigma
^{b}-V_{BH}(\phi _{\infty }^{b},p^{\Lambda },q_{\Lambda })=4S_{BH}^{2}\left(
\phi _{\infty },p,q\right) T_{BH}^{2}\left( \phi _{\infty },p,q\right) .
\end{equation}
By differentiating with respect to $\phi _{\infty }^{c}$ and recalling the
\textit{Metric Postulate} in the asymptotic scalar manifold $\mathcal{M}%
_{\phi _{\infty }}$, one gets\footnote{%
This corrects Eq. (23) of \cite{gibbons3}.}
\begin{gather}
M\frac{\partial M}{\partial \phi _{\infty }^{c}}+\frac{1}{2}G_{ab}\left(
\phi _{\infty }\right) \Sigma ^{a}\nabla _{c,\infty }\Sigma ^{b}-\frac{1}{2}%
\frac{\partial V_{BH}(\phi _{\infty },p,q)}{\partial \phi _{\infty }^{c}}=
\notag \\
\notag \\
=2c\left[ T_{BH}\left( \phi _{\infty },p,q\right) \frac{\partial S_{BH}(\phi
_{\infty },p,q)}{\partial \phi _{\infty }^{c}}+S_{BH}\left( \phi _{\infty
},p,q\right) \frac{\partial T_{BH}(\phi _{\infty },p,q)}{\partial \phi
_{\infty }^{c}}\right] ,  \label{Wed-6}
\end{gather}
where $\nabla _{c,\infty }$ is the (Christoffel) covariant derivative in $%
\mathcal{M}_{\phi _{\infty }}$:
\begin{equation}
\nabla _{c,\infty }\Sigma ^{b}=\frac{\partial \Sigma ^{b}}{\partial \phi
_{\infty }^{c}}+\Gamma _{ca}^{~~b}\left( \phi _{\infty }\right) \Sigma ^{a}.
\end{equation}
When $c$ is constant \textit{globally} in $\mathcal{M}_{\phi _{\infty }}$,
it trivially holds that
\begin{eqnarray}
\frac{\partial c}{\partial \phi _{\infty }^{a}} &=&0\Leftrightarrow
T_{BH}\left( \phi _{\infty },p,q\right) \frac{\partial S_{BH}(\phi _{\infty
},p,q)}{\partial \phi _{\infty }^{c}}+S_{BH}\left( \phi _{\infty
},p,q\right) \frac{\partial T_{BH}(\phi _{\infty },p,q)}{\partial \phi
_{\infty }^{c}}=0,  \notag \\
&&
\end{eqnarray}
and Eq. (\ref{Wed-6}) becomes
\begin{gather}
\frac{\partial M}{\partial \phi _{\infty }^{c}}=\frac{1}{2M}\left[ \frac{%
\partial V_{BH}(\phi _{\infty },p,q)}{\partial \phi _{\infty }^{c}}%
-G_{ab}\left( \phi _{\infty }\right) \Sigma ^{a}\nabla _{c,\infty }\Sigma
^{b}\right] =  \notag \\
\notag \\
=\frac{\left[ \frac{\partial V_{BH}(\phi _{\infty },p,q)}{\partial \phi
_{\infty }^{c}}-G_{ab}\left( \phi _{\infty }\right) \Sigma ^{a}\nabla
_{c,\infty }\Sigma ^{b}\right] }{2\sqrt{V_{BH}(\phi _{\infty },p,q)-\frac{1}{%
2}G_{ab}\left( \phi _{\infty }\right) \Sigma ^{a}\Sigma ^{b}+c^{2}}},
\label{Wed-7}
\end{gather}
where in the last step of Eq. (\ref{Wed-7}) we used Eq. (\ref{monday4}). In
the extremal case ($c=0$) Eq. (\ref{Wed-7}) simplifies to
\begin{equation}
c=0:\frac{\partial M}{\partial \phi _{\infty }^{c}}=\frac{\left[ \frac{%
\partial V_{BH}(\phi _{\infty },p,q)}{\partial \phi _{\infty }^{c}}%
-G_{ab}\left( \phi _{\infty }\right) \Sigma ^{a}\nabla _{c,\infty }\Sigma
^{b}\right] }{2\sqrt{V_{BH}(\phi _{\infty },p,q)-\frac{1}{2}G_{ab}\left(
\phi _{\infty }\right) \Sigma ^{a}\Sigma ^{b}}}.  \label{Wed-8}
\end{equation}

By recalling the expression (\ref{tuesday-fin-6}) of the scalar charges and
the relation (\ref{tuesday-fin-2}) between $V_{BH}$ and the generalized
superpotential $\mathcal{W}$ (both holding in the considered framework for $%
c=0$), one consistently obtains the second of Eqs. (\ref{tuesday8-bis}),
giving the general expression of the (squared) ADM mass $M^{2}$ in the
extremal case ($G^{ab}\left( \phi _{\infty }\right) \equiv G_{\infty }^{ab}$%
, $G_{ab}\left( \phi _{\infty }\right) \equiv G_{ab,\infty }$, $\mathcal{W}%
^{2}\left( \phi _{\infty },p,q\right) \equiv \mathcal{W}_{\infty }^{2}$, $%
V_{BH}\left( \phi _{\infty },p,q\right) \equiv V_{BH,\infty }$; see also
footnote 8):
\begin{equation}
c=0:\left\{
\begin{array}{l}
\frac{1}{2}G_{ab,\infty }\Sigma ^{a}\Sigma ^{b}=2G_{\infty }^{ab}\left(
\partial _{a}\mathcal{W}\right) _{\infty }\left( \partial _{b}\mathcal{W}%
\right) _{\infty }; \\
\\
\\
V_{BH,\infty }=\mathcal{W}_{\infty }^{2}+2G_{\infty }^{ab}\left( \partial
_{a}\mathcal{W}\right) _{\infty }\left( \partial _{b}\mathcal{W}\right)
_{\infty };
\end{array}
\right. \Longrightarrow M^{2}=V_{BH,\infty }-\frac{1}{2}G_{ab,\infty }\Sigma
^{a}\Sigma ^{b}=\mathcal{W}_{\infty }^{2}.  \label{Wed-4}
\end{equation}
By differentiating with respect to $\phi _{\infty }^{a}$, from such a result
it trivially follows:
\begin{equation}
\frac{\partial M}{\partial \phi _{\infty }^{a}}=\frac{\partial \mathcal{W}%
_{\infty }}{\partial \phi _{\infty }^{a}}
\end{equation}
which by recalling Eq. (\ref{Wed-8}), consistently with Eqs. (\ref
{tuesday-fin-3}) and (\ref{tuesday-fin-6}), yields an expression relating
the critical points of $V_{BH}(\phi _{\infty },p,q)$, $\mathcal{W}^{2}\left(
\phi _{\infty },p,q\right) \equiv \mathcal{W}_{\infty }^{2}$ and $\Sigma
^{b}\left( \phi _{\infty },p,q\right) $ in $\mathcal{M}_{\phi _{\infty }}$:
\begin{equation}
c=0:\frac{\partial V_{BH}(\phi _{\infty },p,q)}{\partial \phi _{\infty }^{c}}%
=\frac{\partial \mathcal{W}_{\infty }^{2}}{\partial \phi _{\infty }^{a}}%
+G_{ab}\left( \phi _{\infty }\right) \Sigma ^{a}\nabla _{c,\infty }\Sigma
^{b}.
\end{equation}

As mentioned above, since the explicit form of $\mathcal{W}$ has been
explicitly determined for some $\mathcal{N}>2$-extended, $d=4$
supergravities \cite{ADOT-1} and for some examples of $\mathcal{N}=2$, $d=4$
supergravity \cite{Cer-Dal,ADOT-1}, the explicit form of the ADM mass $M$
(for a given, supporting BH charge configuration) for all such cases is
currently known.

It is worth noticing that in \textit{double-extremal} BHs (having $c=0$ and
for which $\Sigma ^{a}=0$ $\forall a=1,...,n$) the relation (\ref{monday4})
simply becomes
\begin{equation}
M^{2}=V_{BH}(\phi _{H}\left( p,q\right) ,p,q),
\end{equation}
expressing the fact that the ADM mass only depends on charges through the
horizon value of $V_{BH}$, a trivial fact by recalling that the scalars are
constant (independent on $r$)!

The most general class for which the above expression holds is the one of
spherically symmetric, static, asymptotically flat BHs specified by the
metric \textit{Ansatz} (\cite{ferrara3}\cite{K3}; see also \cite{ADFT} and
\cite{ADOT-1}):

\begin{equation}
ds^{2}=-\,e^{2U}dt^{2}+e^{-2U}\gamma _{mn}dx^{m}dx^{n}\equiv
-\,e^{2U}dt^{2}+e^{-2U}\left( {\frac{{c^{4}}}{{\sinh }^{4}\left( c{\tau }%
\right) }d\tau ^{2}+\frac{{c^{2}}}{{\sinh }^{2}\left( c{\tau }\right) }%
d\Omega ^{2}}\right) ,  \label{monday2}
\end{equation}
The evolution coordinate $\tau $ is related to the radial coordinate $r$ by
the relation\footnote{%
This corrects a mistake in Eq. (4.1.8) of \cite{AoB-book}, and a typo in Eq.
(4) of \cite{ADOT-1}.}
\begin{equation}
\left( \frac{dr}{d\tau }\right) ^{2}=\frac{c^{4}}{{\sinh }^{4}\left( c{\tau }%
\right) }=\left( r-r_{-}\right) ^{2}\left( r-r_{+}\right) ^{2}=\left(
r-M+c\right) ^{2}\left( r-M-c\right) ^{2}=\left[ \left( r-M\right) ^{2}-c^{2}%
\right] ^{2},  \label{monday1}
\end{equation}
whose integration yields the relation
\begin{equation}
r=-c\text{cotgh}\left( c\tau \right) +M.  \label{Thu-1}
\end{equation}
In the \textit{extremal} case ($c=0$), since $r_{+}=r_{-}=r_{H}=M$, the
relation (\ref{Thu-1}) reduces to the definition of $\tau \equiv \frac{1}{%
r_{H}-r}=\frac{1}{M-r}$ previously introduced, and the metric (\ref{monday2}%
) becomes
\begin{eqnarray}
ds_{c=0}^{2} &=&-\,e^{2U}dt^{2}+e^{-2U}\left[ {dr^{2}+\left( r-r_{H}\right)
^{2}d\Omega ^{2}}\right] =-\,e^{2U}dt^{2}+e^{-2U}\left[ {dr^{2}+\left(
r-M\right) ^{2}d\Omega ^{2}}\right] =  \notag \\
&&  \notag \\
&=&-\,e^{2U}dt^{2}+\frac{e^{-2U}}{\tau ^{2}}\left[ \frac{d\tau ^{2}}{\tau
^{2}}{+d\Omega ^{2}}\right] =-\,e^{2U}dt^{2}+e^{-2U}\left[ {d\rho }^{2}{%
+\rho }^{2}{d\Omega ^{2}}\right] ,  \label{Thu-4}
\end{eqnarray}
where in the last step we recalled the definition (\ref{Thu-3}). Notice that
Eq. (\ref{Thu-4}) is nothing but the static Papapetrou-Majumdar class of
metrics (\ref{papa1})-(\ref{papa2}).

The relation (\ref{monday1}) allows one to rewrite the metric (\ref{monday2}%
) as follows:
\begin{eqnarray}
ds^{2} &=&-\,e^{2U}dt^{2}+e^{-2U}\left[ {dr^{2}+\left( r-r_{-}\right) \left(
r-r_{+}\right) d\Omega ^{2}}\right] =  \notag \\
&=&-\,e^{2U}dt^{2}+e^{-2U}\left\{ {dr^{2}+\left[ \left( r-M\right) ^{2}-c^{2}%
\right] d\Omega ^{2}}\right\} .  \label{Wed-11}
\end{eqnarray}

By comparing such an expression with the expression of $ds^{2}$ given below
Eq. (22) of the first of Refs. \cite{K3}, one obtains that the \textit{%
(squared) effective radial coordinate} $R^{2}$ introduced in the first of
Refs. \cite{K3} can also be written as follows:
\begin{equation}
R^{2}=e^{-2U}{\left( r-r_{-}\right) \left( r-r_{+}\right) =}e^{-2U}{\left[
\left( r-M\right) ^{2}-c^{2}\right] ,\label{CERN-solo-1}}
\end{equation}
which in the particular case of the \textit{Maxwell-Einstein-(axion-)dilaton}
BH treated in \cite{K3} reads (both in \textit{non-extremal} and \textit{%
extremal} cases; see Eq. (25) of the first of Refs. \cite{K3}, and Eq. (\ref
{CERN-sunny-1}))
\begin{equation}
R^{2}=r^{2}-\Sigma ^{2}.  \label{CERN-solo-2}
\end{equation}
Remarkably, in the \textit{multi-dilaton system} such an expression enjoys
the following generalization (see the treatment of Sect. \ref{Sect7}, as
well as Eq. (\ref{CERN-sunny-2})):
\begin{equation}
R^{2}=r^{2}-\frac{1}{2}G_{ab}\left( \phi _{\infty }\right) \Sigma ^{a}\Sigma
^{b}.  \label{CERN-solo-3}
\end{equation}

In the \textit{non-extremal} case ($c\neq 0$), the requirement of \textit{%
non-vanishing}, \textit{finite} $A_{H}$ (\textit{``large''} BH) implies the
near-horizon limit of the scale function $U$ to be \cite{K2} \cite{ADFT}
\begin{eqnarray}
\lim_{\tau \rightarrow -\infty }e^{-2U} &=&\left( \frac{A_{H}}{4\pi }\right)
\frac{\sinh ^{2}\left( c\tau \right) }{c^{2}}\sim R_{+}^{2}\frac{e^{-2c\tau }%
}{c^{2}}=  \notag \\
&=&\left( \frac{A_{H}}{4\pi }\right) \frac{1}{\left( r-r_{-}\right) \left(
r-r_{+}\right) }=\frac{R_{+}^{2}}{\left( r-r_{-}\right) \left(
r-r_{+}\right) },  \label{Thu-2}
\end{eqnarray}
where the relation (\ref{monday1}) was used, and $R_{+}$ is the \textit{%
effective} \textit{BH radius} defined by Eq. (\ref{sunday1}) (which, as
pointed out above, in presence of scalars and/or for $J\neq 0$ does \textit{%
not} coincide with the radius $r_{+}$ of the outer -event- horizon).

Thus, the near-horizon limit of the particular, spherically symmetric,
asymptotically flat, static Papapetrou-Majumdar metric (\ref{monday2}) (or (%
\ref{Wed-11})) reads
\begin{eqnarray}
\lim_{r\rightarrow r_{H}^{+}}\left. ds^{2}\right| _{c\neq 0} &=&-\left(
\frac{4\pi }{A_{H}}\right) \left( r-r_{-}\right) \left( r-r_{+}\right)
dt^{2}+\left( \frac{A_{H}}{4\pi }\right) \frac{{dr^{2}}}{\left(
r-r_{-}\right) \left( r-r_{+}\right) }{+}\left( \frac{A_{H}}{4\pi }\right) {%
d\Omega ^{2}=}  \notag \\
&=&-\frac{\left( r-r_{-}\right) \left( r-r_{+}\right) }{R_{+}^{2}}dt^{2}+%
\frac{R_{+}^{2}}{\left( r-r_{-}\right) \left( r-r_{+}\right) }{%
dr^{2}+R_{+}^{2}d\Omega ^{2}=}  \notag \\
&=&-\frac{\pi \left( r-r_{-}\right) \left( r-r_{+}\right) }{S_{BH}}dt^{2}+%
\frac{S_{BH}}{\pi \left( r-r_{-}\right) \left( r-r_{+}\right) }{%
dr^{2}+R_{+}^{2}d\Omega ^{2}\sim }  \notag \\
&\sim &-\frac{c^{2}}{R_{+}^{2}}e^{2c\tau }dt^{2}+R_{+}^{2}\frac{e^{-2c\tau }%
}{c^{2}}dr^{2}+{R_{+}^{2}d\Omega ^{2}.}
\end{eqnarray}

By defining $\widehat{\rho }\equiv e^{c\tau }$, one obtains that
\begin{equation}
\lim_{\tau \rightarrow -\infty }\left. ds^{2}\right| _{c\neq 0}=-\left(
\frac{\widehat{\rho }c}{R_{+}}\right) ^{2}dt^{2}+R_{+}^{2}\left( d\widehat{%
\rho }^{2}+d\Omega ^{2}\right) .
\end{equation}
Thus, in the non-extremal case, the near-horizon geometry is not flat nor
conformally flat; moreover, the distance in physical coordinate $\widehat{%
\rho }$ of any point from the horizon $\widehat{{\rho }}_{H}{=0}$ is \textit{%
finite}.

On the other hand, in the extremal case ($c=0$, and thus $r_{+}=r_{-}\equiv
r_{H}$), in order to have a non-vanishing, finite $A_{H}$, $U\left( \tau
\right) $ should behave near the event horizon as follows:
\begin{equation}
\lim_{\tau \rightarrow -\infty }e^{-2U}=\frac{R_{H}^{2}}{\left(
r-r_{H}\right) ^{2}}=R_{H}^{2}\tau ^{2}=\frac{R_{H}^{2}}{\rho ^{2}},
\end{equation}
where we recall once again the notation $R_{+,c=0}^{2}\equiv R_{H}^{2}$, and
thus the near-horizon limit of the extremal metric (\ref{Thu-4}) reads
\begin{eqnarray}
\lim_{\tau \rightarrow -\infty }ds_{c=0}^{2} &=&-\,\frac{\left(
r-r_{H}\right) ^{2}}{R_{H}^{2}}dt^{2}+\frac{R_{H}^{2}}{\left( r-r_{H}\right)
^{2}}\left[ {dr^{2}+\left( r-r_{H}\right) ^{2}d\Omega ^{2}}\right] =  \notag
\\
&=&-\,R_{H}^{2}\tau ^{2}dt^{2}+R_{H}^{2}\left[ \frac{d\tau ^{2}}{\tau ^{2}}{%
+d\Omega }^{2}\right] =-\,\frac{\rho ^{2}}{R_{H}^{2}}dt^{2}+R_{H}^{2}\left[
\frac{{d\rho }^{2}}{\rho ^{2}}{+d\Omega ^{2}}\right] .  \notag \\
&&
\end{eqnarray}

\begin{figure}[h!]
\begin{center}
\includegraphics[width=0.48\textwidth,height=0.3\textheight]{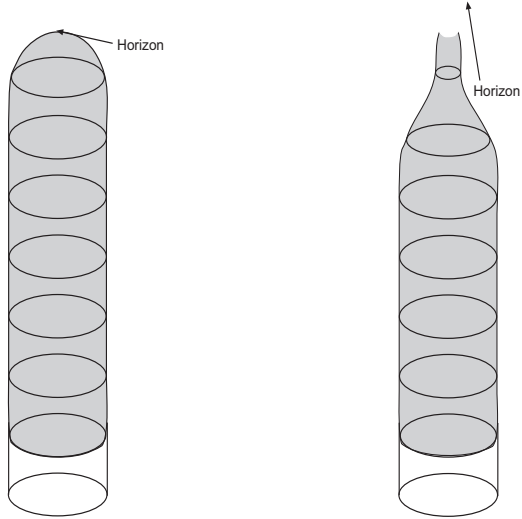}
\end{center}
\caption{\textbf{On the left:} Euclidean section of the near-horizon \textit{%
``cigar''} geometry of a \textit{large} non-extremal BH ($T_{BH}\neq0$).
\textbf{On the right:} Euclidean section of the \textit{infinite throat}
near-horizon geometry of a \textit{large} extremal BH ($T_{BH}=0$).
\protect\cite{K2} }
\end{figure}

As noticed in \cite{K2}, the physical difference between the extremal and
non-extremal class of metrics (\ref{monday2}) gives an hint why in the
former case the \textit{Attractor Mechanism} occurs, contrarily to the
second case. In the extremal case, a scalar field $\phi \left( \omega
\right) $ (where $\omega $ is the physical distance coordinate for $c=0$,
defined above as $\omega \equiv \ln \rho =-\ln \left( -\tau \right) $)
looses memory of the initial conditions of its radial dynamical evolution
when
\begin{equation}
\frac{d\phi \left( \omega \right) }{d\omega }\rightarrow 0,
\end{equation}
and this exactly happens in the near-horizon limit $r\rightarrow
r_{H}^{+}\Leftrightarrow \omega _{H}=-\infty $, meaning that the throat of
an extremal (static, spherically symmetric, asymptotically flat) BH is
\textit{infinite}.

This actually does \textit{not} occur in the non-extremal case, in which the
non-extremal physical coordinate of the event (outer) horizon is $\widehat{%
\rho }_{H}=0$, and
\begin{equation}
\lim_{\widehat{\rho }\rightarrow 0^{+}}\frac{d\phi \left( \widehat{\rho }%
\right) }{d\widehat{\rho }}\nrightarrow 0.
\end{equation}
In other words, the Attractor Mechanism does \textit{not} hold, because the
scalar fields $\phi ^{a}\left( \widehat{\rho }\right) $ \textit{do not have
enough (radial) time to loose memory of the asymptotical data of their
evolution }\cite{K2}.

As shown in \cite{ferrara3} (and related Refs. therein), in general all the
evolution equations for the gravitational degree of freedom ${U(\tau )}$,
scalar fields ${\phi ^{a}(\tau )}$, and electric and magnetic field
strengths (whose unique non-vanishing components respectively are, within
the considered \textit{Ans\"{a}tze} on space-time symmetries, $F_{t\tau
}^{\Lambda }=\partial _{\tau }\psi ^{\Lambda }$ and $G_{t\tau \Lambda
}=\partial _{\tau }\chi _{\Lambda }$) can be derived in a one-dimensional
Lagrangian formulation, obtained by performing a dimensional reduction based
on the space-time symmetries (staticity, spherical symmetry, asymptotical
flatness) of the considered background (\ref{monday2}), yielding the
following (almost geodesic) effective Lagrangian:
\begin{equation}
\mathcal{L}{}\left( {U(\tau ),\;\phi ^{a}(\tau ),p^{\Lambda },\,q_{\Lambda }}%
\right) =\left( {\frac{{dU}\left( \tau \right) }{{d\tau }}}\right) ^{2}+%
\frac{1}{2}G_{ab}\left( \phi \left( \tau \right) \right) \frac{{d\phi ^{a}}}{%
{d\tau }}\frac{{d\phi ^{b}}}{{d\tau }}+e^{2U\left( \tau \right)
}V_{BH}\left( {\phi }\left( \tau \right) {,\,p^{\Lambda },\,q_{\Lambda }}%
\right) ,  \label{tuesday-fin-8}
\end{equation}
along with the constraint
\begin{equation}
\left( {\frac{{dU}\left( \tau \right) }{{d\tau }}}\right) ^{2}+\frac{1}{2}%
G_{ab}\left( \phi \left( \tau \right) \right) \frac{{d\phi ^{a}}}{{d\tau }}%
\frac{{d\phi ^{b}}}{{d\tau }}-e^{2U\left( \tau \right) }V_{BH}\left( {\phi }%
\left( \tau \right) {,\,p^{\Lambda },\,q_{\Lambda }}\right) =c^{2}.
\label{monday3}
\end{equation}

Taking into account that the asymptotical limit of the gravitational scale
function reads
\begin{equation}
\lim_{\tau \rightarrow 0^{-}}U(\tau )\rightarrow M\tau ,  \label{Sun-1}
\end{equation}
the asymptotical limit of the constraint (\ref{monday3}) yields, by
recalling the definition (\ref{monday5}) of the scalar charges, the relation
(\ref{monday4}). On the other hand, by noticing that (under minimal
requirements of regularity) Eq. (\ref{Thu-2}) yields \cite{gibbons3}
\begin{equation}
\lim_{\tau \rightarrow -\infty }e^{2U}\sim c^{2}\frac{e^{2c\tau }}{R_{+}^{2}}%
\Leftrightarrow \lim_{\tau \rightarrow -\infty }U\left( \tau \right) \sim
c\tau \Leftrightarrow \lim_{\tau \rightarrow -\infty }\frac{{dU}\left( \tau
\right) }{{d\tau }}\sim c,
\end{equation}
and considering that the near-horizon limit of $\frac{{d\phi ^{a}}}{{d\tau }}
$ is given by
\begin{equation}
\lim_{\tau \rightarrow -\infty }\frac{{d\phi ^{a}}\left( \tau \right) }{{%
d\tau }}\sim e^{c\tau },
\end{equation}
it is immediate to check that the near-horizon limit of the constraint (\ref
{monday3}) yields nothing but the trivial identity $c^{2}=c^{2}$.

It is worth remarking that the asymptotic relation (\ref{monday4}) does
\textit{not} rely on supersymmetry at all, and holds under the general
assumptions made on the field content, couplings, boundary conditions and
space-time symmetries of the physical system being considered.

On the other hand, the derivation of the so-called \textit{BPS}(-like)
\textit{bound} (\ref{sunday!1-gen}) requires both supersymmetry and duality
invariance \cite{gibbons2}. A conceivable generalization of such a bound in
the considered framework embedded in $\mathcal{N}=2$, $d=4$ (ungauged)
supergravity is
\begin{equation}
M_{ADM}^{2}\geqslant \left| Z\right| ^{2}\left( \phi _{\infty },p^{\Lambda
},q_{\Lambda }\right) ,  \label{monday-eve-1}
\end{equation}
where $Z=Z\left( \phi ,p,q\right) $ is the $\mathcal{N}=2$, $d=4$ \textit{%
central charge function} (in the real parametrization of the scalar
manifold). In particular, under the assumptions made $\left| Z\right|
^{2}\left( \phi _{\infty },p,q\right) $ is the squared absolute value of the
central charge of the $\mathcal{N}=2$, $d=4$ Poincar\'{e} superalgebra.

It is worth pointing out that for $\mathcal{N}>2$-extended, $d=4$ (ungauged)
supergravities $\left| Z\right| $ can be generally replaced by the absolute
value of the largest eigenvalue of the central charge matrix; indeed,
generally the absolute values of the eigenvalues of such a matrix can be
ordered \textit{uniquely}, \textit{i.e.} as functions of the scalars, with
the ordering eventually depending only on the considered supporting BH
charge configuration. In $\mathcal{N}=2$, $d=4$ (ungauged) supergravity
coupled to $n_{V}$ vector multiplets, the bound (\ref{monday-eve-1}) can
eventually be extended by a chain of (not necessarily strict) inequalities,
involving the squared absolute values of the (covariant) derivatives of the
\textit{central charge function} $Z$, \textit{i.e.} the so-called \textit{%
matter charges.} By disregarding $M_{\left( ADM\right) }^{2}$, such a chain
of inequalities generally hold for any value of the radial coordinate $r$,
the ordering eventually depending only on the considered supporting BH
charge configuration.

Now, by using the (supersymmetry-independent) asymptotical relation (\ref
{monday4}), Eq. (\ref{monday-eve-1}) would imply that in $\mathcal{N}=2$, $%
d=4$ supergravity (under the assumptions made) it holds that
\begin{equation}
V_{BH}(\phi _{\infty },p^{\Lambda },q_{\Lambda })-\frac{1}{2}G_{ab}\left(
\phi _{\infty }\right) \Sigma ^{a}\Sigma ^{b}+c^{2}\geqslant \left| Z\right|
^{2}\left( \phi _{\infty },p^{\Lambda },q_{\Lambda }\right) .
\end{equation}
By also recalling that the BH effective potential in $\mathcal{N}=2$, $d=4$
(ungauged) supergravity reads\footnote{%
We use covariant derivatives, because the real parametrization of the
scalars inherits the \textit{K\"{a}hler} structure of the complex
parametrization. Indeed, vector multiplets' scalars of $\mathcal{N}=2$, $d=4$
span \textit{(special) K\"{a}hler} manifolds.} (in the real parametrization
of the scalars)
\begin{equation}
V_{BH}(\phi ,p^{\Lambda },q_{\Lambda })=\left| Z\right| ^{2}\left( \phi
,p^{\Lambda },q_{\Lambda }\right) +G^{ab}\left( \phi \right) \left(
D_{a}Z\right) \left( \phi ,p^{\Lambda },q_{\Lambda }\right) D_{b}Z\left(
\phi ,p^{\Lambda },q_{\Lambda }\right) ,
\end{equation}
one thus would achieve the following result\footnote{%
The extremal limit ($c=0$) of inequality (\ref{monday-fin-1}) corrects and
clarifies Eq. (2.11) of \cite{K2}.}:
\begin{equation}
\begin{array}{l}
G^{ab}\left( \phi _{\infty }\right) \left( D_{a}Z\right) \left( \phi
_{\infty },p^{\Lambda },q_{\Lambda }\right) D_{b}Z\left( \phi _{\infty
},p^{\Lambda },q_{\Lambda }\right) -\frac{1}{2}G_{ab}\left( \phi _{\infty
}\right) \Sigma ^{a}\Sigma ^{b}+c^{2}= \\
\\
=G^{ab}\left( \phi _{\infty }\right) \left[ D_{a}Z\left( \phi _{\infty
},p,q\right) D_{b}Z\left( \phi _{\infty },p,q\right) -\frac{1}{2}\Sigma
_{a}\Sigma _{b}\right] +c^{2}\geqslant 0,
\end{array}
\label{monday-fin-1}
\end{equation}
where $\Sigma _{a}\equiv G_{ab}\left( \phi _{\infty }\right) \Sigma ^{b}$.

In the treatment given further below, the \textit{BPS}(-like) \textit{bound}
(\ref{monday-eve-1}), or equivalently the bound (\ref{monday-fin-1}), will
be shown to hold for the ($\left( U\left( 1\right) \right) ^{6}\rightarrow
\left( U\left( 1\right) \right) ^{2}$ truncation of the)
Maxwell-Einstein-(axion-)dilaton supergravity, corresponding to a particular
$1$-modulus model of $\mathcal{N}=2$, $d=4$ supergravity (see Sects. \ref
{Sect6} and \ref{Sect7}). Furthermore, it is worth pointing out that in \cite
{K2} the bounds (\ref{monday-eve-1}) and (\ref{monday-fin-1}) are claimed to
hold in general for $c=0$.

Extremal BPS BHs have $c=0$ and they saturate the bounds (\ref{monday-eve-1}%
) and (\ref{monday-fin-1}), because $G^{ab}\left( \phi _{\infty }\right) %
\left[ D_{a}Z\left( \phi _{\infty },p,q\right) D_{b}Z\left( \phi _{\infty
},p,q\right) -\frac{1}{2}\Sigma _{a}\Sigma _{b}\right] =0$ due to the
\textit{first order BPS equations} (see \textit{e.g.} \cite{ferrara3}).
However, by assuming the bounds (\ref{monday-eve-1}) and (\ref{monday-fin-1}%
) to hold (also) for $c\neq 0$, it is conceivable that they are saturated in
some particular cases. Consequently, (beside non-extremal non-BPS) also
non-extremal BPS BHs might exist.

In the case of \textit{frozen} scalars (trivial radial dynamics: $\frac{%
d\phi ^{a}\left( r\right) }{dr}=0$ $\forall a=1,...,n$ and $\forall r\in %
\left[ r_{H},\infty \right) $) $\Sigma _{a}=0$, and thus the BPS bound (\ref
{monday-fin-1}) simplifies to
\begin{equation}
G^{ab}\left( \phi _{\infty }\right) \left( D_{a}Z\right) \left( \phi
_{\infty },p^{\Lambda },q_{\Lambda }\right) D_{b}Z\left( \phi _{\infty
},p^{\Lambda },q_{\Lambda }\right) +c^{2}\geqslant 0.
\label{monday-fin-1-frozen}
\end{equation}
For $c=0$ one can have the well known \textit{double-extremal BPS} BHs for
which (see \textit{e.g.} \cite{ferrara3})
\begin{equation}
D_{a}Z\left( \phi \left( r\right) ,p^{\Lambda },q_{\Lambda }\right)
=0\forall a=1,...,n,~\forall r\in \left[ r_{H},\infty \right) ,
\end{equation}
but also \textit{double-extremal non-BPS} BHs, having
\begin{equation}
G^{ab}\left( \phi _{\infty }\right) \left( D_{a}Z\right) \left( \phi
_{\infty },p^{\Lambda },q_{\Lambda }\right) D_{b}Z\left( \phi _{\infty
},p^{\Lambda },q_{\Lambda }\right) >0,  \label{CERN-cloudy2}
\end{equation}
whose first example was explicitly constructed in \cite{K2-bis}). On the
other hand, for $c\neq 0$ one can have \textit{double-non-extremal non-BPS}
BHs for which the strict inequality holds in (\ref{monday-fin-1-frozen}).
Notice that the (strict) positive definiteness of the metric of the scalar
manifold prevents the bound (\ref{monday-fin-1-frozen}) to be saturated for $%
c\neq 0$; thus, \textit{double-non-extremal BHs or non-extremal BHs with no
scalars at all} (such as non-extremal RN BHs) \textit{should never be BPS}%
\footnote{%
As pointed out above, the clash with some results in recent literature%
\textbf{\ (}\cite{Miller}; see also \cite{Cer-Dal,ADOT-1}\textbf{)} is just
apparent, since a different, slightly generalized meaning, is given to
''BPS'' therein, as ''admitting a \textit{first-order} (and thus \textit{%
BPS-like}) formulation''.}.

In the considered framework, the effective (squared) radius relevant for the
computation of the BH entropy is given by the following generalization of
the first of Eqs. (\ref{monday-after-1}):
\begin{equation}
r_{\pm }=M\pm \sqrt{M^{2}-V_{BH}\left( \phi _{\infty },p,q\right) +\frac{1}{2%
}G_{ab}\left( \phi _{\infty }\right) \Sigma ^{a}\Sigma ^{b}}.
\label{Sun-Sun-1}
\end{equation}

Furthermore, as mentioned above, in \cite{Cer-Dal,ADOT-1} a first order
formalism for extremal (and some cases of non-extremal \cite{ADOT-1})
static, spherically symmetric BHs was developed, based on a positive
definite \textit{generalized superpotential} $\mathcal{W}$. In the extremal
case $\mathcal{W}$ is a function only of the conserved BH electric and
magnetic charges and of the scalars, and it does \textit{not} depend
explicitly on the radial evolution coordinate $\tau $:
\begin{equation}
\mathcal{W}=\mathcal{W}\left( \phi \left( \tau \right) p,q\right) .
\label{tuesday8}
\end{equation}
By exploiting the relation between $\mathcal{W}$ and $V_{BH}$, and using the
first order differential Eqs. satisfied by $\mathcal{W}$, in \cite{ADOT-1}%
\textbf{\ }it was proved that in the \textit{extremal }case
\begin{gather}  \label{tuesday8-bis}
\lim_{\tau \rightarrow -\infty }\mathcal{W}\left( \phi \left( \tau \right)
p,q\right) =R_{H}\left( p,q\right) =\sqrt{V_{BH}\left( \phi _{H}\left(
p,q\right) ,p,q\right) }\text{~\textit{(near-horizon limit)}};  \notag \\
\notag \\
\lim_{\tau \rightarrow 0^{-}}\mathcal{W}\left( \phi \left( \tau \right)
p,q\right) =M_{ADM}\left( \phi _{\infty },p,q\right) =r_{H}\left( \phi
_{\infty },p,q\right) ~\text{\textit{(asymptotical limit)}};  \notag \\
\notag \\
\frac{d\mathcal{W}\left( \phi \left( \tau \right) p,q\right) }{d\tau }%
\geqslant 0~\text{(}\tau \text{\textit{-monotonicity of }}\mathcal{W}\text{)}
\notag \\
\Downarrow  \notag \\
r_{H}\left( \phi _{\infty },p,q\right) =M_{ADM}\left( \phi _{\infty
},p,q\right) =\lim_{\tau \rightarrow 0^{-}}\mathcal{W}\left( \phi \left(
\tau \right) p,q\right) \geqslant \sqrt{V_{BH}\left( \phi _{H}\left(
p,q\right) ,p,q\right) }=R_{H}\left( p,q\right) .  \notag \\
\end{gather}
The first two equations give the general expressions of the \textit{%
effective radius} $R_{H}\left( p,q\right) $ (defined by Eq. (\ref{sunday1}))
and of the ADM mass $M_{ADM}\left( \phi _{\infty },p,q\right) =r_{H}\left(
\phi _{\infty },p,q\right) $ (see Eq. (\ref{LNF-1})) for static, spherically
symmetric, asymptotically flat, dyonic \textit{extremal} BHs.

Thus, $\mathcal{W}$ given by Eq. (\ref{tuesday8}) is a positive \textit{%
monotonic} function, \textit{decreasing} from the \textit{moduli-dependent}
value $M_{ADM}\left( \phi _{\infty },p,q\right) =r_{H}\left( \phi _{\infty
},p,q\right) $ at $r\rightarrow \infty $, towards the \textit{%
moduli-independent} value $R_{H}\left( p,q\right) =\sqrt{V_{BH}\left( \phi
_{H}\left( p,q\right) ,p,q\right) }$ at the event BH horizon radius $r=r_{H}$%
. Consequently, in the extremal case $\mathcal{W}^{2}$ appears to be the
suitable candidate for the $C$-function\footnote{%
Notice that, when going beyond the Einstein approximation, the uniqueness of
the $C$-function does not generally hold any more. For example, in \textit{%
Lovelock gravity} \cite{Lovelock} (at least) two independent $C$-functions
can be determined.}, whose existence was shown in \cite{GJMT}.

In Sect. \ref{Sect7} we will prove that the inequality given in the fourth
line of (\ref{tuesday8-bis}) can actually be further specialized in the
\textit{multi-dilaton system} (\textit{i.e.} in $\mathcal{N}=2$ \textit{%
quadratic} $d=4$ supergravity), in which it holds that
\begin{eqnarray}
\frac{S_{BH,c=0}\left( p,q\right) }{\pi } &\equiv &R_{H}^{2}\left(
p,q\right) =r_{H}^{2}\left( \phi _{\infty },p,q\right) -\frac{1}{2}%
G_{ab}\left( \phi _{\infty }\right) \Sigma ^{a}\left( \phi _{\infty
},p,q\right) \Sigma ^{b}\left( \phi _{\infty },p,q\right) =  \notag \\
&&  \notag \\
&=&\mathcal{W}^{2}\left( \phi _{\infty },p,q\right) -2G^{ab}\left( \phi
_{\infty }\right) \lim_{\tau \rightarrow 0^{-}}\left( \partial _{a}\mathcal{W%
}\left( \phi \left( \tau \right) ,p,q\right) \right) \partial _{b}\mathcal{W}%
\left( \phi \left( \tau \right) ,p,q\right) .  \notag \\
&&  \label{CERN-night-3}
\end{eqnarray}
This is nothing but the \textit{extremal limit} $c=0$, expressed in the
framework of \textit{first order formalism}, of Eq. (\ref{sunday-Night-2}),
or equivalently the \textit{many-moduli generalization} of the \textit{%
extremal case} of formula holding for the so-called \textit{(axion-)dilaton
extremal BH} \cite{K3}, given by Eq. (\ref{Lun-1}) or Eq. (\ref{Sun-Sun-2})
below; furthermore, in the last step Eq. (\ref{Wed-4}) was used (for the
assumptions on the limit, see footnotes 8 and 16).

It is here worth pointing out that in the \textit{non-extremal case} (%
\textit{i.e.} $c\neq 0$) the generalization from the \textit{one-modulus}
formula (\ref{Sun-Sun-2}) to the \textit{many-moduli} formula (see Eq. (\ref
{sunday-Night-2}))
\begin{equation}
\frac{S_{BH,c\neq 0}\left( \phi _{\infty },p,q\right) }{\pi }\equiv
R_{+}^{2}\left( \phi _{\infty },p,q\right) =r_{+}^{2}\left( \phi _{\infty
},p,q\right) -\frac{1}{2}G_{ab}\left( \phi _{\infty }\right) \Sigma
^{a}\left( \phi _{\infty },p,q\right) \Sigma ^{b}\left( \phi _{\infty
},p,q\right)
\end{equation}
is only \textit{guessed}, but at present cannot be rigorously proved.
Indeed, for static, spherically symmetric, asymptotically flat dyonic
\textit{non-extremal} BHs a \textit{first order formalism} is currently
unavailable, so there is no way to compute the scalar charges (beside the
direct integration of the Eqs. of motion of the scalars, as far as we know
feasible only for the (axion-)dilaton BH \cite{K3}).

The crucial feature expressed by Eq. (\ref{CERN-night-3}) is the \textit{%
disappearance} of the dependence on the \textit{(asymptotical) moduli} $\phi
_{\infty }$ in the combination $r_{H}^{2}-\frac{1}{2}G_{ab}\Sigma ^{a}\Sigma
^{b}$ of quantities $r_{H}^{2}$ and $G_{ab}\Sigma ^{a}\Sigma ^{b}$, which
separately are \textit{moduli-dependent}.\smallskip

Let us now recall the (guessed generalization of the) BPS bound (\ref
{monday-eve-1}), which in the considered context is equivalent, through the
relation (\ref{monday4}), to the bound (\ref{monday-fin-1}). The first order
formalism for extremal BHs introduced in \cite{Cer-Dal,ADOT-1} might
provide, through the second limit of Eq. (\ref{tuesday8-bis}) and the
knowledge of an explicit expression for $\mathcal{W}$, a proof of the $c=0$
limit of inequalities (\ref{monday-eve-1}) and (\ref{monday-fin-1}) in some
of the $\mathcal{N}\geqslant 2$, $d=4$ (ungauged) supergravities. However,
let us recall once again that currently an expression of $\mathcal{W}$ for a
generic $\mathcal{N}=2$, $d=4$ supergravity is unavailable.\medskip

Summarizing, one ends up with four sets of relations, which can be used to
study the \textit{BH thermodynamics} (at \textit{equilibrium}, for $c=0$)
and its interconnection with \textit{attractors} of the scalar dynamics (and
\textit{supersymmetry}, when the bosonic system under consideration is
embedded in \textit{supergravity}). Let us recall once again that the most
general environment considered in the present work is given by stationary
(rotating with constant angular momentum), spherically symmetric and
asymptotically flat BHs.

The mentioned four sets of relations, whose \textit{interplay} is
(partially) shown by the treatment given above, are:

\textbf{1}) the spatial asymptotic ($r\rightarrow \infty \Leftrightarrow
\tau \rightarrow 0^{-}$) limit (\ref{monday4}) of the \textit{constraint }(%
\ref{monday3}) of the effective one-dimensional Lagrangian density (\ref
{tuesday-fin-8}). It holds true under the assumed asymptotic boundary
conditions for $U\left( \tau \right) $ and $\phi ^{a}\left( \tau \right) $,
for $c\in \mathbb{R}^{+}$ and $J=0$;

\textbf{2}) the general expression of $r_{\pm }^{2}$ (and of the \textit{%
effective squared radius} $R_{+}^{2}$) in terms of the other thermodynamical
quantities (thence, directly yielding the value of the Bekenstein-Hawking
BH\ entropy);

\textbf{3}) the relation given by the \textit{Smarr-integrated}, finite
version of the \textit{(generalized) first law of BH thermodynamics} (in
presence of scalars);

\textbf{4}) the asymptotical inequality (\ref{monday-eve-1}), which is the
proposed generalization of the \textit{BPS} \textit{bound} (\ref
{sunday!1-gen}) \cite{gibbons2}\textbf{\ }to the presence of scalars in $%
\mathcal{N}=2$, $d=4$ ungauged supergravity. In turn, let us recall that the
\textit{BPS} \textit{bound} (\ref{sunday!1-gen}) is the generalization of
the so-called \textit{Theorem of Positivity of Energy} (\ref{sunday!1})-(\ref
{sunday!2}) in the presence of electric charge $q$ and magnetic charge $p$
pertaining to the metric background (absence of scalars). It is worth
remarking that the saturation of the \textit{BPS bound }always implies the
preservation of some (amount of the) supersymmetries, out of the all the
ones pertaining to the asymptotically flat space-time background. The
\textit{BPS bounds} (\ref{sunday!1-gen}) and (\ref{monday-eve-1}) hold for $%
J=0$, and, differently from the relations of points 1, 2 and 3, their
derivation relies on supersymmetry and duality invariance (see \cite
{gibbons2} and Refs. therein). As treated above, by combining the bound (\ref
{monday-eve-1}) with the asymptotic relation (\ref{monday4}), one gets the $%
c $-parametrized bound (\ref{monday-fin-1}), whose extremal limit ($c=0$) in
\cite{K2} is stated to hold true in general in the considered framework. In
Sect. \ref{Sect7} we will prove the extremal limit ($c=0$) of the bounds (%
\ref{monday-eve-1}) and (\ref{monday-fin-1}) to hold in the ($\left( U\left(
1\right) \right) ^{6}\rightarrow \left( U\left( 1\right) \right) ^{2}$
truncation of the) Maxwell-Einstein-(axion-)dilaton supergravity, and we
will further elaborate on the non-extremal case, as well.
\setcounter{equation}0

\section{\label{Sect6}The Maxwell-Einstein-(Axion-)Dilaton Gravity}

The bosonic part of the $d=4$ Lagrangian density is \cite{CSF}

\begin{equation}
-{}\frac{\mathcal{L}}{\sqrt{-g}}=\left( {\partial {}_{\mu }\phi }\right)
^{2}+e^{4\phi }(\partial _{\mu }a)^{2}+e^{-2\phi }\sum\limits_{\Lambda
=1}^{6}{F_{\mu \nu }^{\Lambda }F_{\Lambda }^{\mu \nu }}+\frac{1}{2}%
a\sum\limits_{\Lambda =1}^{6}{F_{\mu \nu }^{\Lambda }\tilde{F}_{\Lambda
}^{\mu \nu }}+\frac{1}{2}\mathcal{R}.  \label{Wed-1}
\end{equation}
It contains $6$ Maxwell fields (for a total $\left( U(1)\right) ^{6}$
gauge-invariance) and a dilaton-axial scalar
\begin{equation}
s\equiv a+ie^{-2\phi }.  \label{Thu-13}
\end{equation}
The Lagrangian density (\ref{Wed-1}) is the bosonic part of $\mathcal{N}=4$,
$d=4$ \textit{pure} supergravity (a sector of heterotic superstring
compactified on a six-torus ${T}^{6}$; see \textit{e.g.} \cite
{Maharana,Duff-stu}). By truncating $\left( {U(1)}\right) {^{6}\rightarrow }%
\left( {U(1)}\right) {^{2}}$, the bosonic part of $\mathcal{N}=2$, $d=4$
supergravity (coupled to one vector multiplet, $n_{V}=1$) with \textit{%
minimal coupling} (the so-called $t^{2}$ model) \cite{Luciani} is recovered.

The coupled set of field equations for such a system in a static,
spherically symmetric, asymptotically flat dyonic BH background described by
the metric (\ref{monday2}) were studied and solved in (the first Ref. of)
\cite{K3}. In such a paper, the considered effective BH potential was
\begin{equation}
V_{BH}\left( \phi ,q,p\right) =e^{2\phi }q^{2}+e^{-2\phi }p^{2}=Q^{2}+P^{2},
\label{Thu-eve-1}
\end{equation}
where $V_{BH}\left( \phi ,q,p\right) \equiv V_{BH}\left( \phi
,a=0,q,p\right) $ (see Eq. (\ref{Thu-14}) below), and the \textit{dressed
charges}
\begin{equation}
Q\equiv qe^{\phi },~~P\equiv pe^{-\phi }  \label{Wed-Wed-1}
\end{equation}
were introduced. The charges $q$ and $p$ respectively are an electric and a
magnetic charge pertaining to two different $U\left( 1\right) $'s inside $%
\left( U\left( 1\right) \right) ^{6}$; all the other charges are chosen to
vanish, and this allows to put the axion field $a$ to zero (in the extremal
case, such a procedure will be clear from the treatment given below, based
on the BH effective potential).

The scalar charge of the dilaton $\phi $ is found to be (see the first Ref.
of \cite{K3}, and \cite{Garfinkle})
\begin{equation}
\Sigma ^{\phi }\equiv \Sigma =\frac{e^{-2\phi _{\infty }}p^{2}-e^{2\phi
_{\infty }}q^{2}}{2M}=\frac{P_{\infty }^{2}-Q_{\infty }^{2}}{2M},
\label{Wed-2}
\end{equation}
where $\phi _{\infty }$ is the value of the dilaton at spatial infinity ($%
r\rightarrow \infty $), and $M$ is the ADM mass of the BH. By recalling Eq. (%
\ref{monday4}), one obtains ($V_{BH,\infty }\equiv V_{BH}\left( \phi
_{\infty },q,p\right) $):
\begin{gather}
M^{2}+\Sigma ^{2}-V_{BH,\infty }=c^{2}\Leftrightarrow M^{2}+\frac{\left(
P_{\infty }^{2}-Q_{\infty }^{2}\right) }{4M^{2}}-Q_{\infty }^{2}-P_{\infty
}^{2}=c^{2};  \notag \\
\Updownarrow  \notag \\
M=\sqrt{V_{BH,\infty }-\Sigma ^{2}+c^{2}}.  \label{Wed-3}
\end{gather}
By noticing that Eq. (\ref{Wed-9}) \cite{gibbons3} or Eqs. (\ref
{tuesday-fin-6}) and (\ref{tuesday-fin-3}) \cite{ADOT-1} imply $\Sigma $ to
vanish at the critical point(s) of $V_{BH(,\infty )}$, Eqs. (\ref{Wed-2})
and (\ref{Wed-3}) allow one to compute the \textit{dilaton charge} as follows%
\footnote{%
We assume
\begin{equation*}
\left( \partial V_{BH}\right) _{\infty }\equiv \lim_{\tau \rightarrow 0^{-}}%
\frac{\partial V_{BH}\left( \phi \left( \tau \right) ,p,q\right) }{\partial
\phi \left( \tau \right) }=\frac{\partial V_{BH}\left( \phi _{\infty
},p,q\right) }{\partial \phi _{\infty }}.
\end{equation*}
Also notice that Eq. (\ref{Wed-10}) is constrained by the \textit{reality
condition}
\begin{equation*}
1-\frac{1}{4}\left( \frac{\left( \partial V_{BH}\right) _{\infty }}{%
V_{BH,\infty }+c^{2}}\right) ^{2}\geqslant 0.
\end{equation*}
\textit{\ }} ($\partial \equiv \frac{\partial }{\partial \phi }$; also
recall Eq. (\ref{Wed-6})):
\begin{equation}
\Sigma ^{2}\left( \phi _{\infty },p,q\right) =\frac{1}{2}\left( V_{BH,\infty
}+c^{2}\right) \left[ 1-\sqrt{1-\frac{1}{4}\left( \frac{\left( \partial
V_{BH}\right) _{\infty }}{V_{BH,\infty }+c^{2}}\right) ^{2}}\right] .
\label{Wed-10}
\end{equation}

In the extremal case ($c=0$) Eq. (\ref{Wed-10}) simplifies to
\begin{equation}
\left. \Sigma ^{2}\left( \phi _{\infty },p,q\right) \right| _{c=0}=\frac{%
V_{BH,\infty }}{2}\left[ 1-\sqrt{1-\left( \frac{1}{2}\left( \partial \ln
V_{BH}\right) _{\infty }\right) ^{2}}\right] .  \label{Thu-18}
\end{equation}

By recalling Eq. (\ref{tuesday1}), the inner (Cauchy) and outer (horizon)
event radii are given for the considered (charge configuration of the)
Maxwell-Einstein-(axion-)dilaton BH read \cite{K3}
\begin{eqnarray}
r_{\pm }\left( \phi _{\infty },p,q\right) &=&M\pm c=M\pm \sqrt{%
M^{2}-V_{BH,\infty }\left( \phi _{\infty },p,q\right) +\Sigma ^{2}\left(
\phi _{\infty },p,q\right) }=  \notag \\
&&  \notag \\
&=&M\pm \sqrt{M^{2}-P_{\infty }^{2}-Q_{\infty }^{2}+\frac{\left( P_{\infty
}^{2}-Q_{\infty }^{2}\right) ^{2}}{4M^{2}}}.  \label{Thu-eve-2}
\end{eqnarray}
The \textit{squared effective radius} reads \cite{K3}
\begin{eqnarray}
R_{+}^{2}\left( \phi _{\infty },p,q\right) &=&r_{+}^{2}\left( \phi _{\infty
},p,q\right) -\Sigma ^{2}\left( \phi _{\infty },p,q\right)
=M^{2}+c^{2}+2cM-\Sigma ^{2}\left( \phi _{\infty },p,q\right) =  \notag \\
&&  \notag \\
&=&2M^{2}-P_{\infty }^{2}-Q_{\infty }^{2}+\sqrt{\left[ 2M^{2}-\left(
P_{\infty }+Q_{\infty }\right) ^{2}\right] \left[ 2M^{2}-\left( P_{\infty
}-Q_{\infty }\right) ^{2}\right] }.  \notag \\
&&  \label{Sun-Sun-2}
\end{eqnarray}
Thus, Eq. (\ref{sunday1}) yields the BH entropy to be
\begin{eqnarray}
S_{BH}\left( \phi _{\infty },p,q\right) &=&\frac{A_{H}\left( \phi _{\infty
},p,q\right) }{4}=\pi R_{+}^{2}\left( \phi _{\infty },p,q\right) =  \notag \\
&=&\pi \left[ 2M^{2}-P_{\infty }^{2}-Q_{\infty }^{2}+\sqrt{\left[
2M^{2}-\left( P_{\infty }+Q_{\infty }\right) ^{2}\right] \left[
2M^{2}-\left( P_{\infty }-Q_{\infty }\right) ^{2}\right] }\right] .  \notag
\\
&&  \label{Thu-eve-3}
\end{eqnarray}

By recalling the definition (\ref{corr1}) of the extremality parameter, one
can compute the BH temperature\footnote{%
Notice we divide by $S_{BH}\left( \phi _{\infty },p,q\right) $, because
\textit{we always assume it to be strictly positive} (by the
Bekenstein-Hawking entropy-area formula (\ref{sunday0}) (or (\ref{sunday1}%
)), this is equivalent to consider \textit{large} BHs, \textit{i.e.} BHs
with non-vanishing $A_{H}$).}:
\begin{eqnarray}
T_{BH}\left( \phi _{\infty },p,q\right) &=&\frac{c}{2S_{BH}\left( \phi
_{\infty },p,q\right) }=  \notag \\
&&  \notag \\
&=&\frac{\sqrt{M^{2}-P_{\infty }^{2}-Q_{\infty }^{2}+\frac{\left( P_{\infty
}^{2}-Q_{\infty }^{2}\right) ^{2}}{4M^{2}}}}{2\pi \left[ 2M^{2}-P_{\infty
}^{2}-Q_{\infty }^{2}+\sqrt{\left[ 2M^{2}-\left( P_{\infty }+Q_{\infty
}\right) ^{2}\right] \left[ 2M^{2}-\left( P_{\infty }-Q_{\infty }\right) ^{2}%
\right] }\right] }.  \notag \\
&&
\end{eqnarray}

Let us now consider the \textit{extremal} case ($c=0$). Thus, one obtains
the following expression of the ADM mass:
\begin{gather}
c=0; \\
\Updownarrow  \notag \\
M^{4}-\left( P_{\infty }^{2}+Q_{\infty }^{2}\right) M^{2}+\frac{1}{4}\left(
P_{\infty }^{2}-Q_{\infty }^{2}\right) ^{2}=0; \\
\Updownarrow  \notag \\
M_{\pm }^{2}=\frac{\left( \left| P_{\infty }\right| \pm \left| Q_{\infty
}\right| \right) ^{2}}{2}\Leftrightarrow M_{\pm }=\frac{\left| \left|
P_{\infty }\right| \pm \left| Q_{\infty }\right| \right| }{\sqrt{2}},
\label{Thu-6}
\end{gather}
and, by using Eq. (\ref{Wed-2}), of the \textit{dilaton scalar charge}:
\begin{gather}
\Sigma _{\pm ,c=0}=\frac{P_{\infty }^{2}-Q_{\infty }^{2}}{2M_{\pm }}=\frac{%
\left( \left| P_{\infty }\right| +\left| Q_{\infty }\right| \right) \left(
\left| P_{\infty }\right| -\left| Q_{\infty }\right| \right) }{\sqrt{2}%
\left| \left| P_{\infty }\right| \pm \left| Q_{\infty }\right| \right| }; \\
\Downarrow  \notag \\
\left| \Sigma _{\pm }\right| _{c=0}=\frac{\left| P_{\infty }^{2}-Q_{\infty
}^{2}\right| }{2M_{\pm }}=\frac{\left| \left| P_{\infty }\right| +\left|
Q_{\infty }\right| \right| \left| \left| P_{\infty }\right| -\left|
Q_{\infty }\right| \right| }{\sqrt{2}\left| \left| P_{\infty }\right| \pm
\left| Q_{\infty }\right| \right| }=\frac{\left| \left| P_{\infty }\right|
\mp \left| Q_{\infty }\right| \right| }{\sqrt{2}}.  \label{Thu-7}
\end{gather}

Consequently, the BH entropy in the extremal case reads ($%
R_{+,c=0}^{2}\equiv R_{H}^{2}$)
\begin{eqnarray}
S_{BH,c=0} &=&\frac{A_{H,c=0}}{4}=\pi R_{H}^{2}=\pi \left[ M_{\pm
}^{2}-\Sigma _{\pm }^{2}\left( \phi _{\infty },p,q\right) \right] _{c=0}=
\notag \\
&&  \notag \\
&=&2\pi \left[ \pm \left| P_{\infty }Q_{\infty }\right| \right] =2\pi \left|
pq\right| ,  \label{Thu-8}
\end{eqnarray}
from which follows that only the branches $M_{+}$\ and $\Sigma _{+}$\ of
Eqs. (\ref{Thu-6}) and (\ref{Thu-7}) are admissible (the branch ``$-$''
yields negative entropy)\textbf{.} Furthermore, let us remark that the
dependence on the asymptotical value $\phi _{\infty }$ of the dilaton drops
out from the expression of the BH entropy $\left. S_{BH}\right| _{c=0}$ in
the extremal case, as it has to be:
\begin{equation}
\frac{\partial S_{BH,c=0}}{\partial \phi _{\infty }}=0.  \label{Thu-9}
\end{equation}

It is worth mentioning that the formula (\ref{Thu-8}) has an $\left(
SU\left( 1,1\right) \times SO\left( n\right) \right) $-invariant
generalization given by
\begin{equation}
S_{BH,c=0}=2\pi \sqrt{p^{2}q^{2}-\left( p\cdot q\right) ^{2}},
\label{Thu-10}
\end{equation}
where $p^{2}\equiv p\cdot p$, $q^{2}\equiv q\cdot q$ and $p\cdot q\equiv
p^{\Lambda }q_{\Lambda }=p^{\Lambda }q^{\Sigma }\delta _{\Lambda \Sigma }$,
here $\Lambda $ ranging $1,...,n$, with the scalar product $\cdot $ defined
by $\delta _{\Lambda \Sigma }$, the $n$-dim. Euclidean metric. In turn,
formula (\ref{Thu-10}) is a particular case ($m=0$, or $n=0$) of the more
general, $\left( SU\left( 1,1\right) \times SO\left( n,m\right) \right) $%
-invariant expression
\begin{equation}
S_{BH,c=0}=2\pi \sqrt{p^{2}q^{2}-\left( p\cdot q\right) ^{2}},
\label{Thu-11}
\end{equation}
with $\Lambda $ ranging $1,...,n,n+1,...n+m$, and the scalar product $\cdot $
defined by $\eta _{\Lambda \Sigma }$, the Lorentzian metric with signature $%
\left( n,m\right) $.

By putting $m=2$ (or $n=2$) in Eq. (\ref{Thu-11}), one obtains the $\left(
SU\left( 1,1\right) \times SO\left( n,2\right) \right) $-invariant formula
of the BH entropy for the $\mathcal{N}=2$, $d=4$ supergravity coupled to $%
n_{V}$ vector multiplets whose scalar manifold is the sequence of reducible
homogeneous symmetric special K\"{a}hler manifolds (with \textit{cubic}
prepotential) $\frac{SU\left( 1,1\right) }{U\left( 1\right) }\times \frac{%
SO\left( 2,n\right) }{SO\left( 2\right) \times SO\left( n\right) }$ ($%
n_{V}=n+1$, $n\in \mathbb{N}$) (see \textit{e.g.} \cite{bellucci1} and Refs.
therein).

On the other hand, by choosing $m=6$ (or $n=6$) in Eq. (\ref{Thu-11}), one
obtains the $\left( SU\left( 1,1\right) \times SO\left( 6,n\right) \right) $%
-invariant formula of the BH entropy for $\mathcal{N}=4$, $d=4$ supergravity
coupled to $n$ matter multiplets, whose scalar manifold is the sequence of
reducible homogeneous symmetric manifolds $\frac{SU\left( 1,1\right) }{%
U\left( 1\right) }\times \frac{SO\left( 6,n\right) }{SO\left( 6\right)
\times SO\left( n\right) }$ (see \textit{e.g.} \cite{ADFT} and Refs.
therein).

When considering \textit{pure} $\mathcal{N}=4$, $d=4$ supergravity (\textit{%
i.e.} $n=0$: no matter multiplets), one thus obtains $\Lambda =1,...,6$ and $%
SU\left( 1,1\right) $ as overall symmetry; the bosonic sector of the theory
is described by the Lagrangian density (\ref{Wed-1}). By further truncating
the gauge group $\left( U\left( 1\right) \right) ^{6}\rightarrow \left(
U\left( 1\right) \right) ^{2}$ (and thus restricting to $\Lambda =1,2$), one
can interpret $SU\left( 1,1\right) $ as the $U$-duality group \cite{hull} of
the $\mathcal{N}=2$, $d=4$ supergravity \textit{minimally coupled} \cite
{Luciani} to $n_{V}=1$ vector multiplet (the so-called $t^{2}$ model), whose
scalar manifold is\footnote{%
Notice that there is another $\mathcal{N}=2$, $d=4$ supergravity coupled to $%
n_{V}=1$ vector multiplet, whose complex scalar spans the rank-$1$
homogeneous symmetric special K\"{a}hler manifold $\frac{SU\left( 1,1\right)
}{U\left( 1\right) }$, endowed with cubic holomorphic prepotential (the
so-called $t^{3}$ model). However, such a manifold is \textit{not} an
element of the sequence $\frac{SU\left( 1,1\right) }{U\left( 1\right) }%
\times \frac{SO\left( 2,n\right) }{SO\left( 2\right) \times SO\left(
n\right) }$, but rather it is an isolated case (see \textit{e.g.} \cite{CFG}%
).
\par
In this sense, the quadratic sequence $\frac{SU\left( 1,n\right) }{U\left(
1\right) \times SU\left( n\right) }$ is the only one (among all the
homogeneous - symmetric - special K\"{a}hler manifolds) to admit a
consistent \textit{pure theory} (\textit{i.e. no matter multiplets}) limit.
\par
On the other hand, the first ($n=1$) element of the cubic sequence $\frac{%
SU\left( 1,1\right) }{U\left( 1\right) }\times \frac{SO\left( 2,n\right) }{%
SO\left( 2\right) \times SO\left( n\right) }$ is the $2$-moduli model based
on $\left( \frac{SU\left( 1,1\right) }{U\left( 1\right) }\right) ^{2}$ (the
so-called $st^{2}$ model).} $\frac{SU\left( 1,1\right) }{U\left( 1\right) }$%
, which is the $n=1$ element of the sequence of irreducible homogeneous
symmetric special K\"{a}hler manifolds (with quadratic prepotential) $\frac{%
SU\left( 1,n\right) }{U\left( 1\right) \times SU\left( n\right) }$ ($n_{V}=n$%
, $n\in \mathbb{N}$) (see \textit{e.g.} \cite{bellucci1} and Refs. therein).
Consistently, this is the theory described by the Lagrangian (\ref{Wed-1})
when $\Lambda =1,2$ (in particular, in the symplectic basis in which the
holomorphic prepotential reads $F\left( X\right) =-iX^{0}X^{1}$, see \textit{%
e.g.} \cite{Cer-Dal}). Indeed, when all BH charges $\left(
p^{1},p^{2},q_{1},q_{2}\right) $ are switched on, the full fledged form of
the BH entropy (\ref{Thu-8}) reads
\begin{eqnarray}
S_{BH,c=0} &=&2\pi \sqrt{\left( \left( p^{1}\right) ^{2}+\left( p^{2}\right)
^{2}\right) \left( q_{1}^{2}+q_{2}^{2}\right) -\left(
p^{1}q_{1}+p^{2}q_{2}\right) ^{2}=}  \notag \\
&=&2\pi \sqrt{\left( p^{1}\right) ^{2}q_{2}^{2}+\left( p^{2}\right)
^{2}q_{1}^{2}-2p^{1}q_{1}p^{2}q_{2}}=2\pi \left|
p^{1}q_{2}-p^{2}q_{1}\right| ,  \label{Thu-12}
\end{eqnarray}
Notice that Eq. (\ref{Thu-12}) reduces to Eq. (\ref{Thu-8}) for
$\left( p\equiv p^{1},q\equiv q_{2},p^{2}=0,q_{1}=0\right) $
or\\$\left( p\equiv p^{2},q\equiv q_{1},p^{1}=0,q_{2}=0\right) $.
Moreover, it corresponds to the $\left( SU\left( 1,1\right) \times
SO\left( n\right) \right) $-invariant formula (\ref{Thu-10}) for
$n=2$. Indeed, this value of $n$\ is the only one for which, as
shown by Eq. (\ref{Thu-12}), the quantity under the square root is a
perfect square for a generic BH charge configuration, thus
reproducing the quadratic invariant $\mathcal{I}_{2}$ of the ($1$-modulus, $%
n=1$ element of the) quadratic $\mathcal{N}=2$, $d=4$ sequence (see \textit{%
e.g.} \cite{bellucci1} and Refs. therein). \setcounter{equation}0

\section{\label{Sect7}Black Hole Attractors in (Axion-)Dilaton (Super)gravity%
}

Let us analyze now the model described \ by the Lagrangian density (\ref
{Wed-1}) in the extremal case ($c=0$) within the formalism based on the BH
effective potential $V_{BH}$. In the general case in which all electric and
magnetic BH charges are switched on, $V_{BH}$ reads\footnote{%
Eq. (\ref{Thu-14}) fixes a typo in Eq. (225) of \cite{ADFT}.} \cite{ADFT} ($%
\Lambda =1,...,6$)
\begin{eqnarray}
V_{BH}\left( {\phi ,\,a,\,p^{\Lambda },\,q_{\Lambda }}\right) &=&e^{2\phi
}(sp_{\Lambda }-q_{\Lambda })(\bar{s}p^{\Lambda }-q^{\Lambda })=  \notag \\
&=&e^{2\varphi }\left[ \left( a+ie^{-2\phi }\right) p_{\Lambda }-q_{\Lambda }%
\right] \left[ \left( a-ie^{-2\phi }\right) p^{\Lambda }-q^{\Lambda }\right]
=  \notag \\
&=&(e^{2\phi }a^{2}+e^{-2\phi })p^{2}+e^{2\phi }q^{2}-2ae^{2\phi }p\cdot q,
\label{Thu-14}
\end{eqnarray}
where the definition (\ref{Thu-13}) was used. By computing the criticality
conditions of $V_{BH}$ given by Eq. (\ref{Thu-14}), one obtains the
following stabilization equations for the axion $a$ and the dilaton $\phi $
\cite{ADFT}:

The critical points are :
\begin{equation}
\left. \frac{{\partial V}_{BH}\left( {\phi ,\,a,\,p,\,q}\right) }{{\partial a%
}}\right| _{\left( \phi ,a\right) =\left( {\phi }_{H}\left( p,q\right) {,\,a}%
_{H}\left( p,q\right) \right) }=0\Longleftrightarrow a_{H}\left( p,q\right) =%
\frac{{p\cdot q}}{{p^{2}}};  \label{Thu-16}
\end{equation}
\begin{gather}
\left. \frac{{\partial V}_{BH}\left( {\phi ,\,a,\,p,\,q}\right) }{{\partial
\phi }}\right| _{\left( \phi ,a\right) =\left( {\phi }_{H}\left( p,q\right) {%
,\,a}_{H}\left( p,q\right) \right) }=  \notag \\
=-e^{-4\phi }p^{2}+q^{2}-a_{H}\left( p,q\right) p\cdot q=-e^{-4\phi
}p^{2}+q^{2}-\frac{\left( p\cdot q\right) ^{2}}{p^{2}}=0;  \notag \\
\Updownarrow  \notag \\
e^{-2\phi _{H}\left( p,q\right) }=\frac{\sqrt{p^{2}q^{2}-(p\cdot q)^{2}}}{{%
p^{2}}}.  \label{Thu-17}
\end{gather}
\newline
Thus, the Bekenstein-Hawking BH entropy can be computed to be
\begin{equation}
S_{BH}\left( p,q\right) =\frac{A_{H}\left( p,q\right) }{4}=\pi V_{BH}\left(
\phi _{H}\left( p,q\right) ,a_{H}\left( p,q\right) ,p,q\right) =2\pi {\sqrt{%
p^{2}q^{2}-(p\cdot q)^{2}},}
\end{equation}
which corresponds to formula (\ref{Thu-10}) for $n=6$, and whose truncation
to $n=2$ gives formula (\ref{Thu-12}). By further putting $p^{1}=0,q_{2}=0$
(or $p^{2}=0,q_{1}=0$) and defining $p^{2}\equiv p,q_{1}\equiv q$ (or $%
p^{1}\equiv p,q_{2}\equiv q$), the expression (\ref{Thu-8}) for the BH
entropy is obtained.

In other words, the Maxwell-Einstein-dilaton $d=4$ gravity is given by the $%
\left( U(1)\right) ^{6}\rightarrow \left( U(1)\right) ^{2}$ truncation of
the (bosonic sector of) $\mathcal{N}=4$, $d=4$ supergravity, in a charge
configuration in which
\begin{equation}
p\cdot q\equiv p^{\Lambda }q_{\Lambda }=p^{1}q_{1}+p^{2}q_{2}=0.
\label{Thu-15}
\end{equation}
Such a constraint implies, by the stabilization Eqs. (\ref{Thu-16})-(\ref
{Thu-17}), the axion to be \textit{frozen out} ($a_{H}\left( p,q\right) =0$%
), and the dilaton to be stabilized as follows:
\begin{equation}
\left. e^{-2\phi _{H}\left( p,q\right) }\right| _{p\cdot q=0}=\left| \frac{q%
}{p}\right| .
\end{equation}
Notice that for the BH charge configuration (\ref{Thu-15}) the \textit{%
vanishing axion} solution $a(r)=0~\forall r\in \left[ r_{H},\infty \right) $
is a particular solution of the axionic Eq. of motion (actually also for $%
c\neq 0$, in which case $r_{H}$ is usually understood as the radius of the
outer - event - horizon). In general, the condition (\ref{Thu-15}) is
necessary but not sufficient for the choice of the \textit{vanishing axion }%
solution.

Furthermore, as anticipated in Sect. \ref{Sect2} (see Eq. (\ref
{sunday-Night-2})), Eq. (\ref{Sun-Sun-2}) can be generalized (\textit{at
least}) for the \textit{multi-dilaton system} whose scalar manifold is $%
\frac{SU\left( 1,n\right) }{U\left( 1\right) \times SU\left( n\right) }$ ($%
n_{V}=n$, $n\in \mathbb{N}$) (which is nothing but the $\mathcal{N}=2$, $d=4$
\textit{quadratic sequence} introduced above) as follows (here and below $%
\phi $ denotes the whole set of scalars; see also Eq. (\ref{sunday-Night-2}%
)):
\begin{eqnarray}
R_{+}^{2} &=&\left. R^{2}\right| _{r=\sqrt{\frac{1}{2}G_{ab}\left( \phi
_{\infty }\right) \Sigma ^{a}\Sigma ^{b}}}^{r=r_{+}}=r_{+}^{2}-\frac{1}{2}%
G_{ab}\left( \phi _{\infty }\right) \Sigma ^{a}\Sigma ^{b}=  \notag \\
&=&2M^{2}-V_{BH}\left( \phi _{\infty },p,q\right) +2M\sqrt{%
M^{2}-V_{BH}\left( \phi _{\infty },p,q\right) +\frac{1}{2}G_{ab}\left( \phi
_{\infty }\right) \Sigma ^{a}\Sigma ^{b}}=  \notag \\
&=&R_{+}^{2}\left( \phi _{\infty },p,q\right) ,  \label{tuesday1}
\end{eqnarray}
where the \textit{effective radial coordinate} $R$ is given by Eqs. (\ref
{CERN-solo-3}) (or (\ref{CERN-sunny-2})). Thus, the BH entropy is
\begin{eqnarray}
S_{BH}\left( \phi _{\infty },p,q\right) &=&\frac{A_{H}\left( \phi _{\infty
},p,q\right) }{4}=\pi R_{+}^{2}\left( \phi _{\infty },p,q\right) =\pi \left[
r_{+}^{2}\left( \phi _{\infty },p,q\right) -\frac{1}{2}G_{ab}\left( \phi
_{\infty }\right) \Sigma ^{a}\Sigma ^{b}\right] =  \notag \\
&=&\pi \left[ 2M^{2}-V_{BH}\left( \phi _{\infty },p,q\right) +2M\sqrt{%
M^{2}-V_{BH}\left( \phi _{\infty },p,q\right) +\frac{1}{2}G_{ab}\left( \phi
_{\infty }\right) \Sigma ^{a}\Sigma ^{b}}\right] .  \notag \\
&&  \label{Thu-5}
\end{eqnarray}
Due to the positive definiteness of $G_{ab}$, Eq. (\ref{tuesday1}) yields $%
R_{+}\leqslant r_{+}$ (see also Eq. (\ref{sunday-Night-2})). Thus, the
presence of scalars actually \textit{decreases} the BH radius relevant for
the computation of the entropy through the Bekenstein-Hawking
(semi)classical formula (\ref{sunday1}). In particular,
\begin{equation}
R_{+}\left( =R_{H}\right) =r_{+}\left( =r_{-}=r_{H}\right)
\label{CERN-cloudy1}
\end{equation}
only for \textit{double-extremal} (static, spherically symmetric,
asymptotically flat) dyonic BHs. On the other hand, in the \textit{%
double-non-extremal} case (see comment below Eq. (\ref{CERN-cloudy2}))
within the same class of BHs the expressions in brackets in Eq. (\ref
{CERN-cloudy1}) do \textit{not} hold.

In the extremal case:
\begin{equation}
c=0\Leftrightarrow M^{2}=V_{BH}\left( \phi _{\infty },p,q\right) -\frac{1}{2}%
G_{ab}\left( \phi _{\infty }\right) \Sigma ^{a}\Sigma ^{b};
\end{equation}
consequently
\begin{eqnarray}
r_{+,c=0}^{2} &=&r_{-,c=0}^{2}\equiv r_{H}^{2}=V_{BH}\left( \phi _{\infty
},p,q\right) -\frac{1}{2}G_{ab}\left( \phi _{\infty }\right) \Sigma
^{a}\Sigma ^{b}=M^{2}; \\
&&  \notag \\
R_{+,c=0}^{2} &\equiv &R_{H}^{2}=\left[ r_{+}^{2}-\frac{1}{2}G_{ab}\left(
\phi _{\infty }\right) \Sigma ^{a}\Sigma ^{b}\right] _{c=0}=\left[ M^{2}-%
\frac{1}{2}G_{ab}\left( \phi _{\infty }\right) \Sigma ^{a}\Sigma ^{b}\right]
_{c=0}=  \notag \\
&&  \notag \\
&=&\left[ 2M^{2}-V_{BH}\left( \phi _{\infty },p,q\right) \right] _{c=0}=%
\left[ V_{BH}\left( \phi _{\infty },p,q\right) -G_{ab}\left( \phi _{\infty
}\right) \Sigma ^{a}\Sigma ^{b}\right] _{c=0}.  \notag \\
&&  \label{tuesday6}
\end{eqnarray}
By recalling Eqs. (\ref{tuesday-fin-2}) and (\ref{tuesday-fin-6}), one
obtains
\begin{eqnarray}
R_{+,c=0}^{2} &\equiv &R_{H}^{2}=\mathcal{W}^{2}\left( \phi _{\infty
},p,q\right) -2G^{ab}\left( \phi _{\infty }\right) \left( \partial _{a}%
\mathcal{W}\right) \left( \phi _{\infty },p,q\right) \left( \partial _{b}%
\mathcal{W}\right) \left( \phi _{\infty },p,q\right) =  \notag \\
&&  \notag \\
&=&\mathcal{W}^{2}\left( z_{\infty },\overline{z}_{\infty },p,q\right) -4G^{i%
\overline{j}}\left( z_{\infty },\overline{z}_{\infty }\right) \left(
\partial _{i}\mathcal{W}\right) \left( z_{\infty },\overline{z}_{\infty
},p,q\right) \left( \overline{\partial }_{\overline{j}}\mathcal{W}\right)
\left( z_{\infty },\overline{z}_{\infty },p,q\right) =  \notag \\
&&  \notag \\
&=&R_{H}^{2}\left( p,q\right)  \label{Lun-2}
\end{eqnarray}
where $\mathcal{W}$ is the superpotential of the first order formulation for
extremal BHs \cite{Cer-Dal,ADOT-1} (see also the treatment of Sects. \ref
{Sect4} and \ref{Sect5}). Notice that in the second line a complex
parametrization of the scalar manifold has been introduced, which is
convenient for the \textit{multi-dilaton system} under consideration, whose
scalar manifold is the homogeneous symmetric special K\"{a}hler space $\frac{%
SU\left( 1,n\right) }{U\left( 1\right) \times SU\left( n\right) }$.

By using the explicit expressions of $\mathcal{W}_{\frac{1}{2}-BPS}$ and $%
\mathcal{W}_{non-BPS(,Z=0)}$ obtained in \cite{ADOT-1} (for the case $n=1$
see also \cite{Cer-Dal}), recalling that $C_{ijk}=0$ for the case at hand,
and using the differential relations of special K\"{a}hler geometry (see
\textit{e.g.} \cite{4}, and Refs. therein), one obtains the following
results:

\textbf{1)} for the $\frac{1}{2}$-BPS attractor flow:
\begin{eqnarray}
r_{H,\frac{1}{2}-BPS}^{2}\left( z_{\infty },\overline{z}_{\infty
},p,q\right) &=&M_{ADM,\frac{1}{2}-BPS}^{2}\left( z_{\infty },\overline{z}%
_{\infty },p,q\right) =\mathcal{W}_{\frac{1}{2}-BPS}^{2}\left( z_{\infty },%
\overline{z}_{\infty },p,q\right) =\lim_{\tau \rightarrow 0^{-}}\left|
Z\right| ^{2};  \notag \\
&& \\
&&  \notag \\
\Sigma _{i,\frac{1}{2}-BPS}\left( z_{\infty },\overline{z}_{\infty
},p,q\right) &=&2\lim_{\tau \rightarrow 0^{-}}\left[ \partial _{i}\mathcal{W}%
_{\frac{1}{2}-BPS}\left( z\left( \tau \right) ,\overline{z}\left( \tau
\right) ,p,q\right) \right] =  \notag \\
&&  \notag \\
&=&2\lim_{\tau \rightarrow 0^{-}}\left[ \nabla _{i}\mathcal{W}_{\frac{1}{2}%
-BPS}\left( z\left( \tau \right) ,\overline{z}\left( \tau \right)
,p,q\right) \right] =  \notag \\
&&  \notag \\
&=&\lim_{\tau \rightarrow 0^{-}}\left[ \sqrt{\frac{\overline{Z}}{Z}}D_{i}Z%
\right] =\lim_{\tau \rightarrow 0^{-}}\left[ \sqrt{\frac{\overline{Z}}{Z}}%
\left( \partial _{i}Z+\frac{1}{2}\left( \partial _{i}K\right) Z\right) %
\right] ; \\
&&  \notag \\
&&  \notag \\
R_{+,c=0,\frac{1}{2}-BPS}^{2} &\equiv &  \notag \\
&&  \notag \\
R_{H,\frac{1}{2}-BPS}^{2} &=&\lim_{\tau \rightarrow 0^{-}}\left[
\begin{array}{l}
\mathcal{W}_{\frac{1}{2}-BPS}^{2}\left( z\left( \tau \right) ,\overline{z}%
\left( \tau \right) ,p,q\right) + \\
\\
-4G^{i\overline{j}}\left( z\left( \tau \right) ,\overline{z}\left( \tau
\right) \right) \left( \partial _{i}\mathcal{W}_{\frac{1}{2}-BPS}\right)
\left( z\left( \tau \right) ,\overline{z}\left( \tau \right) ,p,q\right)
\cdot \\
\\
\cdot \left( \overline{\partial }_{\overline{j}}\mathcal{W}_{\frac{1}{2}%
-BPS}\right) \left( z\left( \tau \right) ,\overline{z}\left( \tau \right)
,p,q\right)
\end{array}
\right] =  \notag \\
&&  \notag \\
&&  \notag \\
&=&\mathcal{I}_{2}\left( p,q\right) =\left. V_{BH}\right| _{\left( z_{H},%
\overline{z}_{H}\right) _{\frac{1}{2}-BPS}}=  \notag \\
&&  \notag \\
&=&R_{H,\frac{1}{2}-BPS}^{2}\left( p,q\right) =\frac{S_{BH,\frac{1}{2}%
-BPS}\left( p,q\right) }{\pi };  \notag \\
&&  \label{Lun-3}
\end{eqnarray}

\textbf{2)} for the non-BPS ($Z=0$) attractor flow
\begin{eqnarray}
r_{H,non-BPS}^{2}\left( z_{\infty },\overline{z}_{\infty },p,q\right)
&=&M_{ADM,non-BPS}^{2}\left( z_{\infty },\overline{z}_{\infty },p,q\right) =
\notag \\
&&  \notag \\
&=&\mathcal{W}_{non-BPS}^{2}\left( z_{\infty },\overline{z}_{\infty
},p,q\right) =\lim_{\tau \rightarrow 0^{-}}\left[ G^{i\overline{j}}\left(
D_{i}Z\right) \overline{D}_{\overline{j}}\overline{Z}\right] ;  \notag \\
&& \\
&&  \notag \\
\Sigma _{i,non-BPS}\left( z_{\infty },\overline{z}_{\infty },p,q\right)
&=&2\lim_{\tau \rightarrow 0^{-}}\left[ \partial _{i}\mathcal{W}%
_{non-BPS}\left( z\left( \tau \right) ,\overline{z}\left( \tau \right)
,p,q\right) \right] =  \notag \\
&&  \notag \\
&=&2\lim_{\tau \rightarrow 0^{-}}\left[ \nabla _{i}\mathcal{W}%
_{non-BPS}\left( z\left( \tau \right) ,\overline{z}\left( \tau \right)
,p,q\right) \right] =  \notag \\
&&  \notag \\
&=&\lim_{\tau \rightarrow 0^{-}}\left[ \frac{\overline{Z}D_{i}Z}{\sqrt{G^{i%
\overline{j}}\left( D_{i}Z\right) \overline{D}_{\overline{j}}\overline{Z}}}%
\right] =  \notag \\
&&  \notag \\
&=&\lim_{\tau \rightarrow 0^{-}}\left[ \frac{\overline{Z}\left( \partial
_{i}Z+\frac{1}{2}\left( \partial _{i}K\right) Z\right) }{\sqrt{G^{i\overline{%
j}}\left( \partial _{i}Z+\frac{1}{2}\left( \partial _{i}K\right) Z\right)
\left( \overline{\partial }_{\overline{j}}\overline{Z}+\frac{1}{2}\left(
\overline{\partial }_{\overline{j}}K\right) \overline{Z}\right) }}\right] ;
\notag \\
&&  \notag \\
&& \\
&&R_{+,c=0,non-BPS}^{2}  \notag \\
&&  \notag \\
&\equiv &R_{H,non-BPS}^{2}=  \notag \\
&=&\lim_{\tau \rightarrow 0^{-}}\left[
\begin{array}{l}
\mathcal{W}_{non-BPS}^{2}\left( z\left( \tau \right) ,\overline{z}\left(
\tau \right) ,p,q\right) + \\
\\
-4G^{i\overline{j}}\left( z\left( \tau \right) ,\overline{z}\left( \tau
\right) \right) \left( \partial _{i}\mathcal{W}_{non-BPS}\right) \left(
z\left( \tau \right) ,\overline{z}\left( \tau \right) ,p,q\right) \cdot \\
\\
\cdot \left( \overline{\partial }_{\overline{j}}\mathcal{W}_{non-BPS}\right)
\left( z\left( \tau \right) ,\overline{z}\left( \tau \right) ,p,q\right)
\end{array}
\right] =  \notag \\
&&  \notag \\
&=&-\mathcal{I}_{2}\left( p,q\right) =\left. V_{BH}\right| _{\left( z_{H},%
\overline{z}_{H}\right) _{non-BPS}}=  \notag \\
&&  \notag \\
&=&R_{H,non-BPS}^{2}\left( p,q\right) =\frac{S_{BH,non-BPS}\left( p,q\right)
}{\pi }.  \label{Lun-4}
\end{eqnarray}
Here $D_{i}$ denotes the K\"{a}hler-$U\left( 1\right) $ and $H=U\left(
1\right) \times SU\left( n\right) $- covariant derivative, whereas $\mathcal{%
I}_{2}\left( p,q\right) $ is the invariant of the $U$-duality group $%
SU\left( 1,n\right) $, \textit{quadratic} in charges (see \textit{e.g.} \cite
{ADFT}, and Refs. therein). Notice that $\left| Z\right| ^{2}$ and $G^{i%
\overline{j}}\left( D_{i}Z\right) \overline{D}_{\overline{j}}\overline{Z}$
are the only independent ($H=U\left( 1\right) \times SU\left( n\right) $%
)-invariants.

Eqs. (\ref{Lun-3}) and (\ref{Lun-4}) are consistent, because the $\frac{1}{2}
$-BPS- and non-BPS ($Z=0$)- supporting BH charge configurations in the
(extremal) \textit{multi-dilaton system} are respectively defined by the
quadratic constraints $\mathcal{I}_{2}\left( p,q\right) >0$ and $\mathcal{I}%
_{2}\left( p,q\right) <0$.

Notice that in the extremality regime ($c=0$) the \textit{effective radius} $%
R_{H}$, and thus $A_{H}$ and the Bekenstein-Hawking entropy $S_{BH}$ are
\textit{independent} on the particular vacuum or ground state of the
considered theory, \textit{i.e.} on $\phi _{\infty }^{a}$ (or equivalently
on $\left( z_{\infty },\overline{z}_{\infty }\right) $), but rather they
depend \textit{only} on the electric and magnetic charges $q_{\Lambda }$ and
$p^{\Lambda }$, which are \textit{conserved} due to the overall $\left(
U(1)\right) ^{n+1}$ gauge-invariance. The independence on $\phi _{\infty
}^{a}$ is of crucial importance for the consistency of the \textit{%
microscopic state counting interpretation} of $S_{BH}$, as well as for the
overall consistency of the macroscopic thermodynamic picture of the BH.
However, it is worth recalling that the ADM mass $M_{\left( ADM\right) }$
generally does depend on $\phi _{\infty }^{a}$ \textit{also in the extremal
case}.

\subsection*{Black Hole Attractors and Supersymmetry}

Let us now analyze the supersymmetry-preserving features of (extremal) BH
attractors of the Maxwell-Einstein-(axion-)dilaton system, when it is
embedded in supergravity.

We start by defining the two following quantities \cite{K3}\cite{Duff-stu}

\begin{eqnarray}
Z_{1}\left( \phi ,p,q\right) &\equiv &\frac{(Q+P)}{\sqrt{2}}=\frac{qe^{\phi
}+pe^{-\phi }}{\sqrt{2}},  \label{Thu-eve-8} \\
Z_{2}\left( \phi ,p,q\right) &\equiv &\frac{(Q-P)}{\sqrt{2}}=\frac{qe^{\phi
}-pe^{-\phi }}{\sqrt{2}}.  \label{Thu-eve-9}
\end{eqnarray}

In $\mathcal{N}=4$, $d=4$ supergravity \cite{CSF}, $Z_{1}$ and $Z_{2}$ are
nothing but the eigenvalues of the skew-diagonal(ized) central charge matrix
$Z^{AB}$, arising in the $\mathcal{N}=4$, $d=4$ superalgebra ($\alpha ,\beta
=1,2$, $A,B=1,...,4$; ${\epsilon }$ is the $2$-dim. symplectic metric):
\begin{eqnarray}
\left\{ {Q_{\alpha }^{A}\;,\;Q_{\beta }^{B}}\right\} &=&\epsilon _{\alpha
\beta }Z^{AB}; \\
&&  \notag \\
Z_{SD}^{AB} &=&\left(
\begin{array}{cc}
Z_{1} & 0 \\
0 & Z_{2}
\end{array}
\right) \otimes \epsilon ,~~{\epsilon \equiv \left(
\begin{array}{cc}
0 & 1 \\
-1 & 0
\end{array}
\right) .}
\end{eqnarray}
When truncating down to the \textit{minimally coupled} $1$-modulus $\mathcal{%
N}=2$, $d=4$ supergravity (and thus after dropping two gravitinos out from
the supergravity multiplet) $Z_{1}$ and $Z_{2}$ can respectively be
interpreted as follows:

\begin{equation}
Z_{1}=Z,\quad Z_{2}=\partial _{\phi }Z,  \label{Thu-21}
\end{equation}
where $Z=Z\left( \phi ,p,q\right) $ is the \textit{central charge function}
of $\mathcal{N}=2$, $d=4$ supergravity. Its asymptotical values for $%
r\rightarrow r_{H}^{+}$ and $r\rightarrow \infty $ respectively define the
central charges of the $\mathfrak{psu}(1,1\left| 2\right) $ (only for $c=0$)
and $\mathcal{N}=2$, $d=4$ Poincar\'{e} superalgebras (here $A,B=1,2$;
recall Eq. (\ref{Thu-19}), which actually holds for any static, spherically
symmetric, asymptotically flat, extremal BH)

\begin{eqnarray}
c &=&0:\left\{ {Q_{\alpha }^{A},Q_{\beta }^{B}}\right\} _{\mathfrak{psu}%
(1,1\left| 2\right) }=\varepsilon _{\alpha \beta }\varepsilon ^{AB}Z\left(
\phi _{H}\left( p,q\right) ,p,q\right) ; \\
&&  \notag \\
c &\in &\mathbb{R}^{+}:\left\{ {Q_{\alpha }^{A},Q_{\beta }^{B}}\right\} _{%
\mathcal{N}=2,d=4~Poincar\acute{e}}=\varepsilon _{\alpha \beta }\varepsilon
^{AB}Z\left( \phi _{\infty },p,q\right) .
\end{eqnarray}

Notice that the central charges $Z_{1}$ and $Z_{2}$ are real, because we
have dropped the axion field $a$, consistently with the charge constraint $%
p\cdot q=0$.

In terms of the central charges (\ref{Thu-21}), Eqs. (\ref{Thu-eve-1}), (\ref
{Wed-2}), (\ref{Thu-eve-2}) and (\ref{Thu-eve-3}) can respectively be recast
in the following form:
\begin{eqnarray}
V_{BH}\left( \phi ,q,p\right) &=&Z_{1}^{2}\left( \phi ,q,p\right)
+Z_{2}^{2}\left( \phi ,q,p\right) =Z^{2}\left( \phi ,q,p\right) +\left(
\partial _{\phi }Z\right) ^{2}\left( \phi ,q,p\right) ;  \label{Thu-eve-4} \\
&&  \notag \\
\Sigma \left( \phi _{\infty },q,p\right) &=&-\frac{Z_{1}\left( \phi _{\infty
},q,p\right) Z_{2}\left( \phi _{\infty },q,p\right) }{M};  \label{Thu-eve-5}
\\
&&  \notag \\
r_{\pm }\left( \phi _{\infty },q,p\right) &=&M\pm c=  \notag \\
&=&M\pm \left[
\begin{array}{l}
M^{2}-Z_{1}^{2}\left( \phi _{\infty },q,p\right) -Z_{2}^{2}\left( \phi
_{\infty },q,p\right) + \\
\\
+\frac{Z_{1}^{2}\left( \phi _{\infty },q,p\right) Z_{2}^{2}\left( \phi
_{\infty },q,p\right) }{M^{2}}
\end{array}
\right] ^{1/2}=  \notag \\
&&  \notag \\
&=&M\pm \frac{1}{M}\sqrt{\left[ M^{2}-Z_{1}^{2}\left( \phi _{\infty
},q,p\right) \right] \left[ M^{2}-Z_{2}^{2}\left( \phi _{\infty },q,p\right)
\right] };  \label{Thu-eve-6} \\
&&  \notag \\
S_{BH}\left( \phi _{\infty },p,q\right) &=&\pi R_{+}^{2}\left( \phi _{\infty
},p,q\right) =\pi \left[ r_{+}^{2}\left( \phi _{\infty },p,q\right) -\Sigma
^{2}\left( \phi _{\infty },p,q\right) \right] =  \notag \\
&&  \notag \\
&=&\pi \left[
\begin{array}{l}
2M^{2}-Z_{1}^{2}\left( \phi _{\infty },q,p\right) -Z_{2}^{2}\left( \phi
_{\infty },q,p\right) + \\
\\
+2\sqrt{\left[ M^{2}-Z_{1}^{2}\left( \phi _{\infty },q,p\right) \right]
\left[ M^{2}-Z_{2}^{2}\left( \phi _{\infty },q,p\right) \right] }
\end{array}
\right] .  \label{Thu-eve-7}
\end{eqnarray}

Let us now consider the \textit{extremal} case ($c=0$). Thus, one obtains
the following expression of the ADM mass:
\begin{gather}
c=0; \\
\Updownarrow  \notag \\
M^{4}-\left[ Z_{1}^{2}\left( \phi _{\infty },q,p\right) +Z_{2}^{2}\left(
\phi _{\infty },q,p\right) \right] M^{2}+Z_{1}^{2}\left( \phi _{\infty
},q,p\right) Z_{2}^{2}\left( \phi _{\infty },q,p\right) =0; \\
\Updownarrow  \notag \\
\left[ M^{2}-Z_{1}^{2}\left( \phi _{\infty },q,p\right) \right] \left[
M^{2}-Z_{2}^{2}\left( \phi _{\infty },q,p\right) \right] =0; \\
\Updownarrow  \notag \\
M=\left\{
\begin{array}{c}
\left| Z_{1}\right| \left( \phi _{\infty },q,p\right) =\left| Z\right|
\left( \phi _{\infty },q,p\right) ; \\
or \\
\left| Z_{2}\right| \left( \phi _{\infty },q,p\right) =\left| \partial
_{\phi }Z\right| \left( \phi _{\infty },q,p\right) .
\end{array}
\right.  \label{Thu-night-1}
\end{gather}
and, by using Eq. (\ref{Thu-eve-6}), of the dilaton charge:
\begin{equation}
\Sigma \left( \phi _{\infty },q,p\right) =-\frac{Z_{1}\left( \phi _{\infty
},q,p\right) Z_{2}\left( \phi _{\infty },q,p\right) }{M}=\left\{
\begin{array}{c}
-sgn\left[ Z_{1}\left( \phi _{\infty },q,p\right) \right] Z_{2}\left( \phi
_{\infty },q,p\right) ; \\
or \\
-sgn\left[ Z_{2}\left( \phi _{\infty },q,p\right) \right] Z_{1}\left( \phi
_{\infty },q,p\right) .
\end{array}
\right. ;  \label{Thu-night-2}
\end{equation}

Consequently, for both branches of Eqs. (\ref{Thu-night-1}) and (\ref
{Thu-night-2}) (whose supersymmetry-preserving features will be discussed
further below) the BH entropy reads ($R_{+,c=0}^{2}\equiv R_{H}^{2}$)
\begin{eqnarray}
S_{BH,c=0} &=&\frac{A_{H,c=0}}{4}=\pi R_{H}^{2}=\pi \left[ M^{2}-\Sigma
^{2}\left( \phi _{\infty },p,q\right) \right] _{c=0}=  \notag \\
&&  \notag \\
&=&\pi \left[ Z_{1}\left( \phi _{\infty },q,p\right) +Z_{2}\left( \phi
_{\infty },q,p\right) \right] \left[ Z_{1}\left( \phi _{\infty },q,p\right)
-Z_{2}\left( \phi _{\infty },q,p\right) \right] =  \notag \\
&&  \notag \\
&=&2\pi \left| pq\right| ,
\end{eqnarray}
consistently with the result (\ref{Thu-8}). Let us remark once again that
\textit{the dependence on the asymptotical value }$\phi _{\infty }$\textit{\
of the dilaton drops out from the expression of the BH entropy }$\left.
S_{BH}\right| _{c=0}$\textit{!}

For the chosen BH charge configuration, the considered extremal BH is a
\textit{non-degenerate} $\frac{1}{4}$-BPS state in \textit{pure} $\mathcal{N}%
=4$, $d=4$ supergravity \cite{Ferrara-Maldacena} (which actually is the
unique \textit{non-degenerate} state in such a theory), for which it holds
\begin{eqnarray}
S_{BH,c=0}\left( p,q\right) &=&\pi Z_{1}^{2}\left( \phi _{H}\left(
p,q\right) ,p,q\right) ; \\
Z_{2}\left( \phi _{H}\left( p,q\right) ,p,q\right) &=&0.
\end{eqnarray}
It is worth pointing out that such a result can be generalized for the
\textit{non-degenerate} $\frac{1}{\mathcal{N}}$-BPS states in all $\mathcal{N%
}>2$-extended, $d=4$ pure supergravities as follows \cite{ferrara2}:
\begin{eqnarray}
S_{BH,c=0}\left( p,q\right) &=&\pi \left| Z_{1}\right| ^{2}\left( \phi
_{H}\left( p,q\right) ,p,q\right) ; \\
Z_{2}\left( \phi _{H}\left( p,q\right) ,p,q\right) &=&...=Z_{\left[ \mathcal{%
N}/2\right] }\left( \phi _{H}\left( p,q\right) ,p,q\right) =0,
\end{eqnarray}
where $Z_{1},Z_{2},...,Z_{\left[ \mathcal{N}/2\right] }$ are the eigenvalues
of the central charge matrix $Z_{AB}$ ($A,B=1,...,\mathcal{N}$), and the
square brackets denote the integer part.

In the $\mathcal{N}=2$, $d=4$, $n_{V}=1$ \textit{minimally coupled}
supergravity interpretation, one obtains that the $\frac{1}{4}$-BPS $%
\mathcal{N}=4$, $d=4$ extremal BH treated above corresponds to both $\frac{1%
}{2}$-BPS and non-BPS ($Z=0$) $\mathcal{N}=2$, $d=4$ extremal BHs. Indeed,
the criticality condition for $V_{BH}$ reads
\begin{eqnarray}
\partial _{\phi }V_{BH} &=&2Z\partial _{\phi }Z+2\left( \partial _{\phi
}Z\right) \partial _{\phi }^{2}Z=4Z\partial _{\phi }Z=  \notag \\
&=&2\left( P+Q\right) \left( P-Q\right) =2\left( pe^{-\phi }+qe^{\phi
}\right) \left( pe^{-\phi }-qe^{\phi }\right) =0,
\end{eqnarray}
where the relation
\begin{equation}
\partial _{\phi }^{2}Z=Z,
\end{equation}
yielded by the definitions (\ref{Thu-eve-8}) and (\ref{Thu-eve-9}) and by
the identifications (\ref{Thu-21}), was used. Thus, the extremal BH
attractors can be classified as follows:
\begin{gather}
\mathcal{N}=2,d=4,~\frac{1}{2}\text{\textit{-BPS}}:  \notag \\
\notag \\
\left\{
\begin{array}{l}
Z\left( \phi _{H,\frac{1}{2}-BPS}\left( p,q\right) ,p,q\right) \equiv Z_{%
\frac{1}{2}-BPS}\left( p,q\right) \neq 0; \\
\\
\left( \partial _{\phi }Z\right) \left( \phi _{H,\frac{1}{2}-BPS}\left(
p,q\right) ,p,q\right) =0\Leftrightarrow \left\{
\begin{array}{l}
\phi _{H,\frac{1}{2}-BPS}\left( p,q\right) =\frac{1}{2}\ln \left( \frac{p}{q}%
\right) ; \\
\\
pq>0;
\end{array}
\right.
\end{array}
\right.  \label{Thu-fin-1} \\
\Downarrow  \notag \\
Z_{\frac{1}{2}-BPS}\left( p,q\right) =\sqrt{2}pe^{-\phi _{H,\frac{1}{2}%
-BPS}\left( p,q\right) }=sgn\left( p\right) \sqrt{2pq}=sgn\left( p\right)
\sqrt{2\left| pq\right| }; \\
\Downarrow  \notag \\
S_{BH,\frac{1}{2}-BPS}\left( p,q\right) =\pi V_{BH}\left( \phi _{H,\frac{1}{2%
}-BPS}\left( p,q\right) ,p,q\right) =\pi Z_{\frac{1}{2}-BPS}^{2}\left(
p,q\right) =2\pi pq=2\pi \left| pq\right| ;  \label{Thu-fin-2}
\end{gather}
\begin{equation*}
\end{equation*}
\begin{gather}
\mathcal{N}=2,d=4,~\text{\textit{non-BPS (}}Z=0\text{\textit{)}}:  \notag \\
\notag \\
\left\{
\begin{array}{l}
\left( \partial _{\phi }Z\right) \left( \phi _{H,non-BPS}\left( p,q\right)
,p,q\right) \equiv \left( \partial _{\phi }Z\right) _{non-BPS}\left(
p,q\right) \neq 0; \\
\\
Z\left( \phi _{H,non-BPS}\left( p,q\right) ,p,q\right) =0\Leftrightarrow
\left\{
\begin{array}{l}
\phi _{H,non-BPS}\left( p,q\right) =\frac{1}{2}\ln \left( -\frac{p}{q}%
\right) ; \\
\\
pq<0;
\end{array}
\right.
\end{array}
\right.  \label{Thu-fin-3} \\
\Downarrow  \notag \\
\left( \partial _{\phi }Z\right) _{non-BPS}\left( p,q\right) =sgn\left(
p\right) \sqrt{-2pq}=sgn\left( p\right) \sqrt{2\left| pq\right| }; \\
\Downarrow  \notag \\
S_{BH,non-BPS}\left( p,q\right) =\pi V_{BH}\left( \phi _{H,non-BPS}\left(
p,q\right) ,p,q\right) =\pi \left( \partial _{\phi }Z\right)
_{non-BPS}^{2}\left( p,q\right) =-2\pi pq=2\pi \left| pq\right| .
\label{Thu-fin-4}
\end{gather}
Thus, for the considered $\mathcal{N}=2$ BH charge configuration $\left(
p^{1},p^{2},q_{1},q_{2}\right) :p\cdot q=0$, one obtains that:
\begin{gather}
\exp \left[ -2\phi _{H,\frac{1}{2}-BPS}\left( p,q\right) \right] =\exp \left[
-2\phi _{H,non-BPS}\left( p,q\right) \right] =\left| \frac{q}{p}\right| ; \\
\notag \\
Z_{\frac{1}{2}-BPS}\left( p,q\right) =\left( \partial _{\phi }Z\right)
_{non-BPS}\left( p,q\right) =sgn\left( p\right) \sqrt{2\left| pq\right| }; \\
\notag \\
S_{BH,\frac{1}{2}-BPS}\left( p,q\right) =S_{BH,non-BPS}\left( p,q\right)
=2\pi \left| pq\right| .
\end{gather}
The unique discrimination between the two classes of attractors is due to
the range of the supporting BH charges: $pq>0$ supports $\frac{1}{2}$-BPS
attractors, whereas $pq<0$ supports the non-BPS ($Z=0$) ones. The case $pq=0$
must be disregarded, because it yields $A_{H,\frac{1}{2}%
-BPS}=A_{H,non-BPS}=0 $ (\textit{``small''} extremal BH), and it does not
allow to consistently stabilize the dilaton at the BH horizon, as well.

Now, the first branch of Eq. (\ref{Thu-night-1}) trivially yields the
saturation of the $c=0$ limit of the (guessed generalization of the) BPS
bound (\ref{monday-eve-1}) which, through the asymptotical relation (\ref
{monday4-bis}), yields the $c=0$ limit of the bound (\ref{monday-fin-1}). It
is thus clear that the first branch of Eqs. (\ref{Thu-night-1}) and (\ref
{Thu-night-2}) corresponds to the $\mathcal{N}=2$, $d=4~\frac{1}{2}$-BPS
attractor solution given by Eqs. (\ref{Thu-fin-1}) and (\ref{Thu-fin-2}).

It remains to show that for the second branch, which thus\footnote{%
We assume the \textit{absence} of more than one \textit{basin of attraction}
in the radial dynamics of the dilaton in the considered extremal dilaton BH
supported by electric and magnetic charges constrained by $p\cdot q=0$ \cite
{kallosh1,K3}.} necessarily belongs to the $\mathcal{N}=2,d=4$ non-BPS ($Z=0$%
) attractor solution given by Eqs. (\ref{Thu-fin-3}) and (\ref{Thu-fin-4}),
the strict inequality of the $c=0$ limit of the bound (\ref{monday-eve-1})
(and thus of the bound (\ref{monday-fin-1})) holds. We can do this even
without solving analytically the equation of motion for the dilaton $\phi
\left( r\right) $ in the extremal case (see \textit{e.g.} - the first Ref.
of - \cite{K3}). Indeed, by using the definition (\ref{Wed-Wed-1}) and Eq. (%
\ref{Thu-6}), and recalling that only its ``$+$'' branch is admissible (see
below Eq. (\ref{Thu-8})), one gets the following expression of the extremal
ADM squared mass:
\begin{equation}
M^{2}=\frac{\left( \left| P_{\infty }\right| +\left| Q_{\infty }\right|
\right) ^{2}}{2}=\frac{\left( \left| p\right| e^{-\phi _{\infty }}+\left|
q\right| e^{\phi _{\infty }}\right) ^{2}}{2}.  \label{Thu-Thu-4}
\end{equation}
On the other hand, Eqs. (\ref{Thu-eve-8}) and (\ref{Thu-21}) yield the
following expression for the asymptotical limit of the squared absolute
value of the \textit{central charge function} (\textit{i.e.} for the central
charge of the $\mathcal{N}=2$, $d=4$ Poincar\'{e} superalgebra):
\begin{equation}
\left| Z\right| ^{2}\left( \phi _{\infty },p,q\right) =Z^{2}\left( \phi
_{\infty },p,q\right) =\frac{(P_{\infty }+Q_{\infty })^{2}}{2}=\frac{\left(
pe^{-\phi _{\infty }}+qe^{\phi _{\infty }}\right) ^{2}}{2}.
\label{Thu-Thu-1}
\end{equation}
Thus, it is immediate to conclude that, depending on the (reciprocal) signs
of the BH supporting charges $p$ and $q$, it holds that
\begin{equation}
\begin{array}{l}
pq>0:M^{2}=\left| Z\right| ^{2}\left( \phi _{\infty },p,q\right) ; \\
\\
pq<0:M^{2}>\left| Z\right| ^{2}\left( \phi _{\infty },p,q\right) .
\end{array}
\label{Thu-Thu-2}
\end{equation}
Such a result is consistent with the fact that, as given by Eqs. (\ref
{Thu-fin-1})-(\ref{Thu-fin-2}) and (\ref{Thu-fin-3})-(\ref{Thu-fin-4}), $%
pq>0 $ supports the $\frac{1}{2}$-BPS attractors, whereas $pq<0$ supports
the non-BPS ($Z=0$) ones. Summarizing, one can finally state that the
(first) branch $M=\left| Z\right| \left( \phi _{\infty },p,q\right) $ of Eq.
(\ref{Thu-night-1}) is the $\frac{1}{2}$-BPS one, the (second) branch $%
M=\left| \partial _{\phi }Z\right| \left( \phi _{\infty },p,q\right) $ of
Eq. (\ref{Thu-night-1}) is the non-BPS ($Z=0$) one.

In general, it seems conceivable that (as stated above and in \cite{K2}) the
BPS bounds (\ref{monday-eve-1}) and (\ref{monday-fin-1}) generally hold in
extremal case ($c=0$), \textit{regardless} of the asymptotical values $\phi
_{\infty }^{a}$ of the scalar fields.

As a further illustrative example, let us consider the \textit{%
double-extremal} dilaton BH (which indeed is a possible solution of the
equation of motion \cite{kallosh1,K3}), \textit{i.e.} for the case in which
\begin{gather}
\frac{d\phi }{dr}=0~\forall r\in \left[ r_{H},\infty \right) \\
\Updownarrow  \notag \\
\left\{
\begin{array}{l}
\frac{1}{2}\text{\textit{-BPS}}:\phi \left( r\right) =\phi _{\infty }=\phi
_{H,BPS}\left( p,q\right) =\frac{1}{2}\ln \left( \frac{p}{q}\right) ,~pq>0;
\\
\\
\text{\textit{non-BPS~(}}Z=0\text{\textit{)}}:\phi \left( r\right) =\phi
_{\infty }=\phi _{H,non-BPS}\left( p,q\right) =\frac{1}{2}\ln \left( -\frac{p%
}{q}\right) ,~pq<0.
\end{array}
\right.
\end{gather}
Let us start by considering the $\frac{1}{2}$-BPS case; since $\left|
\partial _{\phi }Z\right| \left( \phi _{\infty },p,q\right) =$ $\left|
\partial _{\phi }Z\right| \left( \phi _{H,BPS},p,q\right) =0$, it is clear
that the second branch of Eq. (\ref{Thu-night-1}) \textit{cannot} be the BPS
one, because it would yield $M=0$, which, within the assumptions formulated
(see \cite{gibbons2} and Refs. therein), holds iff the space-time is
(globally) flat, which is \textit{not} the case. Thus, it holds that
\begin{equation}
M_{\frac{1}{2}-BPS}\left( \phi _{\infty },p,q\right) =M_{\frac{1}{2}%
-BPS}\left( \phi _{H,BPS}\left( p,q\right) ,p,q\right) =\left| Z\right|
\left( \phi _{H,BPS}\left( p,q\right) ,p,q\right) =\sqrt{2pq}.
\end{equation}
The reasoning gets mirrored for the non-BPS ($Z=0$) case, for which $\left|
Z\right| \left( \phi _{\infty },p,q\right) =$ $\left| Z\right| \left( \phi
_{H,non-BPS},p,q\right) =0$. Consequently, since the considered metric
background is not (globally) flat, the second branch of Eq. (\ref
{Thu-night-1}) \textit{cannot} be the non-BPS ($Z=0$) one. Also, the check
of the second relation of (\ref{Thu-Thu-2}) in such a case is trivial, since
\begin{eqnarray}
M_{non-BPS}\left( \phi _{\infty },p,q\right) &=&M_{non-BPS}\left( \phi
_{H,non-BPS}\left( p,q\right) ,p,q\right) =  \notag \\
&=&\left| \partial _{\phi }Z\right| \left( \phi _{H,non-BPS}\left(
p,q\right) ,p,q\right) =  \notag \\
&=&\sqrt{-2pq}>\left| Z\right| \left( \phi _{H,non-BPS}\left( p,q\right)
,p,q\right) =0.
\end{eqnarray}

On the other hand, Eqs. (\ref{Thu-eve-9}) and (\ref{Thu-21}) yield the
following expression for the asymptotical limit of the squared absolute
value of the derivative of the \textit{central charge function} (\textit{i.e.%
} for the so-called asymptotical \textit{matter charge}):
\begin{eqnarray}
\left| D_{\phi }Z\right| ^{2}\left( \phi _{\infty },p,q\right) &=&\left|
\partial _{\phi }Z\right| ^{2}\left( \phi _{\infty },p,q\right) =\left(
\partial _{\phi }Z\right) ^{2}\left( \phi _{\infty },p,q\right) =  \notag \\
&=&\frac{(-P_{\infty }+Q_{\infty })^{2}}{2}=\frac{\left( -pe^{-\phi _{\infty
}}+qe^{\phi _{\infty }}\right) ^{2}}{2}.  \label{Thu-Thu-3}
\end{eqnarray}
By comparing such a result with Eq. (\ref{Thu-Thu-1}), one thus obtains
\begin{equation}
\begin{array}{l}
pq>0:\left| Z\right| ^{2}\left( \phi _{\infty },p,q\right) >\left| D_{\phi
}Z\right| ^{2}\left( \phi _{\infty },p,q\right) ; \\
\\
pq<0:\left| D_{\phi }Z\right| ^{2}\left( \phi _{\infty },p,q\right) >\left|
Z\right| ^{2}\left( \phi _{\infty },p,q\right) .
\end{array}
\end{equation}

Finally, let us consider the generic, \textit{non-extremal} case ($c\neq 0$%
). By recalling Eqs. (\ref{Wed-3}), (\ref{Thu-eve-4}) and (\ref{Thu-eve-5}),
one gets ($\left| Z\right| _{\infty }^{2}\equiv \left| Z\right| ^{2}\left(
\phi _{\infty },q,p\right) $, $\left| D_{\phi }Z\right| _{\infty }^{2}\equiv
\left| D_{\phi }Z\right| ^{2}\left( \phi _{\infty },q,p\right) $; recall
that $\partial _{\phi }Z=D_{\phi }Z$, due to the considered BH charge
configuration such that $p\cdot q=0$)
\begin{gather}
M^{2}=V_{BH}\left( \phi _{\infty },q,p\right) -\Sigma ^{2}+c^{2}=\left|
Z\right| _{\infty }^{2}+\left| D_{\phi }Z\right| _{\infty }^{2}-\frac{\left|
Z\right| _{\infty }^{2}\left| D_{\phi }Z\right| _{\infty }^{2}}{M}+c^{2};
\notag \\
\Updownarrow  \notag \\
M_{\pm }^{2}=\frac{\left| Z\right| _{\infty }^{2}}{2}\left\{ \left[ 1+\frac{%
\left( \left| D_{\phi }Z\right| _{\infty }^{2}+c^{2}\right) }{\left|
Z\right| _{\infty }^{2}}\right] \left[ 1\pm \sqrt{1-\left[ \frac{2\left|
Z\right| _{\infty }\left| D_{\phi }Z\right| _{\infty }}{\left| Z\right|
_{\infty }^{2}+\left| D_{\phi }Z\right| _{\infty }^{2}+c^{2}}\right] ^{2}}%
\right] \right\} ,  \label{Wed-Wed-Wed-1}
\end{gather}
where $\left| Z\right| _{\infty }^{2}>0$ (otherwise, \textit{i.e.} for $%
\left| Z\right| _{\infty }^{2}=0$, the strict inequality of BPS bounds (\ref
{monday-eve-1}) and (\ref{monday-fin-1}) holds) and the necessary condition
\begin{equation}
\left[ \left| Z\right| _{\infty }^{2}+\left| D_{\phi }Z\right| _{\infty
}^{2}+c^{2}\right] ^{2}-4\left| Z\right| _{\infty }^{2}\left| D_{\phi
}Z\right| _{\infty }^{2}\geqslant 0,  \label{Wed-Wed-Wed-4}
\end{equation}
were assumed to hold. Eqs. (\ref{Wed-Wed-Wed-1}) implies that in the generic
\textit{non-extremal} case ($c\neq 0$, within $\left| Z\right| _{\infty
}^{2}>0$ and the consistency condition (\ref{Wed-Wed-Wed-4})) the BPS bounds
(\ref{monday-eve-1}) and (\ref{monday-fin-1}) do hold true depending on the
asymptotical dilaton value $\phi _{\infty }$ and on the value of the
extremality parameter $c$. Notice that, in the extremal limit $c\rightarrow
0 $ one achieves the results obtained above.

In general, it seems conceivable that in the generic \textit{non-extremal}
case ($c\neq 0$, within suitable consistency conditions) the BPS bounds (\ref
{monday-eve-1}) and (\ref{monday-fin-1}) hold true depending, \textit{in a
model-dependent way}, on the asymptotical values $\phi _{\infty }^{a}$ of
the scalar fields and on the very value of the extremality parameter $c$.
\setcounter{equation}0

\section{\label{Sect8}Rotating Attractors}

Let us now briefly report some results about the \textit{extremal} rotating
case.

In the stationary rotating regime ($J\neq 0$ constant), an asymptotically
flat \textit{extremal} BH has a near-horizon geometry $AdS_{2}\times S^{1}$
(see \textit{e.g.} \cite{rotating-attr} and Refs. therein). The most general
\textit{Ansatz} for the near-horizon metric consistent with the $SO\left(
2,1\right) \times SO\left( 2\right) $ symmetry reads
\begin{equation}
ds^{2}=v_{1}\left( \theta \right) (-r^{2}dt^{2}+\frac{{dr^{2}}}{{r^{2}}}%
)+d\theta ^{2}+v_{2}\left( \theta \right) (d\phi -\alpha rdt)^{2},
\end{equation}
where $\alpha \in \mathbb{R}$, and $v_{1}\left( \theta \right) $ and $%
v_{2}\left( \theta \right) $ are functions of the angle $\theta \in \left[
0,\pi \right] $. By a reparametrization of such a coordinate, the metric can
be recast in the following form \cite{rotating-attr}:
\begin{equation}
ds^{2}=\Xi ^{2}\left( \theta \right) e^{2\psi \left( \theta \right)
}(-r^{2}dt^{2}+\frac{{dr^{2}}}{{r^{2}}}+d\theta ^{2})+e^{-2\psi \left(
\theta \right) }(d\phi -\alpha rdt)^{2},
\end{equation}
where $\Xi \left( \theta \right) $ and $\psi \left( \theta \right) $ are
suitably defined functions of $\theta $. The resulting BH entropy reads
\begin{equation}
S_{BH}=16\pi ^{2}a\left( p,q,J\right) ,
\end{equation}
where $a\left( p,q,J\right) $ is a certain function of the BH charges $%
p^{\Lambda }$, $q_{\Lambda }$ and of $J$. By the very definition (\ref
{sunday1}) of \textit{effective radius} (and recalling $R_{+,c=0}^{2}\equiv
R_{H}^{2}$), it thus holds that
\begin{equation}
R_{H}^{2}\left( p,q,J\right) =16\pi a\left( p,q,J\right) .
\end{equation}

The near-horizon \textit{extremal} Kerr BH metric is given by
\begin{eqnarray}
\alpha &=&1; \\
~\Xi \left( \theta \right) &=&\frac{J}{8\pi }\sin \theta ; \\
~e^{-2\psi \left( \theta \right) } &=&\frac{J}{4\pi }\frac{\sin ^{2}\theta }{%
1+\cos ^{2}\theta }; \\
a &=&\frac{J}{8\pi },
\end{eqnarray}
yielding \cite{Kerr-throat}
\begin{equation}
S_{BH,Kerr}=2\pi J\Leftrightarrow R_{H,Kerr}^{2}\left( J\right) =2J.
\end{equation}

On the other hand, the near-horizon \textit{extremal} Kerr-Newman BH metric
(with electric and magnetic charge $q$ and $p$) is given by
\begin{eqnarray}
\alpha &=&\frac{J}{\sqrt{J^{2}+\left( \frac{q^{2}+p^{2}}{8\pi }\right) ^{2}}}%
;  \label{Sat-1} \\
\Xi \left( \theta \right) &=&a\sin \theta ; \\
e^{-2\psi \left( \theta \right) } &=&\frac{2a\sin ^{2}\theta }{1+\cos
^{2}\theta +\frac{\left( q^{2}+p^{2}\right) \sin ^{2}\theta }{64\pi ^{2}a}};
\\
a &=&\frac{\sqrt{J^{2}+\left( \frac{q^{2}+p^{2}}{8\pi }\right) ^{2}}}{8\pi },
\label{Sat-1-bis}
\end{eqnarray}
yielding \cite{Kerr-throat}
\begin{gather}
S_{BH,Kerr-Newman}=2\pi \sqrt{J^{2}+\left( \frac{q^{2}+p^{2}}{8\pi }\right)
^{2}}; \\
\Updownarrow  \notag \\
R_{H,Kerr-Newman}^{2}\left( p,q,J\right) =2\sqrt{J^{2}+\left( \frac{%
q^{2}+p^{2}}{8\pi }\right) ^{2}}.
\end{gather}
Up to a rescaling of BH charges by $\frac{1}{2\sqrt{\pi }}$ and an overall
rescaling of $ds^{2}$ by $\frac{1}{16\pi }$, in the static limit ($%
J\rightarrow 0$), the near-horizon Kerr-Newman metric (\ref{Sat-1})-(\ref
{Sat-1-bis}) gives back the BR metric (\ref{BR-BR-1}) which, as shown above,
is the near-horizon of every static, spherically symmetric, asymptotically
flat, \textit{extremal} BH (and in this particular case, of the RN extremal
BH with electric and magnetic charge $q$ and $p$):
\begin{gather}
\alpha =0,~\Xi \left( \theta \right) =a\sin \theta ,~e^{-2\psi \left( \theta
\right) }=a\sin ^{2}\theta ,a=\frac{\left( q^{2}+p^{2}\right) }{64\pi ^{2}};
\\
\Downarrow  \notag \\
ds^{2}=\frac{\left( q^{2}+p^{2}\right) }{64\pi ^{2}}(-r^{2}dt^{2}+\frac{{%
dr^{2}}}{{r^{2}}}+d\Omega ^{2})=\frac{1}{16\pi }\left. ds_{BR}^{2}\right|
_{M^{2}=\frac{q^{2}+p^{2}}{4\pi }}.
\end{gather}

In \cite{rotating-attr}, beside the \textit{double-extremal} case, various
explicit examples of stationary, asymptotically flat rotating BHs with
non-trivial scalar dynamics were treated. Among them, let us here mention
the generalization of the \textit{extremal} Kerr-Newman BH to supergravity
given by heterotic compactification on a six-torus $T^{6}$. This in general
determines \textit{matter-coupled} $\mathcal{N}=4$, $d=4$ supergravity \cite
{Maharana,Duff-stu}, which can be consistently truncated to \textit{pure} $%
\mathcal{N}=4$, $d=4$ supergravity (see discussion at the start of Sect. \ref
{Sect6}) or also to $\mathcal{N}=2$, $d=4$ reducible symmetric \textit{cubic}
sequence $\frac{SU\left( 1,1\right) }{U\left( 1\right) }\times \frac{%
SO\left( 2,n\right) }{SO\left( 2\right) \times SO\left( n\right) }$, in the
symplectic basis with maximum non-compact manifest symmetry $SO\left(
2,n\right) $ \cite{Calabi-Vis,CDFVP}, in which the prepotential does \textit{%
not} exist at all (see \textit{e.g.} \cite{4} and Refs. therein). In the
so-called \textit{ergo-free} branch, the corresponding near-horizon metric
reads \cite{rotating-attr}

\begin{eqnarray}
ds^{2} &=&\frac{1}{{8\pi }}\sqrt{\mathcal{I}_{4}-J^{2}\cos ^{2}\theta }%
\;(-r^{2}dt^{2}+\frac{{dr^{2}}}{{r^{2}}}+d\theta ^{2})+  \notag \\
&&  \notag \\
&&+\frac{1}{{8\pi }}\frac{\left( \mathcal{I}{_{4}-J^{2}}\right) }{\sqrt{%
\mathcal{I}_{4}-J^{2}\cos ^{2}\theta }}\;\sin ^{2}\theta \;(d\phi +\frac{J}{%
\sqrt{\mathcal{I}_{4}-J^{2}}}rdt)^{2},
\end{eqnarray}
with entropy
\begin{equation}
S_{BH}=2\pi \sqrt{\mathcal{I}_{4}-J^{2}}\Leftrightarrow R_{H}^{2}=2\sqrt{%
\mathcal{I}_{4}-J^{2}},
\end{equation}
where $\mathcal{I}_{4}$ is the (unique) \textit{quartic} invariant of the
relevant $U$-duality group ($SU\left( 1,1\right) \times SO\left( 6,n\right) $
for $\mathcal{N}=4$, $d=4$ theory; see also the discussion in Sect. \ref
{Sect6}):
\begin{equation}
\mathcal{I}_{4}\equiv p^{2}q^{2}-(p\cdot q)^{2}.
\end{equation}
\setcounter{equation}0

\section{\label{Sect9}Black Hole Entropy and Quantum Entanglement}

Sometimes, it happens that two very different areas of theoretical physics
share the same mathematics. This may eventually lead to the realization that
they are, in fact, dual descriptions of the same physical phenomena, or it
may not. Anyway, this fact often leads to new insights in both areas. Recent
papers \cite{duff1} have established an intriguing analogy between the
entropy of certain $d=4$ supersymmetric BHs in string theory and the
entanglement measure in quantum information theory.

\subsection*{\label{Subsect9-1}$\mathcal{N}=2$, $d=4$ $stu$ Black Holes%
\newline
and the Tripartite Entanglement of Three Qubits}

In quantum information theory, the three qubit system (Alice, Bob, Charlie)
is described by the quantum state
\begin{equation}
\left| \Psi \right\rangle =a_{ABC}\left| ABC\right\rangle ,  \label{Sat-2}
\end{equation}
where $A=0,1$, so the corresponding Hilbert space has dimension $2^{3}=8$.
The numbers $a_{ABC}\in \mathbb{C}$ transforms as the fundamental
representation $(\mathbf{2},\mathbf{2},\mathbf{2})$ of $SL(2,\mathbb{C}%
)_{A}\times SL(2,\mathbb{C})_{B}\times SL(2,\mathbb{C})_{C}$ . The so-called
\textit{tripartite entanglement} of the quantum state (\ref{Sat-2}) is
measured by the 3-\textit{tangle} \cite{coffman,miyake}
\begin{equation}
\tau _{3}(ABC)\equiv 4|Det\left( a_{ABC}\right) |,
\end{equation}
where $Det\left( a_{ABC}\right) $ is \textit{Cayley's hyperdeterminant} \cite
{cayley}
\begin{equation}
Det\left( a_{ABC}\right) \equiv -\frac{1}{2}\epsilon ^{A_{1}A_{2}}\epsilon
^{B_{1}B_{2}}\epsilon ^{A_{3}A_{4}}\epsilon ^{B_{3}B_{4}}\epsilon
^{C_{1}C_{4}}\epsilon
^{C_{2}C_{3}}a_{A_{1}B_{1}C_{1}}a_{A_{2}B_{2}C_{2}}a_{A_{3}B_{3}C_{3}}a_{A_{4}B_{4}C_{4}},
\end{equation}
which is invariant under $SL(2,\mathbb{C})_{A}\times SL(2,\mathbb{C}%
)_{B}\times SL(2,\mathbb{C})_{C}$ and under a \textit{triality}
interchanging $A$, $B$ and $C$ (-indices). \newline

In the context of stringy BHs, the $8$ components of $a_{ABC}$ are the $4$
electric and $4$ magnetic charges of an extremal BH in the so-called ($3$%
-moduli) $stu$ model of $\mathcal{N}=2$, $d=4$ supergravity \cite
{Duff-stu,behrndt}, and thus they take real (or integer, at the quantized
level) values. The $U$-duality group of the $stu$ model is $\left( SU\left(
1,1\right) \right) ^{3}\sim \left( SL(2,\mathbb{R})\right) ^{3}$ (or $\left(
SL(2,\mathbb{Z})\right) ^{3}$, at the quantized level).

The Bekenstein-Hawking entropy $S_{BH}$ of such a BH was firstly calculated
in \cite{behrndt}. The connection to quantum information theory arises by
noting (see the first Ref. of \cite{duff1}) that $S_{BH}$ can also be
expressed in terms of \textit{Cayley's hyperdeterminant}:
\begin{equation}
S_{BH}=\pi \sqrt{|Det\left( a_{ABC}\right) |}.
\end{equation}

Thus, one can establish a \textit{dictionary} between the classification of
various \textit{entangled states} (the so-called \textit{separable A-B-C},
\textit{bipartite entangled of type A-BC, B-CA, C-AB}, \textit{tripartite
entangled W}, \textit{tripartite entangled GHZ} ones) and the classification
of various \textit{small} and \textit{large} BPS and non-BPS $stu$ BHs \cite
{duff1}. For example, the \textit{canonical GHZ} state \cite{greenberger}
\begin{equation}
\left| \Psi \right\rangle \sim \frac{1}{\sqrt{2}}\left( \left|
111\right\rangle +\left| 000\right\rangle \right) ,
\end{equation}
having $Det\left( a_{ABC}\right) >0$ corresponds to a \textit{large} non-BPS
$stu$ BH. The \textit{tripartite entangled W} state
\begin{equation}
\left| \Psi \right\rangle \sim \frac{1}{\sqrt{3}}\left( \left|
100\right\rangle +\left| 010\right\rangle +\left| 001\right\rangle \right)
\end{equation}
having $Det\left( a_{ABC}\right) =0$ corresponds to a \textit{small} BPS $%
stu $ BH. The \textit{GHZ} state
\begin{equation}
\left| \Psi \right\rangle =\frac{1}{2}\left( -\left| 000\right\rangle
+\left| 011\right\rangle +\left| 101\right\rangle +\left| 110\right\rangle
\right)
\end{equation}
having $Det\left( a_{ABC}\right) <0$ corresponds to a \textit{large} BPS $%
stu $ BH.

\subsection*{\label{Subsect9-2}$\mathcal{N}=2$, $d=4$ Axion-Dilaton Black
Holes\newline
and the Bipartite Entanglement of Two Qubits}

An even structurally simpler framework (see the second Ref. of \cite{duff1})
in which the analogy between stringy BHs and quantum information theory
works is provided by the two qubit system (Alice and Bob), described by the
quantum state
\begin{equation}
\left| \Psi \right\rangle =a_{AB}\left| AB\right\rangle ,  \label{Sat-3}
\end{equation}
where $A=0,1$, so the corresponding Hilbert space has dimension $2{^{2}}=4$.
The numbers $a_{AB}\in \mathbb{C}$ transform as the fundamental
representation $(\mathbf{2},\mathbf{2})$ of $SL(2,\mathbb{C})_{A}\times SL(2,%
\mathbb{C})_{B}$. The so-called \textit{bipartite entanglement} of the
quantum state (\ref{Sat-3}) is measured by the 2-\textit{tangle}
\begin{equation}
\tau _{2}(AB)\equiv C^{2}(AB),
\end{equation}
where
\begin{equation}
C(AB)\equiv 2|det\left( a_{AB}\right) |
\end{equation}
is the so-called \textit{concurrence}. $det\left( a_{AB}\right) $ is
invariant under $SL(2,\mathbb{C})_{A}\times SL(2,\mathbb{C})_{B}$ and under
a \textit{duality} interchanging $A$ and $B$(-indices).

In the context of stringy BHs, the $4$ components of $a_{AB}$ are the $2$
electric and $2$ magnetic charges of an extremal BH in the so-called ($1$%
-modulus) $t^{2}$ model of $\mathcal{N}=2$, $d=4$ supergravity. Thus, in
such an interpretation the $a_{AB}$'s take real (or integer, at the
quantized level) values. As pointed out in Sects. \ref{Sect6} and \ref{Sect7}%
, the $t^{2}$ model also corresponds to the $\left( U\left( 1\right) \right)
^{6}\rightarrow \left( U\left( 1\right) \right) ^{2}$ truncation of the
so-called Maxwell-Einstein-axion-dilaton BH, with entropy
\begin{equation}
S_{BH}=\pi |det\left( a_{AB}\right) |.
\end{equation}
For example, the so-called \textit{Bell} state
\begin{equation}
\left| \Psi \right\rangle =\frac{1}{\sqrt{2}}\left( \left| 00\right\rangle
+\left| 11\right\rangle \right)
\end{equation}
with $det\left( a_{AB}\right) >0$ corresponds to a \textit{large} non-BPS
axion-dilaton BH.

\subsection*{$\mathcal{N}=8$, $d=4$ Black Holes\newline
and the Tripartite Entanglement of Seven Qubits}

\begin{figure}[t]
\begin{center}
\includegraphics{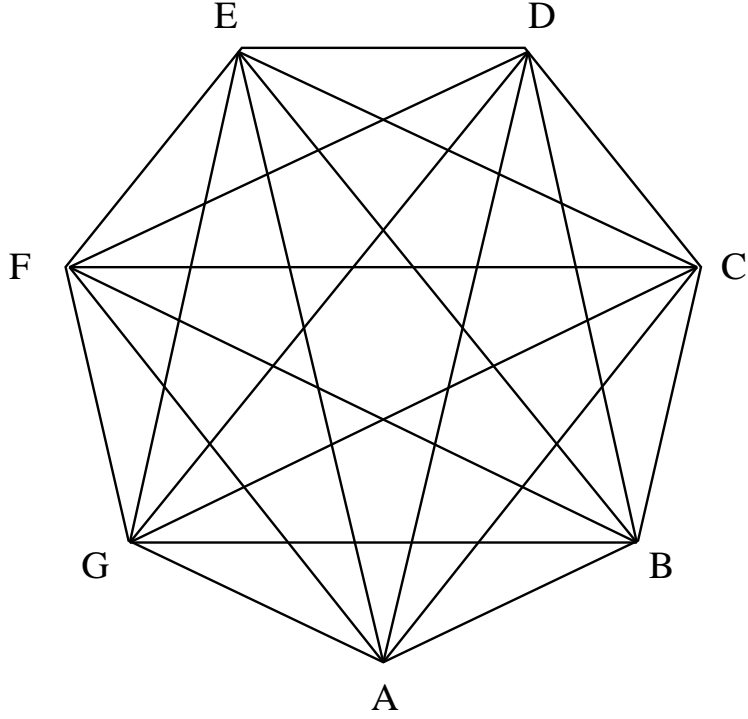}
\end{center}
\caption{\textbf{The $E_{7}$ entanglement diagram.} Each of the seven
vertices A,B,C,D,E,F,G represents a qubit and each of the seven triangles
ABD, BCE, CDF, DEG, EFA, FGB, GAC describes a \textit{tripartite entanglement%
}. \protect\cite{duff1}}
\end{figure}

\begin{figure}[h!]
\begin{center}
\includegraphics[width=0.7\textwidth,height=0.5\textheight]
{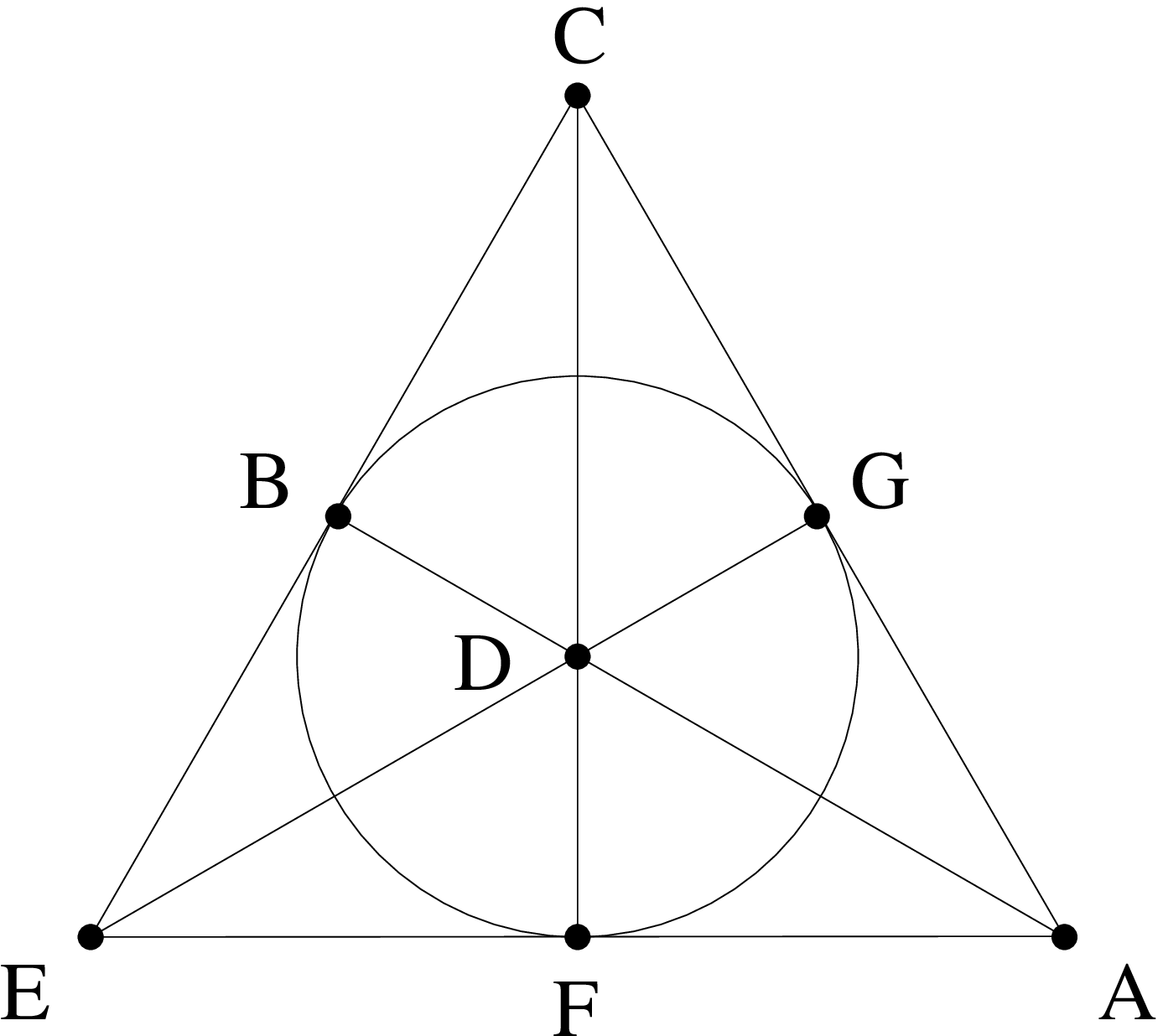}
\end{center}
\caption{The \textit{Fano plane} has seven points, representing the seven
qubits, and seven lines (the circle counts as a line) with three points on
every line, representing the \textit{tripartite entanglement}, and three
lines through every point. \protect\cite{duff1}}
\end{figure}

In $\mathcal{N}=8$, $d=4$ supergravity the (BH) charge vector contains $28$
electric and $28$ magnetic charges, and it sits in the fundamental
representation $\mathbf{56}$ of the $U$-duality group $E_{7(7)}$. The BH
hole entropy reads \cite{kallosh-kol}
\begin{equation}
S_{BH}=\pi \sqrt{|\mathcal{J}_{4}|},
\end{equation}
where $\mathcal{J}_{4}$ is Cartan's quartic $E_{7}$-invariant \cite{cartan},
which can be written as follows:
\begin{equation}
\mathcal{J}_{4}=P^{ij}Q_{jk}P^{kl}Q_{li}-\frac{1}{4}P^{ij}Q_{ij}P^{kl}Q_{kl}+%
\frac{1}{96}(\epsilon ^{ijklmnop}Q_{ij}Q_{kl}Q_{mn}Q_{op}+\epsilon
_{ijklmnop}P^{ij}P^{kl}P^{mn}P^{op}),
\end{equation}
$P^{ij}$ and $Q_{jk}$ being $8\times 8$ antisymmetric charge matrices.

The qubit interpretation of $\mathcal{J}_{4}$ arises from realizing the
existence of the following embedding:
\begin{equation}
E_{7}\left( \mathbb{C}\right) \supsetneq \lbrack SL(2,\mathbb{C})]^{7},
\end{equation}
under which the $\mathbf{56}$ of $E_{7}\left( \mathbb{C}\right) $ branches
as
\begin{equation}
\begin{array}{l}
\mathbf{56}\rightarrow (\mathbf{2},\mathbf{2},\mathbf{1},\mathbf{2},\mathbf{1%
},\mathbf{1},\mathbf{1})+(\mathbf{1},\mathbf{2},\mathbf{2},\mathbf{1},%
\mathbf{2},\mathbf{1},\mathbf{1})+(\mathbf{1},\mathbf{1},\mathbf{2},\mathbf{2%
},\mathbf{1},\mathbf{2},\mathbf{1})+(\mathbf{1},\mathbf{1},\mathbf{1},%
\mathbf{2},\mathbf{2},\mathbf{1},\mathbf{2})+ \\
\\
~~~~~+(\mathbf{2},\mathbf{1},\mathbf{1},\mathbf{1},\mathbf{2},\mathbf{2},%
\mathbf{1})+(\mathbf{1},\mathbf{2},\mathbf{1},\mathbf{1},\mathbf{1},\mathbf{2%
},\mathbf{2})+(\mathbf{2},\mathbf{1},\mathbf{2},\mathbf{1},\mathbf{1},%
\mathbf{1},\mathbf{2}).
\end{array}
\end{equation}
Such a branching decomposition suggests the introduction os a system of
seven qubits (Alice, Bob, Charlie, Daisy, Emma, Fred and George), described
by the quantum state
\begin{eqnarray}
\left| \Psi \right\rangle &=&a_{ABD}\left| ABD\right\rangle +b_{BCE}\left|
BCE\right\rangle +c_{CDF}\left| CDF\right\rangle +d_{DEG}\left|
DEG\right\rangle +  \notag \\
&&+e_{EFA}\left| EFA\right\rangle +f_{FGB}\left| FGB\right\rangle
+g_{GAC}\left| GAC\right\rangle ,  \label{Sat-4}
\end{eqnarray}
where $A=0,1$, yielding that the corresponding Hilbert space has dimension $%
7\times 2^{3}=56$. The $a,b,c,d,e,f,g$ transform as a $\mathbf{56}$ of $E(%
\mathbb{C})$.

The entanglement of the system of seven qubits can be represented by a
heptagon where the vertices A,B,C,D,E,F,G represent the seven qubits and the
seven triangles ABD, BCE, CDF, DEG, EFA, FGB, GAC represent the \textit{%
tripartite entanglement} (see Fig. 4). Alternatively, one can use the
\textit{Fano plane}, (see Fig. 5) corresponding to the multiplication table
of the octonions $\mathbb{O}$.

The so-called \textit{tripartite entanglement of the seven qubits} \textit{%
system} described by the state (\ref{Sat-4}) is measured by the \textit{%
3-tangle}
\begin{equation}
\tau _{3}(ABCDEFG)\equiv 4|\mathcal{J}_{4}|,
\end{equation}
with
\begin{equation}
\mathcal{J}_{4}\sim \left\{
\begin{array}{l}
a^{4}+b^{4}+c^{4}+d^{4}+e^{4}+f^{4}+g^{4}+ \\
\\
+2[a^{2}b^{2}+b^{2}c^{2}+c^{2}d^{2}+d^{2}e^{2}+e^{2}f^{2}+f^{2}g^{2}+g^{2}a^{2}+a^{2}c^{2}+b^{2}d^{2}+c^{2}e^{2}+
\\
\\
+d^{2}f^{2}+e^{2}g^{2}+f^{2}a^{2}+g^{2}b^{2}+a^{2}d^{2}+b^{2}e^{2}+c^{2}f^{2}+d^{2}g^{2}+e^{2}a^{2}+f^{2}b^{2}+g^{2}c^{2}]+
\\
\\
+8[bcdf+cdeg+defa+efgb+fgac+gabd+abce],
\end{array}
\right.
\end{equation}
where products like
\begin{eqnarray}
a^{4} &\equiv &(ABD)(ABD)(ABD)(ABD)=  \notag \\
&=&\epsilon ^{A_{1}A_{2}}\epsilon ^{B_{1}B_{2}}\epsilon
^{D_{1}D_{4}}\epsilon ^{A_{3}A_{4}}\epsilon ^{B_{3}B_{4}}\epsilon
^{D_{2}D_{3}}a_{A_{1}B_{1}D_{1}}a_{A_{2}B_{2}D_{2}}a_{A_{3}B_{3}D_{3}}a_{A_{4}B_{4}D_{4}}
\end{eqnarray}
exclude four individuals (here Charlie, Emma, Fred and George), products
like
\begin{eqnarray}
a^{2}b^{2} &\equiv &\frac{1}{2}(ABD)(ABD)(BCE)(BCE)=  \notag \\
&=&\frac{1}{2}\epsilon ^{A_{1}A_{2}}\epsilon ^{B_{1}B_{3}}\epsilon
^{D_{1}D_{2}}\epsilon ^{B_{2}B_{4}}\epsilon ^{C_{3}C_{4}}\epsilon
^{E_{3}E_{4}}a_{A_{1}B_{1}D_{1}}a_{A_{2}B_{2}D_{2}}b_{B_{3}C_{3}E_{3}}b_{B_{4}C_{4}E_{4}}
\end{eqnarray}
exclude two individuals (here Charlie and Emma), and products like
\begin{eqnarray}
abce &\equiv &(ABD)(BCE)(CDF)(EFA)=  \notag \\
&=&\epsilon ^{A_{1}A_{4}}\epsilon ^{B_{1}B_{2}}\epsilon
^{C_{2}C_{3}}\epsilon ^{D_{1}D_{3}}\epsilon ^{E_{2}E_{4}}\epsilon
^{F_{3}F_{4}}a_{A_{1}B_{1}D_{1}}b_{B_{2}C_{2}E_{2}}c_{C_{3}D_{3}F_{3}}e_{E_{4}F_{4}A_{4}}
\end{eqnarray}
exclude one individual (here George).

Analogously to the cases treated above, \textit{large} non-BPS, \textit{small%
} BPS and \textit{large} ($\frac{1}{8}$-)BPS $\mathcal{N}=8$, $d=4$ BHs
correspond to states with $\mathcal{J}_{4}<0$, $\mathcal{J}_{4}=0$ and $%
\mathcal{J}_{4}>0$, respectively. \setcounter{equation}0

\section{\label{Sect10} Recent Developments:\newline
Moduli Spaces of Attractors}

In $\mathcal{N}=2$ homogeneous (not necessarily symmetric) and $\mathcal{N}%
>2 $-extended, $d=4$ supergravities the Hessian matrix of $V_{BH}$ at its
critical points is in general semi-positive definite, eventually with some
vanishing eigenvalues (\textit{massless Hessian modes}), which actually are
\textit{flat} directions of $V_{BH}$ itself \cite{Ferrara-Marrani-1,ferrara4}%
. Thus, it can be stated that for all supergravities based on homogeneous
scalar manifolds the critical points of $V_{BH}$ which are \textit{%
non-degenerate} (\textit{i.e.} for which it holds $V_{BH}\neq 0$) all are
\textit{stable}, up to some eventual \textit{flat} directions.

The Attractor Equations of $\mathcal{N}=2$, $d=4$ supergravity coupled to $%
n_{V}$ Abelian vector multiplets may have \textit{flat} directions in the
non-BPS cases \cite{Ferrara-Marrani-1,ferrara4}, but \textit{not} in the $%
\frac{1}{2}$-BPS one \cite{ferrara3}. Indeed, in the $\frac{1}{2}$-BPS case
(satisfying $Z\neq 0$, $D_{i}Z=0$ $\forall i=1,...,n_{V}$) the covariant $%
2n_{V}\times 2n_{V}$ Hessian matrix of $V_{BH}$ reads \cite{ferrara3}
\begin{equation}
\left( D_{\widehat{i}}D_{\widehat{j}}V_{BH}\right) _{\mathcal{N}=2,\frac{1}{2%
}-BPS}=\frac{1}{2}\left| Z\right| _{\frac{1}{2}-BPS}\left(
\begin{array}{ccc}
0 &  & G_{i\overline{j}} \\
&  &  \\
G_{j\overline{i}} &  & 0
\end{array}
\right) _{\frac{1}{2}-BPS},  \label{b-sunday}
\end{equation}
where hatted indices can be either holomorphic or anti-holomorphic; thus, as
far as the metric $G_{i\overline{j}}$ of the scalar manifold is strictly
positive definite, Eq. (\ref{b-sunday}) yields that no \textit{massless} $%
\frac{1}{2}$-BPS \textit{Hessian modes} arise out.
\begin{table}[t]
\begin{center}
\begin{tabular}{|c||c|c|c|}
\hline
& $
\begin{array}{c}
\\
\frac{\widehat{H} }{\widehat {h}} \\
~
\end{array}
$ & $
\begin{array}{c}
\\
\text{r} \\
~
\end{array}
$ & $
\begin{array}{c}
\\
\text{\textit{dim}}_{\mathbb{R}} \\
~
\end{array}
$ \\ \hline\hline
$
\begin{array}{c}
\\
\mathbb{R} \oplus \Gamma_{n}, n \in \mathbb{N} \\
~
\end{array}
$ & $SO(1,1)\otimes\frac{SO(1,n-1)}{SO(n-1)}~$ & $
\begin{array}{c}
\\
1 (n=1) \\
2 (n \geqslant 2 ) \\
~
\end{array}
$ & $n~$ \\ \hline
$
\begin{array}{c}
\\
J_{3}^{\mathbb{O} } \\
~
\end{array}
$ & $\frac{E_{6(-26)}}{F_{4(-52)}}~$ & $2~$ & $6~$ \\ \hline
$
\begin{array}{c}
\\
J_{3}^{\mathbb{H} } \\
~
\end{array}
$ & $\frac{SU^{*}(6)}{USp(6)}~$ & $2~$ & $14~$ \\ \hline
$
\begin{array}{c}
\\
J_{3}^{\mathbb{C} } \\
~
\end{array}
$ & $\frac{SL(3,C)}{SU(3)}~$ & $2~$ & $8~$ \\ \hline
$
\begin{array}{c}
\\
J_{3}^{\mathbb{R} } \\
~
\end{array}
$ & $\frac{SL(3, \mathbb{R})}{SO(3)}~$ & $2~$ & $5$ \\ \hline
\end{tabular}
\end{center}
\caption{\textbf{Moduli spaces of non-BPS }$Z\neq 0$ \textbf{\ critical
points of } $V_{BH,\mathcal{N}=2}$ \textbf{in }$\mathcal{N}=2,d=4$ \textbf{%
homogeneous symmetric supergravities. They are the} $\mathcal{N}=2,d=5$
\textbf{homogeneous symmetric real special manifolds} \protect\cite{ferrara4}
}
\end{table}

\begin{table}[h]
\begin{center}
\begin{tabular}{|c||c|c|c|}
\hline
& $
\begin{array}{c}
\\
\frac{\widetilde{H} }{\widetilde {h}} =\frac{\widetilde{H} }{\widetilde
{h}^{\prime}\otimes U(1) } \\
~
\end{array}
$ & $
\begin{array}{c}
\\
\text{r} \\
~
\end{array}
$ & $
\begin{array}{c}
\\
\text{\textit{dim}}_{\mathbb{C}} \\
~
\end{array}
$ \\ \hline\hline
$
\begin{array}{c}
\\
quadratic\; sequence \\
n \in \mathbb{N} \\
~
\end{array}
$ & $\frac{SU(1,n-1)}{U(1)\otimes SU(n-1)}~$ & $1~$ & $n-1~$ \\ \hline
$
\begin{array}{c}
\\
\mathbb{R} \oplus \Gamma_{n}, n \in \mathbb{N} \\
~
\end{array}
$ & $\frac{SO(2,n-2)}{SO(2)\otimes SO(n-2)}, n\geqslant 3~$ & $
\begin{array}{c}
\\
1 (n=3) \\
2 (n \geqslant 4 ) \\
~
\end{array}
$ & $n-2~$ \\ \hline
$
\begin{array}{c}
\\
J_{3}^{\mathbb{O} } \\
~
\end{array}
$ & $\frac{E_{6(-14)}}{SO(10)\otimes U(1)}~$ & $2~$ & $16~$ \\ \hline
$
\begin{array}{c}
\\
J_{3}^{\mathbb{H} } \\
~
\end{array}
$ & $\frac{SU(4,2)}{SU(4)\otimes SU(2)\otimes U(1)}~$ & $2~$ & $8~$ \\ \hline
$
\begin{array}{c}
\\
J_{3}^{\mathbb{C} } \\
~
\end{array}
$ & $\frac{SU(2,1)}{SU(2)\otimes U(1)}\otimes \frac{SU(1,2)}{SU(2)\otimes
U(1)}~$ & $2~$ & $4~$ \\ \hline
$
\begin{array}{c}
\\
J_{3}^{\mathbb{R} } \\
~
\end{array}
$ & $\frac{SU(2,1)}{SU(2)\otimes U(1) }~$ & $1~$ & $2~$ \\ \hline
\end{tabular}
\end{center}
\caption{\textbf{Moduli spaces of non-BPS }$Z=0$ \textbf{\ critical points
of } $V_{BH,\mathcal{N}=2}$ \textbf{in }$\mathcal{N}=2,d=4$ \textbf{%
homogeneous symmetric supergravities. They are (non-special) homogeneous
symmetric K\"{a}hler manifolds} \protect\cite{ferrara4}}
\end{table}

Tables 1 and 2 respectively list the moduli spaces of non-BPS $Z\neq 0$ and
non-BPS $Z=0$ attractors for homogeneous symmetric $\mathcal{N}=2$, $d=4$
special geometries, for which a complete classification is available \cite
{ferrara4}. Notice that the non-BPS $Z\neq 0$ moduli spaces are nothing but
the symmetric real special scalar manifolds of the corresponding $\mathcal{N}%
=2$, $d=5$ supergravity.

Within the symmetric $\mathcal{N}=2$, $d=4$ supergravities, there are some
remarkable models in which no non-BPS \textit{flat }directions exist at all.

The unique $n_{V}=1$ models are the so-called $t^{2}$ and $t^{3}$ models;
they are based on the rank-$1$ scalar manifold $\frac{SU\left( 1,1\right) }{%
U\left( 1\right) }$, but with different holomorphic prepotential functions.
As mentioned at the end of Sect. \ref{Sect6}, the $t^{2}$ model is the first
element ($n=1$) of the sequence of irreducible symmetric special K\"{a}hler
manifolds $\frac{SU\left( 1,n\right) }{U\left( 1\right) \times SU\left(
n\right) }$ ($n_{V}=n$, $n\in \mathbb{N}$) (see \textit{e.g.} \cite
{bellucci1} and Refs. therein), endowed with \textit{quadratic}
prepotential. Let us recall once again that the bosonic sector of the $t^{2}$
model is given by the $\left( U\left( 1\right) \right) ^{6}\rightarrow
\left( U\left( 1\right) \right) ^{2}$ truncation of
Maxwell-Einstein-axion-dilaton (super)gravity, treated in Sects. \ref{Sect6}
and \ref{Sect7}. On the other hand, the $t^{3}$ model has \textit{cubic}
prepotential; it is an \textit{isolated case} in the classification of
symmetric special K\"{a}hler manifolds (see \textit{e.g.} \cite{CFG}), but
can be thought also as the $\mathit{s=t=u}$\textit{\ degeneration }of the $%
stu$ model. It is worth pointing out that the $t^{2}$ and $t^{3}$ models are
based on the same rank-$1$ special K\"{a}hler manifold, with different
constant scalar curvature, which respectively can be computed to be (see
\textit{e.g.} \cite{BFM-SIGRAV06} and Refs. therein)
\begin{equation}
\begin{array}{l}
\frac{SU(1,1)}{U(1)},~t^{2}~\text{\textit{model}}:\mathcal{R}=-2; \\
\\
\frac{SU(1,1)}{U(1)},\text{~}t^{3}~\text{\textit{model}}:\mathcal{R}=-\frac{2%
}{3}.
\end{array}
\end{equation}

Beside the BPS attractors, the $t^{2}$ model admits only non-BPS $Z=0$
critical points of $V_{BH}$ with no \textit{flat} directions. Analogously,
the $t^{3}$ model admits only non-BPS $Z\neq 0$ critical points of $V_{BH}$
with no \textit{flat} directions.

For $n_{V}>1$, the non-BPS $Z\neq 0$ critical points of $V_{BH}$, if any,
all have \textit{flat} directions, and thus a related moduli space (see
Table 1). However, models with no non-BPS $Z=0$ \textit{flat} directions at
all and $n_{V}>1$ exist, namely they are the first and second element ($n=1$%
, $2$) of the sequence of reducible symmetric special K\"{a}hler manifolds $%
\frac{SU\left( 1,1\right) }{U\left( 1\right) }\times \frac{SO\left(
2,n\right) }{SO\left( 2\right) \times SO\left( n\right) }$ ($n_{V}=n+1$, $%
n\in \mathbb{N}$) (see \textit{e.g.} \cite{bellucci1} and Refs. therein),
\textit{i.e.} the so-called $st^{2}$ and $stu$ models, respectively. The $%
stu $ model, which is relevant for the analogy between stringy extremal BHs
and quantum information theory treated in Sect. \ref{Sect9}, has two non-BPS
$Z\neq 0$ \textit{flat} directions, spanning the moduli space $SO\left(
1,1\right) \times SO\left( 1,1\right) $ (\textit{i.e.} the scalar manifold
of the $stu$ model in $d=5$), but \textit{no} non-BPS $Z=0$ \textit{massless
Hessian modes} at all. On the other hand, the $st^{2}$ model (which can be
thought as the $\mathit{t=u}$\textit{\ degeneration} of the $stu$ model) has
one non-BPS $Z\neq 0$ \textit{flat} direction, spanning the moduli space $%
SO\left( 1,1\right) $ (\textit{i.e.} the scalar manifold of the $st^{2}$
model in $d=5$), but \textit{no} non-BPS $Z=0$ \textit{flat} direction at
all. The $st^{2}$ is the \textit{''smallest''} symmetric model exhibiting a
non-BPS $Z\neq 0$ \textit{flat} direction.

Concerning the \textit{''smallest''} symmetric models exhibiting a non-BPS $%
Z=0$ \textit{flat} direction they are the second ($n=2$) element of the
sequence $\frac{SU\left( 1,n\right) }{U\left( 1\right) \times SU\left(
n\right) }$ and the third ($n=3$) element of the sequence $\frac{SU\left(
1,1\right) }{U\left( 1\right) }\times \frac{SO\left( 2,n\right) }{SO\left(
2\right) \times SO\left( n\right) }$. In both cases, the unique non-BPS $Z=0$
\textit{flat} direction spans the non-BPS $Z=0$ moduli space $\frac{SU\left(
1,1\right) }{U\left( 1\right) }\sim \frac{SO\left( 2,1\right) }{SO\left(
2\right) }$ (see Table 2), whose local geometrical properties however differ
in the two cases (for the same reasons holding for the $t^{2}$ and $t^{3}$
models treated above).

\begin{table}[t]
\begin{center}
\begin{tabular}{|c||c|c|c|}
\hline
& $
\begin{array}{c}
\\
\frac{1}{\mathcal{N}}\text{-BPS orbits } \frac{G}{\mathcal{H}} \\
~
\end{array}
$ & $
\begin{array}{c}
\\
\text{non-BPS, }Z_{AB}\neq 0\text{ orbits}~\frac{G}{\widehat{\mathcal{H}}}
\\
~
\end{array}
$ & $
\begin{array}{c}
\\
\text{non-BPS, }Z_{AB}=0\text{ orbits }\frac{G}{\widetilde{\mathcal{H}}} \\
\\
~
\end{array}
$ \\ \hline\hline
$
\begin{array}{c}
\\
\mathcal{N}=3 \\
~
\end{array}
$ & $\frac{SU(3,n)}{SU(2,n)}~$ & $-$ & $\frac{SU(3,n)}{SU(3,n-1)}~$ \\ \hline
$
\begin{array}{c}
\\
\mathcal{N}=4 \\
~
\end{array}
$ & $\frac{SU(1,1)}{U(1)}\otimes \frac{SO(6,n)}{SO(4,n)}~$ & $\frac{SU(1,1)}{%
SO(1,1)}\otimes \frac{SO(6,n)}{SO(5,n-1)}~$ & $\frac{SU(1,1)}{U(1)}\otimes
\frac{SO(6,n)}{SO(6,n-2)}$ \\ \hline
$
\begin{array}{c}
\\
\mathcal{N}=5 \\
~
\end{array}
$ & $\frac{SU(1,5)}{SU(3)\otimes SU\left( 2,1\right) }$ & $-$ & $-$ \\ \hline
$
\begin{array}{c}
\\
\mathcal{N}=6 \\
~
\end{array}
$ & $\frac{SO^{\ast }(12)}{SU(4,2)}~$ & $\frac{SO^{\ast }(12)}{SU^{\ast }(6)}%
~$ & $\frac{SO^{\ast }(12)}{SU(6)}~$ \\ \hline
$
\begin{array}{c}
\\
\mathcal{N}=8 \\
~
\end{array}
$ & $\frac{E_{7\left( 7\right) }}{E_{6\left( 2\right) }}$ & $\frac{%
E_{7\left( 7\right) }}{E_{6\left( 6\right) }}$ & $-~$ \\ \hline
\end{tabular}
\end{center}
\caption{\textbf{Charge orbits of the real, symplectic }$R$ \textbf{%
representation of the }$U$\textbf{-duality group }$G$ \textbf{supporting BH
attractors with non-vanishing entropy in $3\leqslant \mathcal{N}\leqslant 8$%
, $d=4$ supergravities} \protect\cite{bellucci2}}
\end{table}

\begin{table}[t]
\begin{center}
\begin{tabular}{|c||c|c|c|}
\hline
& $
\begin{array}{c}
\\
\frac{1}{\mathcal{N}}\text{-BPS} \\
\text{moduli space }\frac{\mathcal{H}}{\frak{h}}\text{ } \\
~
\end{array}
$ & $
\begin{array}{c}
\\
\text{non-BPS, }Z_{AB}\neq 0 \\
\text{moduli space }\frac{\widehat{\mathcal{H}}}{\widehat{\frak{h}}} \\
~
\end{array}
$ & $
\begin{array}{c}
\\
\text{non-BPS, }Z_{AB}=0 \\
\text{moduli space }\frac{\widetilde{\mathcal{H}}}{\widetilde{\frak{h}}} \\
~
\end{array}
$ \\ \hline\hline
$
\begin{array}{c}
\\
\mathcal{N}=3 \\
~
\end{array}
$ & $\frac{SU(2,n)}{SU(2)\otimes SU\left( n\right) \otimes U\left( 1\right) }%
~$ & $-$ & $\frac{SU(3,n-1)}{SU(3)\otimes SU\left( n-1\right) \otimes
U\left( 1\right) }~$ \\ \hline
$
\begin{array}{c}
\\
\mathcal{N}=4 \\
~
\end{array}
$ & $\frac{SO(4,n)}{SO(4)\otimes SO\left( n\right) }~$ & $SO(1,1)\otimes
\frac{SO(5,n-1)}{SO(5)\otimes SO\left( n-1\right) }~$ & $\frac{SO(6,n-2)}{%
SO(6)\otimes SO\left( n-2\right) }$ \\ \hline
$
\begin{array}{c}
\\
\mathcal{N}=5 \\
~
\end{array}
$ & $\frac{SU\left( 2,1\right) }{SU\left( 2\right) \otimes U\left( 1\right) }
$ & $-$ & $-$ \\ \hline
$
\begin{array}{c}
\\
\mathcal{N}=6 \\
~
\end{array}
$ & $\frac{SU(4,2)}{SU(4)\otimes SU\left( 2\right) \otimes U\left( 1\right) }%
~$ & $\frac{SU^{\ast }(6)}{USp\left( 6\right) }~$ & $-$ \\ \hline
$
\begin{array}{c}
\\
\mathcal{N}=8 \\
~
\end{array}
$ & $\frac{E_{6\left( 2\right) }}{SU\left( 6\right) \otimes SU\left(
2\right) }$ & $\frac{E_{6\left( 6\right) }}{USp\left( 8\right) }$ & $-~$ \\
\hline
\end{tabular}
\end{center}
\caption{\textbf{Moduli spaces of BH attractors with non-vanishing entropy
in $3\leqslant \mathcal{N}\leqslant 8$, $d=4$ supergravities (}$\frak{h}$%
\textbf{, }$\widehat{\frak{h}}$\textbf{\ and }$\widetilde{\frak{h}}$\textbf{%
\ are maximal compact subgroups of }$\mathcal{H}$\textbf{, }$\widehat{%
\mathcal{H}}$\textbf{\ and }$\widetilde{\mathcal{H}}$\textbf{, respectively)}
\protect\cite{bellucci2}}
\end{table}

In $\mathcal{N}>2$-extended, $d=4$ supergravities there are \textit{flat}
directions of $V_{BH}$ at both its \textit{non-degenerate} BPS and non-BPS
critical points. Group-theoretically, this is due to the fact that the
corresponding supporting BH \textit{charge orbits} always have a \textit{%
non-compact} stabilizer \cite{ferrara4,bellucci2}. The BPS \textit{flat}
directions can be interpreted in terms of left-over hypermultiplets' scalar
degrees of freedom in the truncation down to the $\mathcal{N}=2$, $d=4$
theories \cite{ADF-U-duality-d=4,Ferrara-Marrani-1}. In Tables 3 and 4 all
\textit{charge orbits} and the corresponding moduli spaces of attractor
solution in $\mathcal{N}>2$-extended, $d=4$ supergravities are reported \cite
{bellucci2}.

We conclude by pointing out that all the reported results hold at the
classical, Einstein supergravity level. It is conceivable that the \textit{%
flat} directions of classical \textit{non-degenerate} extremal BH attractors
will be removed (\textit{i.e.} lifted) by quantum (perturbative and
non-perturbative) corrections (such as the ones coming from higher-order
derivative contributions to the gravity and/or gauge sector) to the \textit{%
classical} effective BH potential $V_{BH}$. Consequently, \textit{at the
quantum} (perturbative and non-perturbative) \textit{level, no moduli spaces
for attractor solutions might exist at all} (and therefore also \textit{the
actual attractive nature of the critical points of }$V_{BH}$\textit{\ might
be destroyed}). \textit{However, this might not be the case for }$\mathcal{N}%
=8$.

In presence of \textit{quantum} lifts of \textit{classically flat}
directions of the Hessian matrix of $V_{BH}$ at its critical points, in
order to answer to the key question: \textit{'Do extremal BH attractors (in
a strict sense) survive the quantum level?'}, it is thus crucial to
determine whether such lifts originate Hessian modes with \textit{positive}
squared mass (corresponding to \textit{attractive} directions) or with
\textit{negative} squared mass (\textit{i.e.} \textit{tachyonic}, \textit{%
repeller} directions).

The fate of the unique non-BPS $Z\neq 0$ flat direction of the $st^{2}$
model in presence of the most general class of quantum perturbative
corrections consistent with the axionic-shift symmetry has been studied in
\cite{BFMS2}, showing that, as intuitively expected, the \textit{classical
solutions get lifted at the quantum level}. Interestingly, in \cite{BFMS2}%
\textbf{\ }it is found the \textit{quantum} lift occurs more often towards
\textit{repeller} directions (thus destabilizing the whole critical
solution, and \textit{destroying the attractor in strict sense}), rather
than towards \textit{attractive} directions.\textbf{\ }The same behavior may
be expected for the unique non-BPS $Z=0$\ flat direction of the $n=2$\
element of the quadratic irreducible sequence and the $n=3$\ element of the
cubic reducible sequence (see above).

Generalizing to the presence of more than one \textit{flat} direction, this
would mean that \textit{only a (very) few classical attractors do remain
attractors in strict sense at the quantum level}; consequently, \textit{at
the quantum} (perturbative and non-perturbative) \textit{level the
``landscape'' of extremal BH attractors should be strongly constrained and
reduced}.

\section*{Acknowledgments}

The original parts of the contents of these lectures result from
collaborations with L. Andrianopoli, S. Bellucci, R. D'Auria, E. Gimon, M.
Trigiante, and especially M. J. Duff, G. Gibbons, M. G\"{u}naydin, R.
Kallosh and A. Strominger, which are gratefully acknowledged.

A. M. would like to thank the Department of Physics, Theory Unit Group at
CERN, where part of this work was done, for kind hospitality and stimulating
environment.

The work of S.F. has been supported in part by the European Community Human
Potential Programme under contract MRTN-CT-2004-503369 ``\textit{%
Constituents, Fundamental Forces and Symmetries of the Universe''} in
association with LNF-INFN and in part by DOE grant DE-FG03-91ER40662, Task C.

The work of K.H. and A.M. has been supported by Museo Storico della Fisica e
Centro Studi e Ricerche ``\textit{Enrico Fermi''}, Rome, Italy, in
association with INFN-LNF.

\end{document}